\documentstyle[12pt]{article}
\begin{document}
\title{{\bf MAXIMAL ACCELERATION IS NONROTATING}
\thanks{Alberta-Thy-11-97, gr-qc/9706029,
accepted 1998 Feb. 27 for Classical and Quantum Gravity}}
\author{
Don N. Page
\thanks{Internet address:
don@phys.ualberta.ca}
\\
CIAR Cosmology Program, Institute for Theoretical Physics\\
Department of Physics, University of Alberta\\
Edmonton, Alberta, Canada T6G 2J1
}
\date{(1998 March 2)}

\maketitle
\large

\newcommand{\bk}{\mbox{${\bf k}$}}
\newcommand{\bK}{\mbox{${\bf K}$}}
\newcommand{\bl}{\mbox{${\bf l}$}}
\newcommand{\bA}{\mbox{${\bf A}$}}
\newcommand{\bd}{\mbox{${\bf d}$}}
\newcommand{\bu}{\mbox{${\bf u}$}}
\newcommand{\bU}{\mbox{${\bf U}$}}
\newcommand{\buc}{\mbox{${\bf u}_C$}}
\newcommand{\ba}{\mbox{${\bf a}$}}
\newcommand{\bap}{\mbox{${\bf a'}$}}
\newcommand{\bapp}{\mbox{${\bf a''}$}}
\newcommand{\bah}{\mbox{${\bf \hat{a}}$}}
\newcommand{\bat}{\mbox{${\bf \tilde{a}}$}}
\newcommand{\bv}{\mbox{${\bf v}$}}
\newcommand{\bF}{\mbox{${\bf F}$}}
\newcommand{\bff}{\mbox{${\bf f}$}}
\newcommand{\bL}{\mbox{${\bf L}$}}
\newcommand{\bLp}{\mbox{${\bf L'}$}}
\newcommand{\bb}{\mbox{${\bf b}$}}
\newcommand{\btu}{\mbox{${\bf \tilde{u}}$}}
\newcommand{\bo}{\mbox{${\bf \omega}$}}
\newcommand{\bvp}{\mbox{${\bf v'}$}}
\newcommand{\bp}{\mbox{${\bf p}$}}
\newcommand{\ber}{\mbox{${\bf e}^r$}}
\newcommand{\bez}{\mbox{${\bf e}^z$}}
\newcommand{\bxi}{\mbox{${\bf \xi}$}}

\begin{abstract}
\baselineskip 15.5 pt

	In a stationary axisymmetric spacetime,
the angular velocity of a stationary observer
that Fermi-Walker transports its acceleration vector
is also the angular velocity that locally extremizes 
the magnitude of the acceleration of such an observer, 
and conversely if the spacetime is also symmetric under 
reversing both $t$ and $\varphi$ together.
Thus a congruence of
Nonrotating Acceleration Worldlines (NAW)
is equivalent to a
Stationary Congruence Accelerating Locally Extremely
(SCALE).
These congruences are defined completely locally,
unlike the case of Zero Angular Momentum Observers
(ZAMOs), which requires knowledge
around a symmetry axis.
The SCALE subcase of a
Stationary Congruence Accelerating Maximally (SCAM)
is made up of stationary worldlines that may
be considered to be locally most nearly at rest
in a stationary axisymmetric gravitational field.
Formulas for the angular velocity and other properties
of the SCALEs are given explicitly on
a generalization of an equatorial plane,
infinitesimally near a symmetry axis, and in
a slowly rotating gravitational field, including
the far-field limit, where the SCAM
is shown to be counterrotating relative to infinity.
These formulas are evaluated in particular detail
for the Kerr-Newman metric.
Various other congruences are also defined,
such as a Stationary Congruence Rotating at Minimum
(SCRAM), and Stationary Worldlines Accelerating
Radially Maximally (SWARM),
both of which coincide with a SCAM
on an equatorial plane of reflection symmetry.
Applications are also made to the gravitational fields
of maximally rotating stars, the Sun, and the Solar System.

\end{abstract}
\normalsize
\baselineskip 15.5 pt
\newpage

\section{Introduction}

	A stationary axisymmetric spacetime has various
preferred congruences of stationary observers,  
such as those whose four-velocities are parallel
to the Killing vector field
that is timelike at radial infinity 
(wordlines nonrotating relative to infinity),
or those whose four-velocities are perpendicular
to the Killing vector field
which has closed orbits and which vanishes on 
the symmetry axis (Zero Angular Momentum Observers,
or ZAMOs) \cite{Bardeen,MTW}.
Here a new preferred congruence is defined
(SCAM, a special case of SCALE = NAW) in terms of
the purely local properties of the commuting
Killing vector fields, without reference to what
they do elsewhere (e.g., at radial infinity
or around the symmetry axis).

	Using the MTW sign conventions \cite{MTW} ---
in particular, the metric sign convention (-+++) ---
and the same boldface symbols 
for vectors and for the corresponding one-forms
that have components obtained by using 
the metric tensor to lower the vector components, 
consider a region of spacetime with two independent 
Killing vector fields, vector fields
 \begin{equation}
 \bk = k^{\alpha}\partial /\partial x^{\alpha}, \;\; 
 \bl = l^{\alpha}\partial /\partial x^{\alpha},
 \label{eq:1}
 \end{equation}
that are independent (not obeying $a\bk+b\bl=0$ 
for any constants $a$ and $b$ not both zero) and 
whose corresponding 1-form components
 \begin{equation}
 k_{\alpha} = g_{\alpha\beta}k^{\beta}, \;\; 
 l_{\alpha} = g_{\alpha\beta}l^{\beta},
 \label{eq:2}
 \end{equation}
obey Killing's equation,
 \begin{equation}
 k_{\alpha ;\beta} = -k_{\beta ;\alpha}, \;\; 
 l_{\alpha ;\beta} = -l_{\beta ;\alpha}.
 \label{eq:3}
 \end{equation}

	Assume that these two Killing vector fields $\bk$ and $\bl$
also have the following three additional properties
(though only the first two properties are necessary for
the first part of the theorem to be proved):

	(1) The 2-form
 \begin{equation}
 \bA = \bk \wedge \bl
 \label{eq:4}
 \end{equation}
corresponding to the Killing bivector is timelike, obeying
 \begin{equation}
 A^{\alpha\beta}A_{\alpha\beta}
 = 2(\bk\!\cdot\!\bk)(\bl\!\cdot\!\bl) - 2(\bk\!\cdot\!\bl)^2 < 0.
 \label{eq:5}
 \end{equation}
Then in each sufficiently small neighborhood
one can redefine, if necessary, $\bk$ and $\bl$ 
to be two new independent linear combinations 
of the original Killing vectors such
that $\bl$ is spacelike and the orthogonal
vector field $(\bl\!\cdot\!\bl)\bk - (\bk\!\cdot\!\bl)\bl$,
which by Eq. (\ref{eq:5}) is necessarily timelike,
is future pointing (by choosing the appropriate sign
for $\bk$).  Then the linear combination
 \begin{equation}
 \bK = \bk + \Omega \bl
 \label{eq:6}
 \end{equation}
is future-pointing timelike at each point in the
neighborhood for some finite range 
of the constant $\Omega$,
say $\Omega_{-1} < \Omega < \Omega_{+1}$.

	(2)  The two Killing vector fields commute,
 \begin{equation}
 [\bk,\bl] = 0
 \label{eq:7},
 \end{equation}
or, in component form,
 \begin{equation}
 k^{\alpha}l^{\beta}_{\;\; ;\alpha}
 = l^{\alpha}k^{\beta}_{\;\; ;\alpha}.
 \label{eq:8}
 \end{equation}
This implies that one can choose two of the four 
coordinates, say $x^0=t$ and $x^1=\varphi$, such that
 \begin{equation}
 \bk = \partial /\partial t, \;\; 
 \bl = \partial /\partial \varphi.
 \label{eq:9}
 \end{equation}

	(3)  The 2-form $\bA=\bk\wedge\bl$ obeys
 \begin{equation}
 \ast\bA\wedge\ast\bd\bA = 0,
 \label{eq:10}
 \end{equation}
which, when $\bA\neq 0$ holds, 
as is implied by Eq. (\ref{eq:5}),
is equivalent to the orthogonally transitive
or circularity condition \cite{Carter72}
 \begin{equation}
 \bk\wedge\bl\wedge\bd\bk = \bk\wedge\bl\wedge\bd\bl = 0.
 \label{eq:11}
 \end{equation}
In component form it is equivalent to
 \begin{equation}
 A_{\mu[\alpha}A_{\beta\gamma ;\delta]} = 0
 \label{eq:12}
 \end{equation}
or
 \begin{equation}
 k_{[\alpha}l_{\beta}k_{\gamma ;\delta]} =
 l_{[\alpha}k_{\beta}l_{\gamma ;\delta]} = 0.
 \label{eq:13}
 \end{equation}
This condition is equivalent to the condition that
one may construct in a local neighborhood
a family of two-surfaces orthogonal
to both Killing vector fields
\cite{Carter72,KSMH,Wald}.
One may define two coordinates,
say $x^a = (r,\theta)$ for $a = 2,3$,
on these two-surfaces such that orbits 
of the Killing vectors $\bk$ and $\bl$,
and hence of all stationary observers, 
each stay at fixed $x^a$.
Then $\partial/\partial r$ and
$\partial/\partial \theta$
are both orthogonal to
$\bk = \partial /\partial t$ and to 
$\bl = \partial /\partial \varphi$,
so in this coordinate basis the metric tensor
has no components mixing the first two (0 or 1)
and the last two (2 or 3) indices.
I.e., it is block diagonal.
As a result, $\bA$ may be written as
 \begin{equation}
 \bA = -D \bd t \wedge \bd \varphi,
 \label{eq:14}
 \end{equation}
where
 \begin{equation}
 -D \equiv g_{00}g_{11}-g_{01}g_{10}
 \equiv g_{tt}g_{\varphi\varphi}-g_{t\varphi}g_{\varphi t}
 = {1 \over 2}A^{\alpha\beta}A_{\alpha\beta}
 = (\bk\!\cdot\!\bk)(\bl\!\cdot\!\bl) - (\bk\!\cdot\!\bl)^2 < 0
 \label{eq:14b}
 \end{equation}
is the determinant of the first two-dimensional
block of the metric.

	Another simple way to state this
third condition is to say \cite{Chandra}
that the spacetime is invariant under the
simultaneous reversal of both coordinates
$t$ and $\varphi$.  This follows
from the block diagonality of the metric,
and it implies that each of the quantities
in Eq. (\ref{eq:11}) are zero,
since they are odd under this transformation.

	The most important examples
of spacetimes with two independent commuting
Killing vectors obeying these two properties
are asymptotically flat stationary axisymmetric spacetimes
\cite{Carter70}
with the Ricci tensor obeying
 \begin{equation}
 k^{\mu}R_{\mu[\alpha}k_{\beta}l_{\gamma]}
 = l^{\mu}R_{\mu[\alpha}k_{\beta}l_{\gamma]} = 0,
 \label{eq:15}
 \end{equation}
which implies that property (3) above holds
\cite{KundtTrumper66,Pap,Carter69},
though Eq. (\ref{eq:15}) just by itself
does not imply property (3).
In such spacetimes
one may uniquely choose the Killing vector fields
such that $\bk = \partial /\partial t$
is a unit timelike vector field at radial infinity
and $\bl = \partial /\partial \varphi$
is a spacelike vector that vanishes on
the symmetry axes (e.g., at $\theta = 0$ or $\theta = \pi$)
and has closed orbits with period $\Delta\varphi = 2\pi$.
However, the results below apply more generally,
assuming only that $\bk$ and $\bl$ have
$\bk\wedge\bl$ timelike and obey 
Killing's Eq. (\ref{eq:3}),
the commutativity condition $[\bk,\bl] = 0$, and
the two-surface-orthogonality condition
$\ast(\bk\wedge\bl)\wedge\ast\bd(\bk\wedge\bl) = 0$.

	An observer whose four-velocity is
 \begin{equation}
 \bu = (-\bK\!\cdot\!\bK)^{-1/2}\bK
 \label{eq:16}
 \end{equation}
with $\bK = \bk + \Omega\bl$ with fixed 
$\Omega = d\varphi/dt$
(which shall be called the angular velocity,
since that is what it for a stationary axisymmetric spacetime)
may be defined to be a stationary observer (SO).
A stationary congruence of observers (SCO)
is a space-filling family of observers
(one crossing each point of each local spatial hypersurface
in the region of spacetime under consideration)
with four-velocities
 \begin{equation}
 \buc = {(\bk + \Omega(x^a)\bl)\over |\bk + \Omega(x^a)\bl|}
 \label{eq:17}
 \end{equation}
that have the angular velocity $\Omega$
depending only on the two $x^a$ coordinates
that stay fixed along each worldline.

	In order to calculate the acceleration vector
 \begin{equation}
 \ba = \nabla_{\bu}\bu
 \label{eq:18}
 \end{equation}
with $\Omega$ constant along the worldline,
it is convenient to consider the Killing vector field
$\bK = \bk + \Omega\bl$
with this same $\Omega$ fixed as a constant everywhere
(and not having the spatial variation with $x^a$
that a stationary congruence of different worldlines
at different $x^a$ might have).
If for this fixed ($x^a$-independent) $\Omega$ 
one defines the scalar field
 \begin{equation}
 \Phi \equiv \Phi(\Omega) \equiv
 \frac{1}{2}\ln(-\bK\!\cdot\!\bK),
 \label{eq:19}
 \end{equation}
then
 \begin{equation}
 \bu = e^{-\Phi}\bK
 \label{eq:20}
 \end{equation}
is, over the region of spacetime where $\bK$
is (future-pointing) timelike,
the four-velocity of a rigidly rotating 
stationary congruence of observers,
differing from the four-velocities $\buc$ of the
congruence given by Eq. (\ref{eq:17}),
where $\Omega$ is allowed to be a function of $x^a$
(i.e., different angular velocities for different
stationary observers within the congruence,
though I am always taking the angular velocity
$\Omega$ to be fixed for a given observer).

	Then the antisymmetry of the covariant derivative
of the Killing vector field $\bK$ implies that
the covariant components of the acceleration vector are
 \begin{eqnarray}
 a_{\alpha} &=& u^{\beta}u_{\alpha ;\beta}
	= e^{-\Phi}K^{\beta}(e^{-\Phi}K_{\alpha})_{;\beta}
	\nonumber \\
 &=& e^{-2\Phi}(K^{\beta}K_{\alpha ;\beta}
	-K^{\beta}K_{\alpha}\Phi_{;\beta})
	\nonumber \\
 &=& e^{-2\Phi}[-K^{\beta}K_{\beta ;\alpha}
	-K^{\beta}K_{\alpha}\frac{1}{2}
	(-K^{\mu}K_{\mu})_{;\beta}/(-K^{\nu}K_{\nu})]
	\nonumber \\
 &=& e^{-2\Phi}[-\frac{1}{2}(K^{\beta}K_{\beta})_{;\alpha}
	+K_{\alpha}K^{\beta}K^{\mu}K_{\mu ;\beta}/
	(-K^{\nu}K_{\nu})]
	\nonumber \\
 &=& e^{-2\Phi}[-\frac{1}{2}(-e^{2\Phi})_{;\alpha} + 0]
	\nonumber \\
 &=& \Phi_{;\alpha},
 \label{eq:21}
 \end{eqnarray}
which are nonzero only for $\alpha = a = 2$ or 3.  That is,
 \begin{equation}
 \ba = \nabla \Phi
 \label{eq:22}
 \end{equation}
is perpendicular to both $\bk$ and $\bl$.
This fact requires property (2) but not property (3) above.

\section{Stationary Congruence Accelerating \newline
Locally Extremely (SCALE)}

	Now for each value of the pair of coordinates $x^a$,
we would like to find the value of $\Omega$ that extremizes
the magnitude of the acceleration of the corresponding
stationary observer.  For this purpose, it is convenient
to define the following scalar fields (functions of $x^a$):
 \begin{equation}
 A \equiv - \bk \!\cdot\! \bk = -g_{00} \equiv -g_{tt},
 \label{eq:23}
 \end{equation}
 \begin{equation}
 B \equiv - \bk \!\cdot\! \bl = -g_{01} \equiv -g_{t\varphi},
 \label{eq:24}
 \end{equation}
 \begin{equation}
 C \equiv - \bl \!\cdot\! \bl = -g_{11} \equiv -g_{\varphi\varphi},
 \label{eq:25}
 \end{equation}
 \begin{equation}
 F \equiv F(\Omega) \equiv e^{2\Phi} \equiv - \bK \!\cdot\! \bK
	= A + 2\Omega B + \Omega^2 C,
 \label{eq:26}
 \end{equation}
 \begin{equation}
 G \equiv G(\Omega) \equiv
 \frac{1}{2}\frac{\partial}{\partial\Omega}F(\Omega)
 \equiv - \bK \!\cdot\! \bl = B + \Omega C,
 \label{eq:27}
 \end{equation}
where the partial derivative with respect to $\Omega$
is at fixed $x^a$ and so at fixed $A$, $B$, and $C$.
Since we have chosen $\bl$ to be spacelike, $C<0$.
If the source of the axisymmetric gravitational field
is entirely rotating in the same direction, $B$ will
typically have the sign of this direction everywhere,
which by the appropriate choice of the sign of
the coordinate $\varphi$ can be chosen to be positive.
$A$ will be positive outside any ergospheres
but negative inside, if any exist.  However, property (2)
above implies that we are outside any Killing horizon, so
 \begin{equation}
 D\equiv B^2 - AC
 = - {1 \over 2}A^{\alpha\beta}A_{\alpha\beta} > 0.
 \label{eq:27b}
 \end{equation}

	Then
 \begin{equation}
 \Phi = \frac{1}{2}\ln F,
 \label{eq:28}
 \end{equation}
so
 \begin{equation}
 \ba = \nabla\Phi = {\nabla F\over 2F}
 = {\nabla A + 2\Omega \nabla B +\Omega^2 \nabla C
	\over 2(A + 2\Omega B + \Omega^2 C)},
 \label{eq:29}
 \end{equation}
 \begin{equation}
 {\partial \ba \over \partial \Omega} = \nabla(\frac{G}{F})
 = {(A\nabla B - B\nabla A) + \Omega (A\nabla C - C\nabla A)
	+ \Omega^2 (B\nabla C -C\nabla B)
	\over (A + 2\Omega B + \Omega^2 C)^2},
 \label{eq:29b}
 \end{equation}
and thus
 \begin{eqnarray}
 \frac{\partial a^2}{\partial \Omega} \!\!
 &=&\!\! 2\ba\!\cdot\!\frac{\partial\ba}{\partial\Omega}
	=\frac{1}{F}\nabla F\!\cdot\!\nabla(\frac{G}{F})
	=F^{-3}\nabla F\!\cdot\!(F\nabla G - G\nabla F)
	\nonumber \\
 &=& \!\! 2\nabla\Phi\!\cdot\!\
	\nabla(\bK\!\cdot\!\bl/\bK\!\cdot\!\bK)
	= (\bK\!\cdot\!\bK)^{-1}
	\nabla(\bK\!\cdot\!\bK)\!\cdot\!
	\nabla(\bK\!\cdot\!\bl/\bK\!\cdot\!\bK)
	\nonumber \\
 &=& \!\! e^{-6\Phi}(K_{\alpha}l_{\beta}-K_{\beta}l_{\alpha})
	K^{\beta}K^{\gamma}
	K_{\gamma ;\delta}K^{\alpha ;\delta}
	\nonumber \\
 &=&\!\!{[\nabla A\!\! +\!\! 2\Omega \nabla B
	 \!\!+\!\!\Omega^2 \nabla C]\!\cdot\!
	[(A\nabla B\!\! -\!\! B\nabla A)
	\!\!+\!\! \Omega (A\nabla C\!\! -\!\! C\nabla A)
	\!\!+\!\! \Omega^2 (B\nabla C\!\! -\!\! C\nabla B)]
	\over (A + 2\Omega B + \Omega^2 C)^3},
	\nonumber\\
 &&
 \label{eq:30}
 \end{eqnarray}
using $[\bK,\bl] = 0$.

	Setting $\partial a^2/\partial \Omega$ to zero
at each point gives an extremum of 
the magnitude of the acceleration
of a stationary observer there,
with $\Omega$ thus obeying the equation
 \begin{equation}
 [\nabla A + 2\Omega \nabla B +\Omega^2 \nabla C]\cdot
	[(A\nabla B\! -\! B\nabla A)
	 \!+\! \Omega (A\nabla C\! -\! C\nabla A)
	 \!+ \Omega^2 (B\nabla C\! -\! C\nabla B)]
 \!=\! 0.
 \label{eq:30a}
 \end{equation}
Expanded out in powers of the angular velocity $\Omega$,
this is a quartic equation for $\Omega(x^a)$,
with coefficients that are combinations of
$A\equiv -g_{00}$, $B\equiv -g_{01}$, $C\equiv -g_{11}$,
and dot products of their gradients.

	A congruence of stationary worldlines corresponding
to one of the roots of Eq. (\ref{eq:30a})
for $\Omega(x^a)$ at each $x^a$
might be called a Stationary Congruence Accelerating
Locally Extremely (SCALE), and if the local extremum
of the acceleration (as a function of $\Omega$)
is a (local) maximum, the congruence might be called
a Stationary Congruence Accelerating Maximally (SCAM).
Since in the frame of the device (e.g., a rocket)
accelerating the observer,
the magnitude of the acceleration
may be interpreted as the apparent weight
or heaviness of the observer
(e.g., as in saying that an observer in free fall
along a geodesic is ``weightless''),
one might say that an observer moving along
a SCALE is ``extremely heavy,''
taking extreme to mean either a local maximum
(for a SCAM) or a minimum (for the other SCALEs).

	(Generically there is no
global maximum for the acceleration,
since it can be made arbitrarily large by making
$\Omega$ arbitrarily near one of the two endpoints
$\Omega_{-1}$ and $\Omega_{+1}$
of its allowed range,
 \begin{equation}
 \Omega_{\pm 1} = {B\pm\sqrt{B^2 - AC}\over -C},
 \label{eq:30b}
 \end{equation}
where $F$ goes to zero and hence
$\bK = \bk + \Omega\bl$ becomes null,
unless this endpoint corresponds to a null geodesic
where $\nabla F$ also goes to zero,
in which case the acceleration stays finite.
This last fact uses property (1)
in the form of Eq. (\ref{eq:27b}),
$D = B^2 - AC > 0$,
so that the two endpoints have a nonzero
separation in $\Omega$, and hence $F$,
as a quadratic polynomial of $\Omega$
given by Eq. (\ref{eq:26}),
has only a simple zero at each end
and cannot give infinity when divided
into $\nabla F$ if the latter also has a zero
at the corresponding endpoint.)

	Although typically $a^2$ thus has no global
maximum within the allowed range of $\Omega$
where $\bK$ is timelike, there are usually
(at least in weak gravitational fields)
two local minima for $a^2$ and one local maximum
between these two minima, though it is also possible
in strong gravitational fields to have
only one local minimum and no local maximum.

	For a stationary axisymmetric spacetime
in a region of weak gravity, one of the roots
of the quartic Eq. (\ref{eq:30a}) in $\Omega(x^a)$
corresponds to an imaginary spatial four-velocity
$\bu = i(\bK\!\cdot\!\bK)^{-1/2}\bK$ (unphysical),
and the other three roots correspond to real
timelike four-velocities, with the two outer
roots (say $\Omega_-$ and $\Omega_+$)
giving local minima of the acceleration and
the root in between (say $\Omega_0$)
giving a local maximum.
In a static spacetime, the local maximum 
of the acceleration occurs for a static worldline,
at $\Omega_0 = 0$, accelerating against the pull
of gravity to stay at a fixed position.
In a Newtonian description
in which $\bl = \partial/\partial \varphi$
vanishes along the $z$-axis, 
the two local minima of the acceleration 
occur at the angular velocities $\Omega_{\pm}$ 
at which the centrifugal acceleration (in the $x$-$y$ plane)
balances the component of the gravitational acceleration
that is anti-parallel to it, leaving only an unbalanced
$z$-component of the gravitational acceleration.
For a stationary axisymmetric spacetime that has 
a reflection symmetry about an equatorial plane,
in that plane there is no other component
of the gravitational acceleration, 
so in the equatorial plane the local minima actually have
zero acceleration and correspond to stationary geodesics
or circular Keplerian orbits
at the corresponding $\Omega_{\pm}$.

	For a nonstatic (e.g., rotating)
stationary axisymmetric spacetime,
the value $\Omega_0$ of the angular velocity that gives
the local maximum of the acceleration
($\Omega_- < \Omega_0 < \Omega_+$)
gives a local definition of a congruence, the SCAM,
that in a local sense can be considered to be
the most nearly at rest.
Any slightly different rotation rate $\Omega$
would give a change in the centrifugal acceleration
and/or gravitational acceleration that would
reduce the total acceleration.
(In the Newtonian limit in which the gravitational
acceleration is independent of the velocity,
the reduction of the acceleration needed to balance
gravity would be provided purely by the centrifugal
acceleration, and in that limit, $\Omega_0 = 0$
is the angular velocity giving no centrifugal acceleration.)

	As one enters regions of strong gravity
(e.g., near a black hole, $r < 3M$ for a Schwarzschild
black hole), one of the roots $\Omega_-$ or $\Omega_+$
may reach the corresponding endpoint
$\Omega_{-1}$ or $\Omega_{+1}$, or it may merge
with $\Omega_0$ and thence go complex,
in either case disappearing from
the allowed region of the real $\Omega$ line
(or all three roots may merge simultaneously
in the nonrotating case), leaving only a single physical
extremum (a minimum) for the acceleration.
The fact that the acceleration then increases for a change
in the velocity can be attributed to a reversal of
the direction of the centrifugal acceleration
\cite{AbrLas,ACL,AbrPra,AbrMil,Abr90,All,Abr92,Abr93,
AbrSzu,ANW,ANW2,SM}
if one assumes (or defines) the gravitational acceleration
to be independent of the velocity,
or it can be attributed to a greater increase
in the gravitational acceleration than that in
the centrifugal acceleration
\cite{deF91,deFUT91,deFUT93,PageSciAm,deF94,
Sem94,Sem95,BBI,Sem96a,SB}
if one assumes (or defines) the gravitational acceleration
to increase with velocity in the way
(i.e., proportional to the square of the relativistic
gamma factor, $\propto \gamma^2 = 1/(1-v^2)$)
that one would get
for an object moving transversely across a spatially flat
horizontal floor in a rocket whose vertical acceleration
in flat spacetime simulates gravity inside
by the equivalence principle.

\section{Local definitions of stationary congruences}

	The definition of the SCAM has
the advantage of locality over the definition
of the congruence with $\Omega = 0$
(nonrotating with respect to infinity),
which requires the definition of which linear combination
(ignoring the overall normalization,
which is irrelevant for the present purpose)
of the two Killing vector fields is $\bk$
(usually made by choosing the combination
that remains timelike at spatial infinity),
a definition that cannot be made locally
but instead requires a knowledge of the behavior
of the Killing vector fields out to spatial infinity.

	The SCAM also has this same advantage,
though to a lesser degree, over the ZAMOs
(Zero Angular Momentum Observers),
which are defined to be orthogonal to the $\bl$
Killing vector field and so require that that vector
be uniquely picked out, again only up to normalization
(typically by choosing the Killing vector field
with closed orbits, which usually vanishes on a symmetry axis).
This again requires nonlocal knowledge,
unless one is at the symmetry axis where $\bl$ vanishes.
Since property (1) implies that $\bl$ not vanish,
I am explicitly assuming that one is not at a symmetry axis,
except for some discussions below where I take the limit
of going there.

	ZAMOs were originally called
``locally nonrotating observers'' \cite{Bardeen, MTW},
because they have angular velocities
midway between $\Omega_{-1}$ and $\Omega_{+1}$,
so that if two photons (in null but generically nongeodesic
stationary orbits, say skimming along mirrors)
were sent around both opposite directions from a ZAMO,
they would both return to the ZAMO at the same time.
This definition (essentially equivalent to defining
$\bl$ to be the combination of the Killing vector fields
with closed orbits) is quasilocal in that it does
not require a knowledge of the Killing vector fields
out to spatial infinity, but it is still nonlocal in that
it requires a knowledge of the fields along
the stationary null orbits until they return to the ZAMO
(i.e., all the way around the symmetry axis).
On the other hand, the definition of the SCAM
(when it exists) is completely local.

	To state more precisely what the conditions
are for a quantity to be local if it depends on the two Killing
vector fields $\bk$ and $\bl$, note that locally (away
from spatial infinity and from a symmetry axis)
one has nothing that determines which vector
in the entire $\bk\wedge\bl$ plane of vectors
is $\bk$ and which other vector is $\bl$.
In fact, one can make a global redefinition by the constant
linear transformation
 \begin{equation}
 \left (\matrix{
\bk\cr
\bl\cr
}\right ) \rightarrow
\left (\matrix{
\tilde{\bk}\cr
\tilde{\bl}\cr
}\right )
=
\pmatrix{
\alpha & \beta \cr
\gamma & \delta \cr
}
\left (\matrix{
\bk\cr
\bl\cr
}\right )
=
\left (\matrix{
\alpha\bk + \beta\bl\cr
\gamma\bk + \delta\bl\cr
}\right )
 \label{eq:30c}
 \end{equation}
with constants $\alpha, \beta, \gamma, \delta$.

	If the four-velocity $\bu$ given in Eqs. (\ref{eq:16}),
(\ref{eq:17}), and (\ref{eq:20}) is to remain invariant
under this redefinition of the Killing vector fields,
the angular velocity $\Omega$ must transform by
the fractional linear transformation
 \begin{equation}
 \Omega \rightarrow \tilde{\Omega}
 = {\alpha\Omega - \beta \over \delta - \gamma\Omega}.
 \label{eq:30d}
 \end{equation}

	Conversely, any four-velocity defined in terms
of an angular velocity $\Omega$ which does not transform
by Eq. (\ref{eq:30d}),
when the Killing vectors are transformed
by Eq. (\ref{eq:30c}), is not locally determined.
For example, the congruence that is nonrotating relative
to infinity has $\Omega = 0$ when $\bk$ is chosen to
be the Killing vector field that remains timelike
at spatial infinity.  But local information does not determine
this $\bk$, and when one considers the transformation
given by Eq. (\ref{eq:30c}), $\Omega = 0$ is not invariant
under Eq. (\ref{eq:30d}) as it would need to be
to be a locally determined condition.

	The ZAMO is defined by the condition that
 \begin{equation}
 0 = \bu\!\cdot\!\bl = e^{-\Phi}(\bk + \Omega_Z\bl)\!\cdot\!\bl
 = e^{-\Phi}(-B - \Omega_Z C),
 \label{eq:30e}
 \end{equation}
or
 \begin{equation}
 \Omega_Z = - {B\over C}
 \equiv - {g_{01}\over g_{11}}
 \equiv - {g_{t\varphi}\over g_{\varphi\varphi}}.
 \label{eq:30f}
 \end{equation}
However, this angular velocity, like $\Omega=0$,
does not transform
according to Eq. (\ref{eq:30d}) when $\bl$
is transformed to a linear combination of itself
and of $\bk$, so it also does not locally determine
a unique four-velocity $\bu$.

	To see this in detail and to prepare the way
for recognizing which four-velocities $\bu$ determined
by the behavior of $A$, $B$, and $C$ are locally
determined and independent of the transformation
(\ref{eq:30c}), one should note that
Eqs. (\ref{eq:23}) - (\ref{eq:25}) defining these
quantities imply that they transform as
 \begin{eqnarray}
 \pmatrix{
A & B \cr
B & B \cr
}
&\rightarrow&
 \pmatrix{
\tilde{A} & \tilde{B} \cr
\tilde{B} & \tilde{C} \cr
}
=
\pmatrix{
\alpha & \beta \cr
\gamma & \delta \cr
}
 \pmatrix{
A & B \cr
B & B \cr
}
\pmatrix{
\alpha & \gamma \cr
\beta & \delta \cr
}
\nonumber\\
&=&
\pmatrix{
\alpha^2 A + 2\alpha\beta B + \beta^2 C
& \alpha\gamma A + (\alpha\delta+\beta\gamma)B
 + \beta\delta C\cr
\alpha\gamma A + (\alpha\delta+\beta\gamma)B
 + \beta\delta C
& \gamma^2 A + 2\gamma\delta B + \delta^2 C \cr }\!.
 \label{eq:30g}
 \end{eqnarray}
Therefore, if one makes a linear transformation
of the Killing vector fields by
Eq. (\ref{eq:30c}) and then defines
$\hat{\Omega}_Z \equiv -\tilde{B}/\tilde{C}$,
one finds that when $\gamma \neq 0$,
this $\hat{\Omega}_Z$ does not generally agree with
the $\tilde{\Omega}_Z$ one would get from applying
Eq. (\ref{eq:30d}) to $\Omega_Z = -B/C$.

	In contrast, other quantities and conditions
are invariant under the transformations of
Eq. (\ref{eq:30c}) and (\ref{eq:30d}),
such as $\bu$ (by construction, since the
transformation of $\Omega$ was designed
to keep $\bu$ invariant),
$\ba$ given by Eq. (\ref{eq:22}),
the normalized two-form
$(-{1\over 2}A^{\alpha\beta}A_{\alpha\beta})^{-1/2}\bA$
obtained from the two-form $\bA = \bk\wedge\bl$
(which itself, like $D^{1/2}$ from Eq. (\ref{eq:14b}),
gets multiplied by the determinant of the
transformation matrix, $\alpha\delta - \beta\gamma$),
and the condition given by Eq. (\ref{eq:30a})
for the SCALEs.  Thus the SCALEs are indeed locally
determined congruences, unlike the nonrotating
(relative to infinity) observers and the ZAMOs.

	Incidentally, I might point out that there are
other quantities, analogous to $\bA$ and $D^{1/2}$,
that are not quite invariant under the transformations
(\ref{eq:30c}) and (\ref{eq:30d}) but which for
constant $\Omega$ get multiplied by the constant
$(\alpha\delta - \beta\gamma)/(\delta - \gamma\Omega)$,
namely $\bK$ and $e^{\Phi} \equiv F^{1/2}$.
Thus $\Phi = (1/2)\ln{(-\bK\!\cdot\!\bK)}$
is not quite locally determined,
but for a Killing vector field $\bK$ that is fixed
up to normalization (e.g., by being set parallel
at some location to a stationary four-velocity $\bu$
that is locally determined),
$\nabla \Phi$ (which is the acceleration of $\bu$)
is locally determined.

\section{Nonrotating Acceleration Worldlines (NAW)
\newline form a SCALE}

	Another way of locally selecting a preferred
stationary congruence of observers is to require
that their acceleration vectors be Fermi-Walker
transported and hence be nonrotating relative
to an ideal system of gyroscopes
carried by the observers.
In other words, the direction that each nonweightless
observer feels to be up stays fixed
in his or her Fermi-Walker transported frame,
so that one might state Delphically that
each such observer is ``fixed up'' or ``stays up.''
Such a congruence might be called
Nonrotating Acceleration Worldlines (NAW).
Here I shall show that a NAW is equivalent to a SCALE
(or SCAM in the case that the local extremum
of the acceleration is a local maximum).

	This equivalence is what one would expect
in the Newtonian limit, since there, as discussed above,
the SCAM has zero angular velocity ($\Omega_0 = 0$)
and hence has an acceleration vector that stays constant.
For observers that rotate, generically the
acceleration vector also rotates, but for the SCALEs
that are the two minima of the acceleration
at angular velocities $\Omega_{\pm}$,
the acceleration vector is purely in the
$z$-direction, which does not rotate.

	The Fermi-Walker derivative of a vector $\bv$
along a worldline with four-velocity $\bu = d/d\tau$
and acceleration $\ba = \nabla_{\bu}\bu$ is \cite{MTW}
 \begin{equation}
 \bvp \equiv \bF_{\bu}[\bv]
 \equiv \nabla_{\bu}\bv - (\bu\wedge\ba)\!\cdot\!\bv
 \equiv D\bv/d\tau - \bu(\ba\!\cdot\!\bv) + \ba(\bu\!\cdot\!\bv),
 \label{eq:31}
 \end{equation}
or, in component form,
 \begin{equation}
 v'_{\alpha} = u^{\beta}v_{\alpha ;\beta}
	- u_{\alpha}a^{\beta}v_{\beta}
	+ a_{\alpha}u^{\beta}v_{\beta}.
 \label{eq:32}
 \end{equation}
This is zero by construction when $\bv$ is 
the four-velocity $\bu$.
When $\bv$ is the acceleration $\ba$, 
the Fermi-Walker derivative is
 \begin{equation}
 \bap \equiv \bF_{\bu}[\ba]
 = \nabla_{\bu}\ba - a^2 \bu
 \label{eq:33}
 \end{equation}
(since $\bu\!\cdot\!\ba = 0$).
A Nonrotating Acceleration Worldline or NAW
(more strictly, in view of my definition above
of a NAW, a member of a NAW congruence)
is a worldline in which $\bap$ is either zero
or is parallel to $\ba$ (i.e., $\ba\wedge\bap=0$), 
so that in a frame carried along with the observer 
by Fermi-Walker transport, the acceleration vector $\ba$
does not change direction.  For a stationary observer,
the magnitude of $\ba$ stays fixed, so $\bap=0$
if the stationary observer is a member of a NAW.

	Using the explicit formulas (\ref{eq:16})
and (\ref{eq:21}) for $\bu$ and $\ba$ with fixed
$\Omega$, which imply
 \begin{equation}
 u_{(\alpha ;\beta)} + u_{(\alpha}a_{\beta)} = 0,
 \label{eq:34}
 \end{equation}
one can write this Fermi-Walker derivative
of the acceleration of a stationary observer as
 \begin{equation}
 \bap = \nabla_{\ba}\bu
 \label{eq:35}
 \end{equation}
or
 \begin{equation}
 a'_{\alpha} = a^{\beta}u_{\alpha ;\beta} =
 e^{-5\Phi}K^{\beta}K_{\gamma}K^{\gamma ;\delta}
 (K_{\beta}K_{\alpha ;\delta}-K_{\alpha}K_{\beta ;\delta}).
 \label{eq:36}
 \end{equation}
Now a comparison with Eq. (\ref{eq:30})
shows that
 \begin{equation}
 \frac{\partial a^2}{\partial \Omega}
 = -4e^{-\Phi}l^{\alpha}a'_{\alpha}
 = -4e^{-\Phi}\bl\!\cdot\!\bap.
 \label{eq:37}
 \end{equation}

	Thus $\bap=0$ implies that the acceleration
is an extremum with respect to $\Omega$, so
NAW$\:\Rightarrow\:$SCALE
(stationary worldlines whose acceleration
is nonrotating also have their acceleration
a local extremum).  The proof of the implication
in this direction used only the assumptions
that the four-velocity is a linear combination
$e^{-\Phi}\bK = e^{-\Phi}(\bk + \Omega \bl)$
of two Killing vector fields $\bk$ and $\bl$
that commute.  Semer\'{a}k \cite{Sem96},
after announcing that he had heard of my result
by personal communication,
has independently given a concise proof
in this direction, NAW$\:\Rightarrow\:$SCALE.

	A mnemonic for remembering this result,
using the somewhat cryptic shorthand phrases
given above, is to say,
``A stationary observer fixed up is extremely heavy,''
or, ``A stationary observer who stays up
is extremely heavy.''

\section{A SCALE is a NAW}

	To show the converse, that SCALE$\:\Rightarrow\:$NAW
(stationary worldlines with locally extreme acceleration
also have the acceleration nonrotating,
which was not proved in  \cite{Sem96}),
we need to invoke not only the assumed commutativity
of the two Killing vector fields $\bk$ and $\bl$,
but also the assumed property (3),
$\bk\wedge\bl\wedge\bd\bk = \bk\wedge\bl\wedge\bd\bl = 0$
for the corresponding 1-forms.

	For this purpose it is convenient to define
the Killing vector field $\bL$
(a linear combination of $\bk$ and $\bl$)
that at the position of the stationary worldline
under consideration has unit length and is
orthogonal to the observer's four-velocity
$\bu = e^{-\Phi} \bK$ there,
 \begin{equation}
 \bL = \gamma\bk + \delta\bl,
 \label{eq:38}
 \end{equation}
where $\gamma$ and $\delta$
are constants that at the position of
the stationary observer have the values
 \begin{equation}
 \gamma = -{e^{-\Phi} \over \sqrt{D}}G
 = {- B - \Omega C \over
 \sqrt{(B^2 - AC)(A+2\Omega B + \Omega^2 C)}},
 \label{eq:39}
 \end{equation}
 \begin{equation}
 \beta = {e^{-\Phi} \over \sqrt{D}}(F - \Omega G)
 = {A + \Omega B \over
 \sqrt{(B^2 - AC)(A+2\Omega B + \Omega^2 C)}}.
 \label{eq:40}
 \end{equation}
One can readily check from these formulas that
at the position of the stationary observer,
$\bL\!\cdot\!\bL = 1$ and $\bL\!\cdot\!\bu = 0$.
Thus $\bu$ and $\bL$ are orthonormal
vectors (or 1-forms, since I am using the same
symbols for both vectors and the corresponding
1-forms, with the distinction, if necessary, being clear
from the context) in the $\bk\wedge\bl$ plane.
Also, clearly $[\bk,\bl]=0 \Rightarrow [\bK,\bL]=0$, or
 \begin{equation}
 K^{\alpha}L^{\beta}_{\;\; ;\alpha}
 = L^{\alpha}K^{\beta}_{\;\; ;\alpha}.
 \label{eq:41}
 \end{equation}

	A 1-form in the (2,3) plane orthogonal to
the $\bk\wedge\bl$ or $\bu\wedge\bL$ or (0,1) plane
is $\ba$ (which I will now assume is nonzero;
otherwise it trivial that the acceleration is nonrotating).
One may normalize it to get the unit 1-form
 \begin{equation}
 \bah = (\ba\!\cdot\!\ba)^{-1/2}\ba
 \label{eq:42}
 \end{equation}
An orthonormal 1-form in this same (2,3) plane is
 \begin{equation}
 \bb = \ast(\bu\wedge\bL\wedge\bah).
 \label{eq:42b}
 \end{equation}

	Thus $(\bu,\bL,\bah,\bb)$ form an orthonormal
frame or basis of 1-forms
at the position of the stationary observer
that is defined purely in terms of the two commuting
Killing vector fields and hence is Lie transported
by the action of either one of them.
However, this orthonormal frame is generically rotating
relative to a frame that is Fermi-Walker transported
along the stationary observer's worldline.
In fact, the Fermi-Walker derivative of any
of the basis 1-forms above, or of any linear
combination of them with constant coefficients,
say $\bv$, is given by
 \begin{equation}
 \bvp = \ast(\bu\wedge\bo\wedge\bv),
 \label{eq:43}
 \end{equation}
where
 \begin{equation}
 \bo = \frac{1}{2} \ast(\bu\wedge\bd\bu)
 = \frac{1}{2} e^{-2\Phi}\ast(\bK\wedge\bd\bK)
 \label{eq:44}
 \end{equation}
is the normalized twist or rotation or vorticity
1-form of the Killing vector field  $\bK$
along one of whose integral curves
the stationary observer moves.
In particular,
 \begin{equation}
 \bap = \ast(\bu\wedge\bo\wedge\ba).
 \label{eq:44b}
 \end{equation}

	The normalization of $\bo$ makes it independent
of the constant linear transformation
(\ref{eq:30c}) of the Killing vector fields
and corresponding fractional linear transformation
(\ref{eq:30d}) of the angular velocity $\Omega$,
so long as the direction of $\bK$ is kept fixed,
as it indeed will be if one uses both transformations
(\ref{eq:30c}) and (\ref{eq:30d}).  In particular, $\bo$ is
constructed to be independent of the normalization
of $\bK$.  Therefore, if the four-velocity $\bu$
is locally determined, so is $\bo$ and hence also $\bap$.

	Eq. (\ref{eq:43}) is the analogue for 1-forms
of the four-dimensional form of the equation
that in the subspace of the
tangent space orthogonal to the four-velocity $\bu$,
the Fermi-Walker derivative $\bvp$
is the three-dimensional cross-product
of the rotation vector $\bo$ with $\bv$.
In component notation,
 \begin{equation}
 v'_{\alpha} = -\epsilon_{\alpha\beta\gamma\delta}
   u^{\beta} \omega^{\gamma} v^{\delta},
 \label{eq:45}
 \end{equation}
where
 \begin{equation}
 \omega_{\alpha}
 = \frac{1}{2} \epsilon_{\alpha\beta\gamma\delta}
   u^{\beta} u^{\gamma ;\delta}
 = \frac{1}{2} \epsilon_{\alpha\beta\gamma\delta}
   e^{-2\Phi}K^{\beta} K^{\gamma ;\delta}.
 \label{eq:46}
 \end{equation}

	The Fermi-Walker rotation is thus in the two-plane
orthogonal to the $\bu\wedge\bo$ plane,
so both $\bu$ and $\bo$ have zero
Fermi-Walker derivative.
Also, it is clear from Eq. (\ref{eq:44}) or (\ref{eq:46})
that $\bo$ is orthogonal to $\bu$.

	Note that in the definition (\ref{eq:44}) or (\ref{eq:46})
of $\bo$, I am implicitly assuming that
the four-velocity $\bu$ which is differentiated there
has a fixed angular velocity $\Omega$ and hence
is proportional to a fixed combination
$\bK = \bk + \Omega \bl$ of the two Killing vector fields,
with $\Omega$ held constant during
the exterior or covariant differentiation.
In other words, $\bo$ is the rotation 1-form for
a congruence that has four-velocity everywhere
parallel to a single Killing vector field $\bK$
and hence may be considered to be rigidly rotating,
since its shear and expansion are zero.

	When one considers a congruence, such as a
SCALE or a NAW, that has $\Omega$ varying as a
function of the two $x^a$ coordinates ($a = 2$ or 3),
then $\bo$ as I am generally using it in this paper
is {\it not} the rotation 1-form that one
would get from inserting that $\bu(x^a)$ into
Eq. (\ref{eq:44}), but rather the 1-form that one
gets at each $x^a$ by instead using the auxiliary
congruence with constant $\Omega$
(and hence with four-velocities parallel to
a single Killing vector field $\bK$)
that is chosen to match that
of the original congruence at that point.
(Of course, for a stationary congruence
$\Omega$ does not depend
on the proper time along each worldline.
Unless the context makes it clear that I am
assuming otherwise,
I actually make the stronger
assumption that the congruences
I am considering have $\Omega$
and other scalar quantities
independent of both of the (0,1)
coordinates, so their derivatives
in both the $\bk$ and $\bl$ directions vanish.)

	The stationarity of the worldlines
tangent to the Killing vector field $\bK$
imply that {\it any} covariantly determined
vector $\bv$ that is orthogonal to the worldline
and determined by the local properties
of the worldline and curvature
(i.e., contractions of derivatives
of the four-velocity and of the
curvature tensor and its derivatives)
is Lie transported by the Killing vector field $\bK$
and so has itself a Fermi-Walker derivative
that obeys Eq. (\ref{eq:43}) or (\ref{eq:45}).
In other words, the stationarity of metric
and of the worldline under translations along
the integral curves of $\bK$ imply that all
locally determined vectors have components
that stay constant in the locally rotating frame
with basis 1-forms obeying Eq. (\ref{eq:43})
or (\ref{eq:45}) with respect to
a locally nonrotating
(Fermi-Walker transported) frame.

	If $\bo = 0$, then one would obviously get
$\bap = 0$ by inserting $\ba$
instead of $\bv$ in Eq. (\ref{eq:43}) or (\ref{eq:45}),
so to show that SCALE$\:\Rightarrow\:$NAW,
it remains only to check the case in which
$\bo \neq 0$, which will henceforth be assumed.

	Now $\bK\wedge\bL\propto\bk\wedge\bl$,
so Eq. (\ref{eq:11}) implies
the corresponding equation for $\bK$ and $\bL$,
 \begin{equation}
 \bK\wedge\bL\wedge\bd\bK
 = \bK\wedge\bL\wedge\bd\bL = 0.
 \label{eq:61}
 \end{equation}
In terms of the rotation or normalized twist 1-form,
the first of these equations is simply
 \begin{equation}
 \bo\!\cdot\!\bL=0.
 \label{eq:62}
 \end{equation}

	Therefore, $\bu$, $\bo$, $\bL$, and
 \begin{equation}
 \bLp = \ast(\bu\wedge\bo\wedge\bL)
 \label{eq:63}
 \end{equation}
form an orthogonal set of 1-forms
at the position of the stationary observer,
with $\bu$ and $\bL$ having unit magnitude
and hence with $\bo$ and $\bLp$ having
the same magnitude.   Thus one may
write
 \begin{equation}
 \bo = \ast(\bu\wedge\bLp\wedge\bL),
 \label{eq:64}
 \end{equation}
and then using this in Eq. (\ref{eq:44b}) gives
 \begin{equation}
 \bap = \ast(\bu\wedge\bo\wedge\ba)
 = - (\ba\!\cdot\!\bLp)\bL.
 \label{eq:65}
 \end{equation}

	Now inserting $\bo$ from Eq. (\ref{eq:44})
into Eq. (\ref{eq:63}) for $\bLp$,
evaluating, and comparing with Eq. (\ref{eq:29b})
shows that
 \begin{equation}
 \bLp = {e^{2\Phi} \over 2\sqrt{D}}
	\frac{\partial \ba}{\partial \Omega}
 = {(A + 2\Omega B + \Omega^2 C)
	\over 2 \sqrt{B^2 - AC}}
	\frac{\partial \ba}{\partial \Omega}.
 \label{eq:66}
 \end{equation}
Putting this expression into Eq. (\ref{eq:65})
and comparing with Eq. (\ref{eq:30}),
or comparing with Eq. (\ref{eq:37}),
one gets
 \begin{equation}
 \bap = - {e^{2\Phi} \over 4\sqrt{D}}
	\frac{\partial a^2}{\partial \Omega}\bL
 = - {(A + 2\Omega B + \Omega^2 C)
	\over 4 \sqrt{B^2 - AC}}
	\frac{\partial a^2}{\partial \Omega}\bL.
 \label{eq:67}
 \end{equation}

	Thus, for a spacetime with two independent
Killing vector fields with properties (1)-(3) above,
${\partial a^2}/{\partial \Omega}=0 \Leftrightarrow \bap=0$.
That is, a 
Stationary Congruence Acceleration Locally Extremely
(SCALE) is equivalent to
a Nonrotating Acceleration Worldline (NAW).
This proof implies the title statement,
``Maximal Acceleration Is Nonrotating''
(so it might be called the ``MAIN Page proof.'')

	In contrast, one might be tempted to say
of the converse result given earlier,
"NAW, it's a SCAM."
But this would be a scam,
since actually a NAW was proved to be a SCALE,
and not all SCALEs are SCAMs.

	A Delphic way of stating the result,
using phrases explained above, and substituting
``permanent'' and ``always'' for ``stationary,'' would be,
``Someone permanently fixed up is extremely heavy,
and someone always extremely heavy stays up.''

\section{Acceleration and rotation of observers}

\subsection{Stationary congruence parallel to a Killing field}

	As an aside, one can make an analogy between
the acceleration and rotation vectors for
a stationary observer corresponding
to a particular timelike Killing vector field,
and the electric and magnetic fields
of an electromagnetic field as seen
by a particular observer.
In particular, if the timelike Killing vector
field is $\bK$ with squared magnitude
 \begin{equation}
 F\equiv e^{2\Phi}\equiv -\bK\!\cdot\!\bK,
 \label{eq:77}
 \end{equation}
then one may regard the normalized 2-form
 \begin{equation}
 \bff \equiv - {1 \over 2} e^{-\Phi} \bd\bK
	= -{1 \over 2}\bd\bu + {1 \over 2}\bu\wedge\ba
 \label{eq:78}
 \end{equation}
with antisymmetric tensor components
 \begin{equation}
 f_{\alpha\beta}\equiv e^{-\Phi}K_{\alpha ;\beta}
	= u_{[\alpha ;\beta]} + u_{[\alpha}a_{\beta]}
 \label{eq:79}
 \end{equation}
as being algebraically analogous to an electromagnetic
field 2-form with tensor components $F_{\alpha\beta}$,
though it will not in general obey the analogues
$\bd\bff = 0$ and $\bd\ast\bff = 0$
of the vacuum Maxwell equations.

	(The normalization of $\bff$ is chosen to
make it independent of the constant transformations
(\ref{eq:30c}) and (\ref{eq:30d}) of the Killing vector
fields $\bk$ and $\bl$ and of $\Omega$ that keep $\bK$
pointing in the same spacetime direction but which
may change its overall magnitude.)

	Then in the frame of the stationary observer
with normalized four-velocity $\bu = e^{-\Phi}\bK$,
the acceleration 1-form with components
 \begin{equation}
 a_{\alpha} = f_{\alpha\beta}u^{\beta}
 \label{eq:80}
 \end{equation}
is analogous to the electric field in a frame
with four-velocity $\bu$, which has components
 \begin{equation}
 E_{\alpha} = F_{\alpha\beta}u^{\beta}.
 \label{eq:81}
 \end{equation}
Similarly, the rotation 1-form with components
 \begin{equation}
 \omega_{\alpha}
 = {1 \over 2}\epsilon_{\alpha\beta\gamma\delta}
	f^{\beta\gamma}u^{\delta}
 \label{eq:82}
 \end{equation}
is analogous to the negative of
the magnetic field in the observer's frame,
which has components
 \begin{equation}
 B_{\alpha}
 = -{1 \over 2}\epsilon_{\alpha\beta\gamma\delta}
	F^{\beta\gamma}u^{\delta}.
 \label{eq:83}
 \end{equation}

	One can then write the Fermi-Walker derivative
of the components of any 1-form $\bv$
which is Lie transported by
the Killing vector field $\bK$ as
 \begin{equation}
 v'_{\alpha} = \omega_{\alpha\beta} v^{\beta},
 \label{eq:84}
 \end{equation}
where
 \begin{eqnarray}
 \omega_{\alpha\beta}
 &=& - \epsilon_{\alpha\beta\gamma\delta}
   u^{\gamma} \omega^{\delta}
	\nonumber \\
 &=& (\delta_{\alpha}^{\;\;\mu}+u_{\alpha}u^{\mu})
	u_{\mu;\nu} (\delta^{\nu}_{\;\;\beta}+u^{\nu}u_{\beta})
	\nonumber \\
 &=& (\delta_{\alpha}^{\;\;\mu}+u_{\alpha}u^{\mu})
	f_{\mu\nu} (\delta^{\nu}_{\;\;\beta}+u^{\nu}u_{\beta})
	\nonumber \\
 &=& u_{\alpha;\beta} + a_{\alpha}u_{\beta}
	\nonumber \\
 &=& f_{\alpha\beta} - u_{\alpha}a_{\beta}
 + a_{\alpha}u_{\beta}
 \label{eq:85}
 \end{eqnarray}
are the components of the rotation 2-form
 \begin{equation}
 \tilde{\bo} = -\ast(\bu\wedge\bo)
 = - {1 \over 2}e^{\Phi}\bd(e^{-2\Phi}\bK)
 = - {1 \over 2}e^{\Phi}\bd(e^{-\Phi}\bu)
 \label{eq:85b}
 \end{equation}
which is the part of the 2-form $-{1\over 2}\bd\bu$,
or of the 2-form $\bff = -{1\over 2}e^{-\Phi}\bd\bK$,
that is orthogonal to
$\bu = e^{-\Phi}\bK$.
As usual, the normalization is chosen to make $\tilde{\bo}$
independent of the transformations
(\ref{eq:30c}) and (\ref{eq:30d}).

	In particular, one can see from Eq. (\ref{eq:44b})
that the Fermi-Walker derivative of the acceleration 1-form,
$\bap$, is analogous to $4\pi$ times the Poynting
vector flux of the electromagnetic field that is analogous
to $\bff$.  In other words, if one defines a
symmetric second rank tensor
 \begin{equation}
 4\pi t_{\alpha\beta} = f_{\alpha\mu}f_{\beta}^{\;\;\mu}
	- {1 \over 4} g_{\alpha\beta}f_{\mu\nu}f^{\mu\nu},
 \label{eq:86}
 \end{equation}
that is the analogue of the electromagnetic
stress-energy tensor (generically not conserved here,
since $\bff$ does not obey the vacuum Maxwell equations),
then
 \begin{equation}
 a'_{\alpha} = 4\pi p_{\alpha}
 = [\ast(\bu\wedge\bo\wedge\ba)]_{\alpha}
 = -4\pi (\delta_{\alpha}^{\;\;\beta}+u_{\alpha}u^{\beta})
	t_{\beta\gamma}u^{\gamma}
 = (\delta_{\alpha}^{\;\;\beta}+u_{\alpha}u^{\beta})
	f_{\beta}^{\;\;\gamma}f_{\gamma}^{\;\;\delta}u_{\delta}.
 \label{eq:87}
 \end{equation}

	One can easily see that there are precisely
three independent Lorentz-invariant scalars
at a point that depend only on the Killing vector field $\bK$
and its first covariant derivative
and are invariant under the transformations
(\ref{eq:30c}) and (\ref{eq:30d}) that change its normalization:
 \begin{equation}
 a^2 \equiv \ba\!\cdot\!\ba
 = - f^{\alpha}_{\;\;\beta} f^{\beta}_{\;\;\gamma}
	u^{\gamma}u_{\alpha},
 \label{eq:88}
 \end{equation}
 \begin{equation}
 \omega^2 \equiv \bo\!\cdot\!\bo
 = f^{\alpha}_{\;\;\beta} f^{\beta}_{\;\;\gamma}
 ({1\over 2}\delta^{\gamma}_{\alpha} - u^{\gamma}u_{\alpha}),
 \label{eq:89}
 \end{equation}
and
 \begin{equation}
 \ba\!\cdot\!\bo
 = -{1 \over 8}\epsilon_{\alpha\beta\gamma\delta}
	f^{\alpha\beta}f^{\gamma\delta}
 = \pm[{1\over 4}f^{\alpha}_{\;\;\beta}f^{\beta}_{\;\;\gamma}
		f^{\gamma}_{\;\;\delta}f^{\delta}_{\;\;\alpha}
	-{1\over 8}
	(f^{\alpha}_{\;\;\beta}f^{\beta}_{\;\;\alpha})]^{1/2}.
 \label{eq:90}
 \end{equation}

	For example, dot products of $\bo$, $\ba$ and of all of
its Fermi-Walker derivatives $\ba^{(n)}$ of order $n$
may be expressed algebraically
in terms of these three scalars.
For positive integers $m$ and $n$,
 \begin{equation}
 a'^2 \equiv \bap\!\cdot\!\bap \equiv \ba^{(1)}\!\cdot\!\ba^{(1)}
 = (\ba\!\cdot\!\ba)(\bo\!\cdot\!\bo) - (\ba\!\cdot\!\bo)^2
 = a^2\omega^2 - (\ba\!\cdot\!\bo)^2,
 \label{eq:91}
 \end{equation}
 \begin{equation}
 \bo\!\cdot\!\ba^{(n)} = 0,
 \label{eq:92}
 \end{equation}
 \begin{equation}
 \ba\!\cdot\!\ba^{(2n-1)} = 0,
 \label{eq:92b}
 \end{equation}
 \begin{equation}
 \ba\!\cdot\!\ba^{(2n)} = \omega^{2n-2}a'^2,
 \label{eq:93}
 \end{equation}
 \begin{equation}
 \ba^{(n)}\!\cdot\!\ba^{(n+2m-1)} = 0,
 \label{eq:94}
 \end{equation}
and
 \begin{equation}
 \ba^{(n)}\!\cdot\!\ba^{(n+2m)}
 = (-1)^m\omega^{2n+2m-2}a'^2.
 \label{eq:95}
 \end{equation}

	One can also get analogous relations between
the vectors or 1-forms themselves:  If one defines
 \begin{equation}
 \ba_{\bot}\equiv\ba - (\ba\!\cdot\!\bo)\bo/\omega^2,
 \label{eq:96}
 \end{equation}
the part of the acceleration vector or 1-form $\ba$
that is orthogonal to the rotation vector or 1-form $\bo$,
which has squared length
 \begin{equation}
 a_{\bot}^2 \equiv \ba_{\bot}\!\cdot\!\ba_{\bot}
 = a'^2/\omega^2 = a^2 - (\ba\!\cdot\!\bo)^2/\omega^2,
 \label{eq:97}
 \end{equation}
then $\bu$, $\bo$, $\ba_{\bot}$, and $\bap$ form an
orthogonal set of vectors or 1-forms, and each
Fermi-Walker derivative $\ba^{(n)}$ is parallel
(or anti-parallel, in alternate cases) either to
$\ba_{\bot}$ or to $\bap$:
 \begin{equation}
 \ba^{(2n)} = (-1)^n\omega^{2n}\ba_{\bot},
 \label{eq:98}
 \end{equation}
 \begin{equation}
 \ba^{(2n+1)} = (-1)^n\omega^{2n}\bap.
 \label{eq:99}
 \end{equation}

	When $\bap\neq 0$, one can solve for
the rotation 1-form
 \begin{equation}
 \bo = \ast(\bu\wedge\bap\wedge\bapp)/a'^2
 \label{eq:100}
 \end{equation}
in terms of the four-velocity $\bu$ and the
first two Fermi-Walker derivatives
$\bap$ and $\bapp$of the acceleration $\ba$.
In particular, the squared magnitude
of this rotation 1-form is then
 \begin{eqnarray}
 \omega^2 &\equiv& \bo\!\cdot\!\bo
 = {\bapp\!\cdot\!\bapp \over \bap\!\cdot\!\bap}
	\nonumber \\
 &=& {[(A\nabla B - B\nabla A)
	+ \Omega (A\nabla C - C\nabla A)
	+ \Omega^2 (B\nabla C -C\nabla B)]^2
	\over 4(B^2-AC)(A + 2\Omega B + \Omega^2 C)^2}.
 \label{eq:101}
 \end{eqnarray}
One can compute directly from Eq. (\ref{eq:30g})
that this formula for $\omega^2$
in terms of $A$, $B$, and $C$
is invariant under the transformations
(\ref{eq:30c}) and (\ref{eq:30d}).

	When one has a stationary observer in a curved spacetime,
one can also get Lorentz-invariant scalars from contracting
combinations of the Killing vector field $\bK$ and its covariant
derivatives with combinations of the Riemann curvature tensor
and its covariant derivatives.  As is well known, one can use
Killing's equation, Eq. (\ref{eq:3}), to eliminate all covariant
derivatives of the Killing vector field of order higher than two.
Therefore, since
 \begin{equation}
 \bff \equiv - {1 \over 2} e^{-\Phi} \bd\bK
 = \bu\wedge\ba + \tilde{\bo}
 = \bu\wedge\ba - \ast(\bu\wedge\bo),
 \label{eq:102}
 \end{equation}
any of these Lorentz-invariant scalars can be obtained
by contracting appropriate combinations of
$\bu$, $\bo$, and $\ba$ with appropriate combinations
of the Riemann curvature tensor and its covariant derivatives,
up to powers of the squared magnitude
$e^{2\Phi} = -\bK\!\cdot\!\bK$ of the Killing vector field
that get multiplied by constants when one
performs the transformations (\ref{eq:30c}) and (\ref{eq:30d}),

	Just as a SCALE extremizes, and a SCAM locally maximizes,
the scalar $a^2$ as a function of the angular velocity $\Omega$,
so one could also define other congruences that extremize,
maximize, minimize, or set to zero other combinations of
the scalars $a^2$, $\omega^2$, or $\ba\!\cdot\!\bo$.
For example, one might define a Stationary Congruence
Rotating At Minimum (SCRAM) as a stationary congruence
that at each point,
as a function of the angular velocity $\Omega$,
minimizes the squared magnitude
$\omega^2$ of the rotation $\bo$.
(Remember that $\bo$ is actually the rotation at each point,
not of the original congruence being considered,
but of an auxiliary rigid congruence,
parallel to a single Killing vector field $\bK$ with $\Omega$
having a constant value that matches the
$x^a$-dependent $\Omega$ of the original congruence
at the point where $\bo$ is being evaluated.)
One can see from extremizing Eq. (\ref{eq:101})
that the equation for this minimum, like that for a SCAM,
is generically a messy quartic equation in $\Omega$.
The resulting four-velocity is independent of the
transformations (\ref{eq:30c}) and (\ref{eq:30d})
and hence is determined purely locally.

\subsection{Nonstationary observers}

	Many of the formulas above apply
or have generalizations
to the case of a nonstationary congruence with $\bu$
defined over the region of spacetime under consideration,
even when it is not proportional to any Killing vector field.
For example, the acceleration vector $\ba$ is still given
by Eq. (\ref{eq:18}), and the rotation 1-form $\bo$
by the first expression on the right hand side
of Eq. (\ref{eq:44}).
Then one can use the last expression
on the right hand side of Eq. (\ref{eq:102})
to define a 2-form $\bff$
that is algebraically analogous to an electromagnetic field
and which gives back the acceleration and rotation 1-forms
by Eqs. (\ref{eq:80}) and (\ref{eq:82}).
One can also use various other definitions,
such as Eq. (\ref{eq:96}) for $\ba_{\bot}$,
the part of the acceleration vector
that is orthogonal to $\bo$.

	For a nonstationary congruence,
the components of quantities defined in terms
of the four-velocity and its covariant derivatives
(e.g., the components of $\ba$ and of $\bo$)
are no longer constant in a frame which
accelerates and rotates with the congruence,
as such components are for a stationary congruence.
Therefore, the Fermi-Walker derivative is no longer
given by Eq. (\ref{eq:43}).  In particular, $\bap$
is no longer given by Eq. (\ref{eq:44b}) or (\ref{eq:87}).
Nevertheless, one can still define an analogue of $4\pi$
times the Poynting flux,
 \begin{equation}
 4\pi\bp \equiv \ast(\bu\wedge\bo\wedge\ba),
 \label{eq:103}
 \end{equation}
whose components are given by the right hand side
of Eq. (\ref{eq:87}).
When $4\pi\bp$ is nonzero,
it may be normalized to define the unit 1-form
 \begin{equation}
 \bL \equiv \bp/(\bp\!\cdot\!\bp)^{1/2}
 \equiv [a^2\omega^2 - (\ba\!\cdot\!\bo)^2]^{-1/2}
	\ast(\bu\wedge\bo\wedge\ba).
 \label{eq:104}
 \end{equation}
Then $\bu$, $\bo/|\bo|$, $\ba_{\bot}/|\ba_{\bot}|$,
and $\bL$ form an orthonormal basis defined
by the congruence.

	In general for a nonstationary congruence
(or even for a stationary congruence in a metric
which does not have a second Killing vector
field obeying properties (2) and (3) above),
$\bL$ will not be parallel to any Killing vector.
However, one can start from an original congruence
with the four-velocity field $\bu$,
construct the unit orthogonal vector $\bL$
by the procedure above (when $4\pi\bp\neq 0$),
and thereby define a new set of congruences
with four-velocities
 \begin{equation}
 \bU = {\bu + v\bL \over \sqrt{1-v^2}},
 \label{eq:105}
 \end{equation}
where $v$ is the velocity of the new congruence
relative to the old one, a parameter analogous
to $\Omega$ in the stationary axisymmetric case.

	Then one can play the same game of
choosing the velocity $v$ at each point
of the region of spacetime under consideration
to extremize, maximize, minimize,
or set to zero various combinations of
such scalars as $a^2$, $\omega^2$, $\ba\!\cdot\!\bo$,
$\ba\!\cdot\!\bap$, $\bo\!\cdot\!\bap$, $\bap\!\cdot\!\bap$,
$\ba\!\cdot\!\bapp$, $\bo\!\cdot\!\bapp$,
$\bap\!\cdot\!\bapp$, $\bapp\!\cdot\!\bapp$, etc.,
an arbitrarily large number of which can be independent
for a nonstationary congruence,
unlike the case of a stationary congruence,
for which these scalars are all algebraically
dependent on the first three.

	For example, one might define $v$
to give a local extremum of $a^2$,
which might be called
a Congruence Accelerating Locally Extremely (CALE),
or a Congruence Accelerating Maximally (CAM)
if the extremum is a local maximum.
Or, one might minimize $\omega^2$, giving
a Congruence Rotating At Minimum (CRAM).
One could still have a Nonrotating Acceleration
Worldline (NAW) if $\bap$ is parallel to $\ba$,
but since this requirement is equivalent to solving two
equations, it generically cannot be satisfied
by any choice of the single parameter $v$
for a nonstationary congruence,
as it could in the stationary axisymmetric case.
One could instead choose $v$ at each point
to minimize
the squared length of the Fermi-Walker
derivative of the direction of the acceleration,
 \begin{equation}
 ({\ba \over |\ba|})'\!\cdot\!({\ba \over |\ba|})'
 = {a^2a'^2 - (\ba\!\cdot\!\bap)^2 \over a^4},
 \label{eq:106}
 \end{equation}
giving a
Congruence Having Acceleration Rotating Minimally
(CHARM).
Clearly a NAW is a special case of a CHARM.

	Because of the independence of the
higher covariant derivatives of the four-velocity
of a nonstationary congruence,
in contrast to the case for a stationary congruence,
it appears that one would not in general be able
to conclude that a CAM is a CHARM, for instance.
It might be of interest to see when this is so,
but I shall not pursue this issue further here.

	Some of the formulas above can be applied
even when there is no congruence but only
a single worldline.  In this case one cannot define
a rotation 1-form $\bo$ by Eq. (\ref{eq:44}),
but one can still define the acceleration $\ba$
and its various Fermi-Walker derivatives
$\bap$, $\bapp$, etc.
If one liked, one could make up a definition
for a rotation 1-form $\bo$, such as Eq. (\ref{eq:100}),
which agrees with Eq. (\ref{eq:44})
for a stationary congruence when $a'^2 \neq 0$.
It might be somewhat better for a general
stationary congruence to replace
$\bap$ by its part that is orthogonal to $\ba$, namely
 \begin{equation}
 \bap_{\bot} \equiv \bap - \ba(\ba\!\cdot\!\bap)/a^2,
 \label{eq:107}
 \end{equation}
with squared magnitude
 \begin{equation}
 (a'_{\bot})^2 \equiv \bap_{\bot}\!\cdot\!\bap_{\bot}
 = \bap\!\cdot\!\bap - (\ba\!\cdot\!\bap)^2/a^2,
 \label{eq:108}
 \end{equation}
thus leaving out a possible change in the magnitude
of the acceleration that is automatically absent
for a stationary worldline, and to replace
$\bapp$ by $(\bap_{\bot })'$, the Fermi-Walker derivative
of $\bap_{\bot}$.
Then one could define a rotation 1-form to be
 \begin{equation}
 \bo = \ast(\bu\wedge\bap_{\bot}
	\wedge(\bap_{\bot })')/(a'_{\bot})^2
 \label{eq:109}
 \end{equation}
One might prefer also to take only the part of
$(\bap_{\bot })'$ that is orthogonal to $\bap_{\bot}$,
but this makes no difference in Eq. (\ref{eq:109}).

	Thus for a single worldline one can say that
it has the property of being a member of a NAW
if and only if $\bap_{\bot} = 0$,
but without having the rest of a congruence,
one cannot say whether
or not it is a member of a SCALE, SCAM,
SCRAM, CALE, CAM, or CHARM (except that it
must have the latter property if it is a NAW).

\section{Stationary Congruence Accelerating
\newline Maximally (SCAM)}

	For a generic location (of the two $x^a$ coordinates
parametrizing the two-surfaces orthogonal to the
two Killing vector fields $\bk$ and $\bl$)
in a generic axisymmetric metric,
the quartic Eq. (\ref{eq:30a})
giving the angular velocity $\Omega_0$
of the SCAM (Stationary Congruence Accelerating Maximally)
is messy to solve, and the explicit form of the solution
is presumably not very perspicuous
(though I have not bothered to write it out in gory detail).
However, there are at least three situations
in which one can get simpler explicit solutions.
Below I shall give an example where the the quartic
factorizes into two fairly simple quadratics.
But first I shall give the angular velocity of
a SCAM in a spacetime with relatively slow rotation.

\subsection{In a spacetime with slow rotation or weak fields}

	For a spacetime with relatively slow rotation,
after a suitable constant linear transformations (\ref{eq:30c})
of the Killing vector fields $\bk$ and $\bl$ if necessary,
the part of the metric in the (0,1) plane is nearly
diagonal, and so are its derivatives.
In particular, $B^2 \ll -AC$, and
$\nabla B\!\cdot\!\nabla B \ll - \nabla A\!\cdot\!\nabla C$.
Then one can readily write down the solution
of Eq. (\ref{eq:30a}) that is first order in $B$ and $\nabla B$,
ignoring higher order corrections in these quantities
(which will start with terms cubic in these quantities,
which I shall label simply as $O(B^3)$):
 \begin{equation}
 \Omega_0 = - {\nabla A\!\cdot\!(A\nabla B - B\nabla A)
 \over \nabla A\!\cdot\!(A\nabla C - C\nabla A)}
 + O(B^3).
 \label{eq:181}
 \end{equation}

	In the far-field limit outside an isolated source,
typically $A\equiv -g_{tt}$ is just
slightly less than 1 (e.g., roughly $1 - 2M/r$ at $r \gg M$)
and has a gradient nearly in
the positive radial direction with coordinate $r = x^2$.
For simplicity, in the following discussion
let the derivative with respect to $r$
be denoted by a prime.
(The context will keep it clear that here
this prime does not mean a Fermi-Walker
derivative, which will usually be zero
for the quantities that I shall be applying
the prime to, such as the stationary
scalar quantities $A$, $B$, and $C$.)
$C\equiv -g_{\varphi\varphi}$ will be negative,
typically going roughly as the negative square of the
distance from the axis of rotation,
so at angle $\theta = x^3$ from the positive axis,
$C$ will go like $-r^2 \sin^2{\theta}$,
and hence off the axes will have $C'<0$.
Furthermore, $AC'-CA'$ will be dominated
by its first term and hence be negative.
Then if the source is rotating in the positive $\varphi$
direction, then in the far-field limit
$B\equiv -g_{t\varphi}$ will be positive,
typically going as $(2J/r) \sin^2{\theta}$,
where $J$ is the intrinsic angular momentum
of the source \cite{MTW}.
Thus $B$ will have a negative radial gradient
$B'$ as it tends to zero at infinity,
and $AB'-BA'$ will be dominated by its first term
and hence be negative.

	Therefore, in the far-field limit outside an
isolated source centered at the origin of
standard spherical polar coordinates,
Eq. (\ref{eq:181}) gives
the approximate angular velocity of the SCAM as
 \begin{equation}
 \Omega_0 \approx - {B'\over C'} \approx -{J\over r^3}.
 \label{eq:182}
 \end{equation}

	This is in contrast to a ZAMO, which has
 \begin{equation}
 \Omega_Z = -{B\over C} \approx +{2J\over r^3}.
 \label{eq:183}
 \end{equation}
That is, a ZAMO corotates with the source
by what is usually called the dragging of inertial frames,
but a SCAM counterrotates.  Why is this?

	A simple physical argument that applies
in the far-field limit is the following:
Consider the source being a ring of nonrelativistic matter
in the the equatorial plane
rotating with $v\ll 1$ in the positive $\varphi$-direction,
and consider a stationary observer just outside this ring.
In the observer's frame, the energy density
of the nearby portion of the ring will be a minimum
if the observer is corotating with the ring.
If the observer is not rotating relative to infinity
($\Omega = 0$),
she will see a higher energy density for the source,
because of a relativistic $\gamma$-factor
($\gamma \equiv 1/\sqrt{1-v^2}$)
for the energy of each particle of the source
(increase of its kinetic energy),
and because of another $\gamma$-factor from
the Lorentz-contraction of the source.
If the observer is counterrotating,
she will have an even greater velocity relative
to the nearby portion of the ring, and so she will
see an even higher energy density
(increasing roughly linearly with $-\Omega$
when it is small compared with the magnitude
of the rotation rate of the ring itself).

	Because energy density is the main source
of the gravitational field in the nonrelativistic limit,
a counterrotating stationary observer will thus have
a greater gravitational attraction to the ring
(greater acceleration radially outward as seen
in a freely falling frame).

	As the observer increases her counterrotation rate,
she will also have a counter-acting
centripetal acceleration inward,
but it will increase at first only quadratically
with her counterrotating angular velocity $-\Omega$.
Thus a small counterrotation rate will increase
the experienced outward acceleration against
gravity more than the increase of the
centripetal acceleration inward,
so the net acceleration
(which is outward, primarily against
the gravitational attraction of the ring)
will at first increase with increasing
the magnitude of the counterrotation
angular velocity $-\Omega$.
(De Felice \cite{deF95} previously noticed this effect
on the equatorial plane of the Kerr metric.)

	Eventually, the (roughly quadratic) increase
in the inward centripetal acceleration will balance
the (roughly linear) increase in the outward
acceleration against gravity, and one will
reach a local maximum of the acceleration,
thus at a counterrotating angular velocity
($\Omega_0 < 0$) of the SCAM.

	This is the picture in a weak gravitational field,
but in a strong field ($M\sim r$), the increase of the
gravitational attraction with velocity
(by the two $\gamma$-factors, i.e., as $1/(1-v^2)$)
can dominate over the increase in the centripetal
acceleration (which goes as $v^2/(1-v^2)$,
but typically with a different coefficient,
roughly $1/r$ rather than roughly $M/r^2$),
so that there is no local maximum of the acceleration
and hence no SCAM at that location.
By taking out the $\gamma$-factor dependence
of the gravitational acceleration
and including it instead with the centripetal acceleration
(which is not strictly forbidden, since only the sum
of the two is a gauge-invariant observable quantity),
Abramowicz and his collaborators
\cite{AbrLas,ACL,AbrPra,AbrMil,Abr90,All,Abr92,Abr93,
AbrSzu,ANW,ANW2,SM}
label this latter effect as
the reversal of the sign of the centripetal acceleration,
though to me it seems more intuitively understandable
to recognize it as the greater increase of the
gravitational acceleration with velocity than
has the centripetal acceleration
\cite{deF91,deFUT91,deFUT93,PageSciAm,deF94,
Sem94,Sem95,BBI,Sem96a,SB}.

\subsection{At an orbital}

	Another situation in which the SCAM has properties
that are simpler than those given by the solution of
the generic quartic Eq. (\ref{eq:30a}),
is at a location which I shall call an orbital.
This I define to be a location where
there are two stationary timelike geodesics
at different values of $\Omega$
(e.g., corotating and counterrotating
circular Keplerian orbits).
There the angular velocity of the SCAM is given
by a quadratic equation rather than a quartic.

	In terms of the (reversed-sign) $(t,\varphi)$
metric components
$A \equiv -g_{tt}$, $B \equiv -g_{t\varphi}$,
and $C \equiv -g_{\varphi\varphi}$ defined above,
Eq. (\ref{eq:29}) gave
 \begin{equation}
 \ba = {\nabla A + 2 \Omega \nabla B + \Omega^2\nabla C
		\over 2(A+ 2 \Omega B +\Omega^2 C)}.
 \label{eq:201}
 \end{equation}
At a generic location,
$\nabla A$, $\nabla B$, and $\nabla C$
are rather arbitrary vectors
in the 2-dimensional (2,3) tangent plane,
so for no value of $\Omega$ is $\ba=0$.

	Only if $-2\nabla B$ is on the two-branched hyperbola
$\Omega^{-1}\nabla A + \Omega\nabla C$
is there a value of $\Omega$ that gives $\ba = 0$
(a stationary geodesic).
If there are two stationary geodesics
at different values of $\Omega$,
then the two branches of the hyperbola must degenerate
to a single straight line
(i.e., $\nabla A$ and $\nabla C$ must point
in the same or opposite direction).
For $\nabla B$ to be on this degenerate hyperbola
(at two different values of $\Omega$),
it must also point in that same direction.
In other words, at an orbital,
$\nabla A$, $\nabla B$, and $\nabla C$
must all be parallel (or anti-parallel).
However, this is a sufficient condition
for an orbital only if both values of $\Omega$
that give $\ba = 0$ give timelike four-velocities.

	Since $\nabla A$, $\nabla B$, and $\nabla C$
lie in the tangent (2,3) plane that orthogonal
to the plane of the two-form $\bA = \bk \wedge \bl$
(which is not to be confused
with the scalar $A = -\bk\!\cdot\!\bk$),
the condition that these three gradients are all parallel
is equivalent to the two scalar equations
 \begin{equation}
 \ast(\bA\wedge\nabla A\wedge\nabla B) = 0,
 \label{eq:200a}
 \end{equation}
 \begin{equation}
 \ast(\bA\wedge\nabla B\wedge\nabla C) = 0.
 \label{eq:200b}
 \end{equation}
In the (2,3) coordinate space, these
two equations generically have solutions
only at isolated points, if at all, so it is
by no means guaranteed that a stationary
axisymmetric metric will have any orbitals.
But if the space is symmetrical with respect to
reflection about an equatorial plane containing $\bl$,
then on that plane $\nabla A$, $\nabla B$, and $\nabla C$
will all lie within its tangent plane and be parallel
(since they are all orthogonal to $\bl$).
Therefore, if a spacetime has such an equatorial
plane, and if both values of $\Omega$
that give $\ba = 0$ give timelike four-velocities
for some region of that plane,
then this region consists of orbitals.

	Assume that the common direction
of $\nabla A$, $\nabla B$, and $\nabla C$
is not orthogonal to
the radial direction with coordinate $r = x^2$.
(I.e., choose the radial direction so that
it is not orthogonal to that common direction.)
As above, let the derivative with respect to $r$
be denoted by a prime.  Then
clearly $\nabla A$, $\nabla B$, and $\nabla C$
will all be proportional to $A'$, $B'$, and $C'$
respectively, with the same constant
of proportionality (at a fixed location).

	Therefore, at an orbital, Eq. (\ref{eq:30a})
for the angular velocity $\Omega$ of a SCALE becomes
 \begin{equation}
 [A' + 2 B' \Omega + C' \Omega^2]
 [(AB' - BA') + (AC' - CA') \Omega + (BC' -CB') \Omega^2]
 = 0.
 \label{eq:202}
 \end{equation}
This thus factorizes into two quadratic equations.
One is the equation for the geodesic
condition, $\ba = 0$, or
 \begin{equation}
 A' + 2 B' \Omega + C' \Omega^2 = 0.
 \label{eq:203}
 \end{equation}
Its solutions are the angular velocities of the
Stationary Congruences Accelerating Locally Extremely
(SCALEs) that have local minima
of the magnitude of the acceleration
(here both at the global minimum of 0).
I have been calling the angular velocities
of the SCALEs that are local minima
$\Omega_{\pm}$,
but here I shall call them $\Omega_{\pm K}$,
since for a stationary axisymmetric spacetime
they are the angular velocities of circular Keplerian orbits:
 \begin{equation}
 \Omega_{\pm K} = {B'\pm\sqrt{B'^2 - A'C'}\over -C'}.
 \label{eq:204}
 \end{equation}

	The other quadratic equation,
 \begin{equation}
 (AB'-BA')+(AC'-CA')\Omega+(BC'-CB')\Omega^2 = 0,
 \label{eq:205}
 \end{equation}
is the equation at an orbital for the
Stationary Congruence Accelerating Maximally (SCAM),
the SCALE with a local maximum of the magnitude
of the acceleration (and for an unphysical root).
The physical root has angular velocity
 \begin{equation}
 \Omega_0 = {AC'-CA'
	+\sqrt{(AC'-CA')^2 - 4(AB'-BA')(BC'-CB')}
 \over 2(CB'-BC')}.
 \label{eq:206}
 \end{equation}

	One can see from Eq. (\ref{eq:101}) that
at an orbital, a SCAM has zero rotation, $\omega^2 = 0$,
so it is obviously also a SCRAM there.
(This fact was previously noted \cite{deF94}
for the Kerr metric, and now we see that it is general.)
Therefore, at an orbital, a SCAM has the same
four-velocity as a rigid congruence with four-velocity
parallel to the Killing vector field $\bK$ with constant
$\Omega$ that at that location has zero vorticity.
(The value of $\Omega$ giving $\omega=0$ on the equatorial
plane of Kerr was first given in \cite{deFUT91},
and later \cite{deF94} de Felice discovered this is also
the angular velocity that extremizes the acceleration there.)

	Again one should be reminded that I have defined
the rotation $\bo$ so that it is the rotation of a congruence
moving along the orbits of a single Killing vector field
$\bK$ with $\Omega$ constant
(with the constant matching the angular velocity $\Omega$
of the original congruence at that location,
but not necessarily at other locations).
That is, if one inserted the $x^a$-dependent four-velocity
$\bu(x^a)$ of a SCAM at an orbital into Eq. (\ref{eq:44}),
$\bo = \frac{1}{2} \ast(\bu\wedge\bd\bu)$,
one would generically not get zero, but only if one used
in that formula the four-velocity
$\bu = e^{-\Phi}(\bk + \Omega \bl)$
with constant $\Omega$ chosen to make this $\bu$
match that of the SCAM at the position where
$\bo$ is being evaluated.
Thus a SCAM at an orbital is not generically
part of a congruence that itself has zero vorticity.
Only the rigid congruence moving along the orbits
of $\bK$ has zero vorticity there (and that congruence
generically has nonzero vorticity at other locations).

	(When we have two Killing vector fields obeying the
properties (1) - (3) above, stationary congruences
with zero vorticity are those with a constant ratio
of angular momentum to energy,
 \begin{equation}
 {\mbox{\rm angular momentum}\over
 \mbox{\rm energy}}
 = {u_{\varphi} \over -u_0} = {-B-C\Omega \over A+B\Omega}
 = \mbox{\rm const.},
 \label{eq:207}
 \end{equation}
such as the ZAMOs with $\Omega$ obeying
Eq. (\ref{eq:183}), $\Omega = -B/C$, so that
its angular momentum, and hence the ratio above, is zero.
Thus a congruence of ZAMOs has the local property of
zero vorticity \cite{NF,Sem93},
though it generically does not
have $\bo=0$ by my indirect method of defining
$\bo$ in terms of an auxiliary congruence for
each location that rotates rigidly with constant $\Omega$.
Incidentally, though the zero vorticity of a ZAMO
is a local property of that congruence, it is not sufficient
to distinguish locally that particular congruence
from the other zero-vorticity stationary congruences
with different constant values of the ratio (\ref{eq:207})
of angular momentum to energy.  The problem is that
locally there is no way to distinguish the angular
momentum connected with the Killing vector field $\bl$
with closed orbits from a combination of angular
momentum and energy connected with a different
Killing vector field, since the property only $\bl$ has,
of having closed orbits, is not a local property
that can be determined without knowing the metric
in a loop around the symmetry axis.)

	At an orbital one has,
for a fixed choice of the Killing vector fields
$\bk$ and $\bl$,
seven special values (at least)
of the angular velocity $\Omega$:
$\Omega_{+1}$ and $\Omega_{-1}$
given by Eq. (\ref{eq:30b})
(those of the speed of light in the forward
and backward directions respectively),
$\Omega_{+K}$ and $\Omega_{-K}$
given by Eq. (\ref{eq:204})
(those of the stationary geodesics
or circular Keplerian orbits
in the forward and backward directions respectively),
$\Omega_0$ given by Eq. (\ref{eq:206})
(that of the Stationary Congruence Accelerating
Maximally, or SCAM, which has the local
maximum of the magnitude of the acceleration
as a function of $\Omega$, and which,
like the geodesics with $\Omega_{\pm K}$
that are members of the other
Stationary Congruences Accelerating Locally Extremely,
or SCALEs,
are also Nonrotating Acceleration Worldlines,
or members of a NAW congruence that Fermi-Walker
transports the acceleration vector),
$\Omega_Z$ given by Eq. (\ref{eq:30f})
(that of the ZAMO, which has zero angular momentum),
and $\Omega_{NR} \equiv 0$
(that of a congruence nonrotating relative to infinity).

	The four-velocities corresponding to the first
five of these angular velocities are determined locally
and are invariant under the linear transformations
(\ref{eq:30c}) of the Killing vector fields $\bk$ and $\bl$,
but, as discussed above, that is not true of the last two,
since they require the nonlocal knowledge of which
Killing vector field is $\bk$
(e.g., the one that is timelike at infinity,
usually normalized to have unit timelike length there)
and of which one is $\bl$ (e.g., the one with closed orbits,
usually normalized to give period $2\pi$ around the orbit).
(For the first two angular velocities $\Omega_{\pm 1}$,
there are no normalized four-velocities
with those angular velocities,
so for them I mean instead the corresponding null vectors,
which are locally determined only up to normalization.)
Of course, the particular values of all but the last of these
angular velocities $\Omega$ depends on the particular
choice of $\bk$ and $\bl$, but my point is that the
corresponding $\bK$ (up to normalization) at each point
does not, for the first five angular velocities.

	Now, as one might expect, there are a number of
algebraic relations between these angular velocities
and between the corresponding four-velocities.
For example, it is well known \cite{MTW} that
the ZAMO has the average of the angular velocities
of the two null orbits,
 \begin{equation}
 \Omega_Z = {1 \over 2} (\Omega_{+1} + \Omega_{-1}),
 \label{eq:208}
 \end{equation}
which implies that if a ZAMO sent two photons around
an orbital in opposite direction (using, say, a tube to
deflect the photons into these nongeodesic null orbits
by an infinite number of glancing collisions
that each transfer an infinitesimal momentum
to the corresponding photon),
they would return to the ZAMO at the same time
\cite{MTW}.
Of course, this procedure requires the nonlocal information
of the metric around the orbital, the same
nonlocal information that is required to pick out $\bl$
as the Killing vector field with closed orbits,
a choice that is necessary before one can
define the angular velocities and get formulas
such as Eq. (\ref{eq:208}),
which is not invariant under the transformations
(\ref{eq:30c}) and (\ref{eq:30d}) if there the transformation
constant $\gamma \neq 0$.
However, one may at least note that Eq. (\ref{eq:208})
is independent of the choice of the Killing vector $\bk$
(determined by the transformation constants
$\alpha$ and $\beta$)
or of the normalization of the Killing vector $\bl$
(determined by the transformation constant $\delta$
if $\gamma=0$), even though changing
these will change the $\Omega$'s appearing
in Eq. (\ref{eq:208}).

	Another more complicated relation one may find is
 \begin{equation}
 \Omega_0 = {\sqrt{(\Omega_{+1}-\Omega_{-K})
	(\Omega_{-K}-\Omega_{-1})}\Omega_{+K}
	+\sqrt{(\Omega_{+1}-\Omega_{+K})
	(\Omega_{+K}-\Omega_{-1})}\Omega_{-K} \over
	\sqrt{(\Omega_{+1}-\Omega_{-K})
	(\Omega_{-K}-\Omega_{-1})}
	+\sqrt{(\Omega_{+1}-\Omega_{+K})
	(\Omega_{+K}-\Omega_{-1})}}.
 \label{eq:209}
 \end{equation}
Thus the SCAM angular velocity $\Omega_0$
at an orbital is a weighted average of the angular
velocities $\Omega_{\pm K}$ of the two
circular Keplerian orbits.  This relationship
{\it is} invariant under the transformations
(\ref{eq:30c}) and (\ref{eq:30d}).

	This relation simplifies greatly if one
considers instead the relative velocities
between these various observers.
One readily finds that in the frame of
an observer having four-velocity $\bu_1$
with angular velocity $\Omega_1$,
the three-velocity of a second observer
having four-velocity $\bu_2$
with angular velocity $\Omega_2$ is
 \begin{equation}
 \bv_2 \equiv {\bu_2 \over -\bu_1\!\cdot\!\bu_2} - \bu_1
 = v(\Omega_2,\Omega_1)\bL_1,
 \label{eq:210}
 \end{equation}
where the (signed) relative speed of observer 2
relative to observer 1 is \cite{Sem96}
 \begin{eqnarray}
 v(\Omega_2,\Omega_1)
 &=& {\sqrt{B^2-AC}(\Omega_2 - \Omega_1) \over
	A + B(\Omega_1 + \Omega_2) + C\Omega_1\Omega_2}
	\nonumber\\
 &=& {(\Omega_{+1}-\Omega_{-1})(\Omega_2 - \Omega_1)
 \over (\Omega_{+1}+\Omega_{-1})(\Omega_2+\Omega_1)
	- 2\Omega_{+1}\Omega_{-1} - 2\Omega_1\Omega_2},
 \label{eq:211}
 \end{eqnarray}
which is invariant under the transformations
(\ref{eq:30c}) and (\ref{eq:30d}).

	Then one can calculate \cite{Sem96} that at an orbital,
the two circular Keplerian orbits have equal and opposite
speeds in the frame of the SCAM, the magnitude
of which may be called the Keplerian orbital speed $v_K$:
 \begin{eqnarray}
 v_K &=& v(\Omega_{+K},\Omega_0)
	= - v(\Omega_{-K},\Omega_0)
	\nonumber\\
 &=& {\sqrt{(\Omega_{+1}-\Omega_{-K})
	(\Omega_{+K}-\Omega_{-1})}
	-\sqrt{(\Omega_{+1}-\Omega_{+K})
	(\Omega_{-K}-\Omega_{-1})} \over
	 \sqrt{(\Omega_{+1}-\Omega_{-K})
	(\Omega_{+K}-\Omega_{-1})}
	+\sqrt{(\Omega_{+1}-\Omega_{+K})
	(\Omega_{-K}-\Omega_{-1})}}
	\nonumber\\
 &=& {(\sqrt{(\Omega_{+1}-\Omega_{-K})
	(\Omega_{+K}-\Omega_{-1})}
	-\sqrt{(\Omega_{+1}-\Omega_{+K})
	(\Omega_{-K}-\Omega_{-1})})^2 \over
	(\Omega_{+1}-\Omega_{-1})
	(\Omega_{+K}-\Omega_{-K})}
	\nonumber\\
 &=& {(B^2-AC)' - \sqrt{(B^2-AC)'^2 - 4(B^2-AC)(B'^2-A'C')}
	\over 2\sqrt{(B^2-AC)(B'^2-A'C')}}
	\nonumber\\
 &=& {D' - \sqrt{\sigma}
	\over 2\sqrt{DH}}
	= {2\sqrt{DH} \over D' + \sqrt{\sigma}},
 \label{eq:212}
 \end{eqnarray}
where Eqs. (\ref{eq:14b}) and (\ref{eq:27b}) give
$D\equiv B^2-AC$ (minus the determinant
of the first two-dimensional block of the metric),
 \begin{equation}
 H \equiv B'^2-A'C',
 \label{eq:212b}
 \end{equation}
and
 \begin{eqnarray}
 \sigma &\equiv& D'^2 - 4DH
	\equiv (B^2-AC)'^2 - 4(B^2-AC)(B'^2-A'C')
	\nonumber\\
 &=& (AC'-CA')^2 - 4(AB'-BA')(BC'-CB')
	\nonumber\\
 &=& C^2C'^2(\Omega_{+1}-\Omega_{+K})
	(\Omega_{+1}-\Omega_{-K})
	(\Omega_{+K}-\Omega_{-1})
	(\Omega_{-K}-\Omega_{-1}).
 \label{eq:213}
 \end{eqnarray}
One may check from Eq. (\ref{eq:30g})
for the transformations
of the quantities $A$, $B$, and $C$
that the final three expressions
on the right hand side of Eq. (\ref{eq:212})
are indeed invariant
under the transformations
(\ref{eq:30c}) and (\ref{eq:30d})
and hence are locally defined.

	Therefore, a SCAM at an orbital has a four-velocity $\bu_0$
that is a normalized average between the four-velocities
$\bu_{+K}$ and $\bu_{-K}$
of the two circular Keplerian orbits:
 \begin{equation}
 \bu_0 = {\bu_{+K} + \bu_{-K} \over |\bu_{+K} + \bu_{-K}|}
	\equiv {\bu_{+K} + \bu_{-K} \over
	\sqrt{2 - 2 \bu_{+K}\!\cdot\!\bu_{-K}} }.
 \label{eq:214}
 \end{equation}
This means that if one has two equal-mass point particles
moving along opposite stationary Keplerian orbits,
and they collide to form a single particle
in a totally inelastic collision,
the velocity of this single particle immediately
after the collision will be that of the SCAM
at that location.  This relation was discovered by
Semar\'{a}k first in the special case
of the Kerr metric \cite{Sem94},
and then later in general \cite{Sem96}.
Thus we see that it is
a feature at an orbital of any stationary
axisymmetric metric invariant under reversing
both $t$ and $\varphi$.

	One can see that even if $\nabla A$, $\nabla B$,
and $\nabla C$ are all parallel (or anti-parallel),
the SCAM is not defined as a real congruence
when $\sigma$ defined by Eq. (\ref{eq:213}) goes negative.
It reaches zero when one of the two
circular Keplerian orbits reaches the speed of light
(usually when $\Omega_{-K}$ becomes as negative
as $\Omega_{-1}$ for a positively rotating source,
so that it is usually the last factor in
the last expression of Eq. (\ref{eq:213})
that goes to zero first).  Then there are no longer
two timelike stationary geodesics, so one is not
really at an orbital as defined above, even if
$\nabla A$, $\nabla B$, and $\nabla C$ are all parallel
(or anti-parallel).

	At an orbital, the two SCALEs that are not the SCAM
give the two circular Keplerian orbits.  Away from an orbital,
these two SCALEs have locally minimal, but not zero,
magnitude of acceleration.
One might think that Eq. (\ref{eq:214}) would generalize
to this case, with $\bu_{\pm K}$ being replaced by
$\bu_{\pm}$, the four-velocities of the SCALEs that
are not the SCAM, but this is not generically the case.
In other words, Eq. (\ref{eq:214}) applies only at
an orbital, the only place where there are stationary
geodesics, the circular Keplerian orbits with four-velocities
parallel to combinations of the two Killing vector fields
$\bk$ and $\bl$.  (I am always implicitly excluding
stationary geodesics that have four-velocities parallel
to any other possible Killing vector fields that might be present,
such as in a spherically symmetric spacetime.)

	Another difference from the case of of an orbital,
where the solution of the quadratic equation
for the SCAM breaks down (becomes complex)
after one of the circular Keplerian orbits
reaches the speed of light
as one enters deeper into a strong gravitational field,
is that away from an orbital,
the solution of the quartic equation
for the SCAM becomes complex
after the four-velocity for the SCAM
merges with that of one of the other SCALEs
as one enters deeper into a strong gravitational field.
(Deeper in the field there is only one extremum,
a minimum, for the magnitude of the acceleration
as a function of the angular velocity
for timelike worldlines.)
Thus one can have the case in which
the relative velocity between the SCAM and
one of the other SCALEs goes to zero
(at the boundary of the region where
the SCAM is defined as a real congruence),
whereas the relative velocity between the
the SCAM and the third SCALE can remain nonzero there.

	In the frame of the SCAM at an orbital,
one can readily show that the 3-velocity
of an observer nonrotating relative to infinity
(i.e, nonrotating relative to the $\bk$ Killing vector,
so $\Omega_{NR} = 0$),
and of a member of a ZAMO are, respectively,
 \begin{equation}
 v_{NR} = {-\Omega_0\sqrt{B^2-AC} \over A + B\Omega_0},
 \label{eq:215}
 \end{equation}
 \begin{equation}
 v_Z = {B + C\Omega_0 \over \sqrt{B^2-AC}}.
 \label{eq:216}
 \end{equation}
However, these quantities are not locally defined
and hence are not invariant under the transformations
(\ref{eq:30c}) and (\ref{eq:30d}).

	For a corotating source, usually $A$, $B$, $-C$, and
$-\Omega_0$ are positive, so both $v_{NR}$ and $v_Z$
are then positive.  In the far-field limit outside an isolated
source centered at the origin of standard spherical
polar coordinates, one gets,
using Eqs. (\ref{eq:182}) and (\ref{eq:183}),
 \begin{equation}
 v_{NR} \approx {\sqrt{-C}B' \over C'}
 \approx {J\over r^2}\sin{\theta},
 \label{eq:217}
 \end{equation}
 \begin{equation}
 v_Z \approx {BC'-CB' \over C'\sqrt{-C}}
 \approx {3J\over r^2}\sin{\theta} \approx 3v_{NR}.
 \label{eq:218}
 \end{equation}
The orbitals of an isolated source,
if any exist, are on or near
the approximate equatorial plane $\theta = 0$,
but Eqs. (\ref{eq:217}) and (\ref{eq:218})
apply at arbitrary $\theta$ outside an isolated
source in the far-field limit.

	The stationary observer of the SCAM at an orbital has
an acceleration that, in an orthogonal ($g_{23}=0$)
coordinate system for the (2,3) plane
in which the gradients of $A$, $B$, and $C$
are purely in the direction of the coordinate $r = x^2$,
has magnitude
 \begin{eqnarray}
 a_0 &=& {D' - \sqrt{\sigma} \over 4D\sqrt{g_{rr}}}
	\nonumber\\
 &=& {(B^2-AC)' - \sqrt{(B^2-AC)'^2 - 4(B^2-AC)(B'^2-A'C')}
	\over 4(B^2-AC)\sqrt{g_{rr}}}.
 \label{eq:219}
 \end{eqnarray}
(Here the subscript 0 does not denote the time component
of the acceleration, which is zero, but rather the acceleration
at the SCAM, for which the subscript 0 has been used.)
The acceleration 1-form of the SCAM is then $\ba = a_0\ber$,
with
 \begin{equation}
 \ber = \sqrt{g_{rr}}\,\bd r = \sqrt{g_{rr}}\,\nabla r
 \label{eq:219b}
 \end{equation}
being the unit 1-form in the outward radial direction.
This acceleration 1-form $\ba$ is clearly locally defined
and invariant under the transformations (\ref{eq:30c}) and
(\ref{eq:30d}).

	Now one can readily calculate that for any stationary
observer (one moving in the $\bk\wedge\bl$ plane
along the orbits of a fixed Killing vector field $\bK$)
that has speed $v$ in the frame of the SCAM at
an orbital, the acceleration 1-form is simply
 \begin{equation}
 \ba = a_0{1 - v^2/v_K^2 \over 1 - v^2}\ber.
 \label{eq:220}
 \end{equation}
In particular, the acceleration is a symmetric function
of the velocity in the SCAM frame.

	One can regard $a_0/(1-v^2)$ as being the (radial outward)
gravitational part of the acceleration,
which is indeed proportional to $\gamma^2 = 1/(1-v^2)$
as one would get by the equivalence principle
for an object moving horizontally
across the flat floor of a rocket that has constant
acceleration perpendicular to the floor.
Then $(a_0/v_K^2) v^2/(1-v^2)$ can be regarded
as the centripetal part of the acceleration,
pointing radial inward, and also having the $\gamma^2$
dependence that it does in flat spacetime.

	For example, suppose that one had an idealized
(fictitious) static, spherically symmetric metric with constant $m$,
 \begin{equation}
 ds^2 = - e^{-2m/R}dt^2 + R^2\sin^2{\theta}d\varphi^2
	+ dR^2 + R^2 d\theta^2,
 \label{eq:221}
 \end{equation}
which is spatially flat but has a spherically symmetric
gravitational potential
 \begin{equation}
 \Phi = - {m\over R}
 \label{eq:222}
 \end{equation}
for a static observer with four-velocity
$\bu = e^{-\Phi}\bK = e^{-\Phi}\bk$.
(I use the radial coordinate $R$ instead of $r$
so that later I can compare with a different
coordinate $r$ in a different metric, such as Kerr-Newman.
The radial metric component is $g_{RR} = 1$, so
the unit radial 1-form $\ber$ is simply $\bd R$ or $\nabla R$.)

	Here $B\equiv g_{t\varphi} = 0$, so the SCAM
consists of nonrotating (static) observers,
which have acceleration
 \begin{equation}
 \ba = \nabla \Phi = {m\over R^2}\ber = a_0 \ber.
 \label{eq:223}
 \end{equation}
In the equatorial plane, $\theta = \pi/2$,
which has orbitals everywhere,
a stationary observer orbiting with velocity $v$
relative to that of the static SCAM has acceleration
 \begin{equation}
 \ba = \ba_g + \ba_c
 = ({m\over R^2 (1-v^2)} - {v^2\over R(1-v^2)})\ber
 = a_0{1 - v^2/v_K^2 \over 1 - v^2}\ber
 \label{eq:224}
 \end{equation}
with $a_0 = m/R^2$ and $v_K = \sqrt{a_0 R} = \sqrt{m/R}$,
so it is natural to split up the total acceleration into
a gravitational piece
 \begin{equation}
 \ba_g = {m\over R^2 (1-v^2)}\ber
 = {a_0 \over 1 - v^2}\ber
 \label{eq:225}
 \end{equation}
and a centripetal piece
 \begin{equation}
 \ba_c = - {v^2\over R(1-v^2)}\ber
 = - {a_0 v^2/v_K^2 \over 1 - v^2}\ber.
 \label{eq:226}
 \end{equation}

	One might use this analysis with the idealized metric
(\ref{eq:221}) to define an `effective orbital radius of curvature'
 \begin{equation}
 R \equiv {v_K^2\over a_0}
 = {(D' - \sqrt{\sigma})\sqrt{g_{rr}} \over H}
 = {4D\sqrt{g_{rr}} \over D' + \sqrt{\sigma}}
 \label{eq:227}
 \end{equation}
and an `effective gravitational mass'
 \begin{equation}
 m \equiv {v_K^4\over a_0}
 = Rv_K^2
 = {R^3H\over g_{rr}D}
 = {(D' - \sqrt{\sigma})^3 \sqrt{g_{rr}} \over H}
 \label{eq:228}
 \end{equation}
for an orbital in any stationary axisymmetric metric.
These formulas are chosen to make Eqs.
(\ref{eq:224}) - (\ref{eq:226}) true for a particular
orbital in any stationary axisymmetric metric.
Then one can write the total acceleration (\ref{eq:220}) as
 \begin{equation}
 \ba = \biggl({v_K^2 - v^2 \over 1 - v^2}\biggr){\ber \over R},
 \label{eq:228ab}
 \end{equation}
which is thus much simpler in the SCAM frame than the
corresponding formula in the ZAMO frame \cite{BBI}.

	Following \cite{BBI}, we can also note that there is
a simple geometrical description of the effective
orbital radius of curvature $R$:
If the Killing vector field $\bl$
has period $\Delta\varphi=2\pi$,
then the circumferential radius
(circumference divided by $2\pi$)
measured by a stationary congruence
with constant $\Omega$ is
 \begin{equation}
 \hat{r} = \sqrt{D/F} = e^{-\Phi}\sqrt{D}.
 \label{eq:228ac}
 \end{equation}
Then if
 \begin{equation}
 ds_r \equiv \sqrt{g_{rr}}dr
 \label{eq:228ad}
 \end{equation}
is an infinitesimal element of proper distance
in the radial direction (an element of proper radius),
one can readily calculate that if $\Omega=\Omega_0$
(to be held constant during the spatial differentiation,
as usual in this paper),
 \begin{equation}
 R = \hat{r}{ds_r\over d\hat{r}},
 \label{eq:228ae}
 \end{equation}
which is what one would calculate
the proper radial distance
to be to a center (where $\hat{r}$ should vanish)
if one assumed that $\hat{r}$
varied linearly with proper distance.
E.g., for a circle of latitude in the northern hemisphere
on the surface of an axisymmetric earth in flat space,
$\hat{r}$ would be the cylindrical radial distance
from the circle to the axis inside the earth
(intersecting it orthogonally),
$ds_r$ would be an infinitesimal proper distance
along a meridian (line of constant longitude)
in the southward direction along the earth's surface,
and $R$ would be the distance along a cone
tangent to the surface of the earth at the circle,
from the circle to the apex of the cone over the
north pole.

	Then if one uses Eqs. (\ref{eq:219b}),
(\ref{eq:228ab}), (\ref{eq:228ad}), (\ref{eq:228ae}),
and the fact that
$\bd \hat{r} \equiv \nabla \hat{r} = (d\hat{r}/ds_r)\ber$,
one gets an alternative simple formula for the acceleration
of a stationary worldline of speed $v$
in the frame of the SCAM,
 \begin{equation}
 \ba = \biggl({v_K^2 - v^2 \over 1 - v^2}\biggr)
 {\nabla \hat{r} \over \hat{r}}
 = {\nabla\Phi - v^2 \nabla\ln{ \hat{r}} \over 1 - v^2},
 \label{eq:228af}
 \end{equation}

	Of course,
unlike the case for the idealized metric (\ref{eq:221}),
for a more general metric not only the effective orbital radius
of curvature $R$, but also the effective gravitational mass $m$,
may vary from orbital to orbital.
For example, for the Schwarzschild metric
 \begin{equation}
 ds^2 = - (1 - {2M\over r})dt^2 + r^2\sin^2{\theta}d\varphi^2
	+ (1 - {2M\over r})^{-1}dr^2 + r^2 d\theta^2
 \label{eq:228b}
 \end{equation}
with constant ADM mass $M$, one has
 \begin{equation}
 a_0 = {M\over r^2} (1 - {2M\over r})^{-1/2}
 \label{eq:228c}
 \end{equation}
and
 \begin{equation}
 v_K = \sqrt{{M\over r}}  (1 - {2M\over r})^{-1/2}
 = \sqrt{{M\over r-2M}},
 \label{eq:228d}
 \end{equation}
so one gets for the effective radius $R$ and mass $m$
 \begin{equation}
 R = r (1 - {2M\over r})^{-1/2}
 \label{eq:228e}
 \end{equation}
and
 \begin{equation}
 m = M (1 - {2M\over r})^{-3/2}.
 \label{eq:228f}
 \end{equation}

	In the Schwarzschild metric,
one can compensate for this dependence
of the effective mass $m$
on the radial coordinate $r$
by defining an `effective Scharzschildean mass'
 \begin{eqnarray}
 \hat{M} &\equiv& {m\over(1+2m/R)^{-3/2}}
 = {v^4_K \over a_0(1+2v^2_K)^{3/2}}
	\nonumber\\
 &=&{16\sqrt{g_{rr}}D^2H \over
	[(D'+\sqrt{\sigma})^2+8DH]^{3/2}},
 \label{eq:228g}
 \end{eqnarray}
which for the Schwarzschild metric is designed to give
precisely $M$.

	Since only the total acceleration $\ba$ is observable,
one is free to divide it up into `gravitational' and `centripetal'
contributions any way one wishes.  Although the split
given by Eqs. (\ref{eq:224}) - (\ref{eq:226}) above
seems most natural to me (and to several others
\cite{deF91,deFUT91,deFUT93,PageSciAm,deF94,
Sem94,Sem95,BBI,Sem96a,SB}),
Abramowicz and his collaborators
\cite{AbrLas,ACL,AbrPra,AbrMil,Abr90,All,Abr92,Abr93,
AbrSzu,ANW,ANW2,SM}
have advocated an alternative split
 \begin{equation}
 \ba = \bat_g + \bat_c
 \label{eq:229}
 \end{equation}
with
 \begin{equation}
 \bat_g = a_0\ber
 \label{eq:230}
 \end{equation}
being independent of velocity and all the velocity
dependence being put into the centripetal piece
 \begin{equation}
 \bat_c = - {a_0(1 - v_K^2)v^2 \over v_K^2(1 - v^2)}\ber.
 \label{eq:231}
 \end{equation}

	If one wished, with this alternative split
one could define an alternative effective radius
$\tilde{R}$ and effective gravitational mass $\tilde{m}$
so that
 \begin{equation}
 \bat_g = a_0\ber = {\tilde{m}\over\tilde{R}^2}\ber
 \label{eq:232}
 \end{equation}
and
 \begin{equation}
 \bat_c = - {v^2\over \tilde{R}(1-v^2)}\ber.
 \label{eq:233}
 \end{equation}
Then the alternative effective radius is
 \begin{equation}
 \tilde{R} \equiv {v_K^2\over a_0 (1-v_K^2)}
 = {R \over 1-v_K^2}
 = {R^2 \over R - m},
 \label{eq:234}
 \end{equation}
and the alternative effective gravitational mass is
 \begin{equation}
 \tilde{m} \equiv {v_K^4\over a_0 (1-v_K^2)^2}
 = {mR^2 \over (R - m)^2}.
 \label{eq:235}
 \end{equation}

	One can also give the same geometrical
interpretation for $\tilde{R}$ as for $R$
if one uses instead of $\hat{r}$ the
optical circumferential radius
 \begin{equation}
 \tilde{r} = e^{-\Phi}\hat{r} = e^{-2\Phi}\sqrt{D}
 \label{eq:236}
 \end{equation}
from the optical metric
 \begin{equation}
 d\tilde{s}^2 = e^{-2\Phi}ds^2.
 \label{eq:237}
 \end{equation}
Then one has
 \begin{equation}
 \tilde{R} = \tilde{r}{ds_r\over d\tilde{r}},
 \label{eq:238}
 \end{equation}
and
 \begin{equation}
 \ba = a_0\ber
 - \biggl({v^2 \over 1 - v^2}\biggr){\bd \tilde{r}
 \over \tilde{r}}
 = \nabla\Phi
 - \biggl({v^2 \over 1 - v^2}\biggr)\nabla \ln{\tilde{r}}.
 \label{eq:239}
 \end{equation}

	When the circular Keplerian orbital velocity $v_K$
exceeds 1 (the speed of light), $\bat_c$, $d\tilde{r}$,
and $\tilde{R}$ reverse sign,
which is what Abramowicz calls the reversal
of the sign of the centrifugal acceleration.
However, this interpreted reversal depends
on the particular splitting of the total acceleration
$\ba$ into gravitational and centripetal parts
given by Eqs. (\ref{eq:229}) - (\ref{eq:231})
so that the gravitational part $\bat_g$ is defined
to be independent of the orbital velocity $v$.
To me it seems more natural
to use the split given by
Eqs. (\ref{eq:224}) - (\ref{eq:226})
so that the gravitational part $\ba_g$
has the $\gamma^2$ velocity dependence
that one would expect if gravity couples to energy
rather than, say, to rest mass.
In an equivalence-principle argument,
one can readily calculate
that the acceleration for objects skimming
horizontally over a spatially flat floor
of a rocket having constant acceleration vertically
in flat spacetime indeed has this
$\gamma^2$ velocity dependence.

	Nevertheless, even in the rocket example,
one could follow Abramowicz and say that part
of the acceleration is centripetal
acceleration from the curvature of the floor
world membrane (the timelike three-surface
that is the world history of the accelerating floor).
Indeed, the floor looks curved
as seen by photon geodesics
(which the floor will intercept twice
if they are not traveling purely vertically),
even though at each moment of time
the instantaneous two surface of the floor
is flat and contains tachyon geodesics that
move at infinite speed.
However, one cost of Abramowicz's interpretation
is that no gravitational acceleration would be
ascribed to photons, so that they would be
interpreted as having no weight,
even though they have energy.

\subsection{Near a rotation axis}

	A third location in which the properties of the SCAM
are more simple than in the generic location
(besides the two discussed above, first of the slow-rotation
and/or far-field limit that one would expect to find
far from an isolated source,
and second at the location
of orbitals where both corotating
and counterrotating circular Keplerian geodesic orbits exist)
is infinitesimally near a rotation axis
where $\bl$ vanishes.
If $\bl$ is given its standard normalization of
having closed orbits with period $2\pi$
(i.e., if $\bl = \partial / \partial \varphi$ with $\varphi$
being periodic with period $2\pi$),
then regularity of the metric near the axis
implies that $C = -\bl\!\cdot\!\bl$ must go as
$-\varpi^2 +O(\varpi^4)$ (with unit coefficient),
to lowest order being
the negative square of the distance $\varpi$ from the axis
along a spatial geodesic that is orthogonal to $\bk$
and $\bl$ and which intersects the axis orthogonally.
We can let $\varpi$ be one of the cylindrical coordinates
for the (2,3) plane orthogonal to $\bk$ and $\bl$
and let $z$ be the other, with the property that $z$
is proper distance along the axis and that each
spatial geodesic mentioned above that intersects the
axis orthogonally has constant $z$, so that
$\nabla \varpi$ and $\nabla z$ are orthogonal
everywhere and the former is normalized
to have unit magnitude.

	Then in a neighborhood of the axis where these
two coordinates are well behaved, the metric may
be written
 \begin{eqnarray}
 ds^2 \!\!&=&\!\! - A(\varpi,z)dt^2 - 2B(\varpi,z)dt d\varphi
 - C(\varpi,z)d\varphi^2 + d\varpi^2 + g_{33}(\varpi,z)dz^2
 \nonumber\\
 &=&\!\! - [A_0(z) \!+\! A_2(z)\varpi^2 \!+\! O(\varpi^4)]dt^2
 \!-\! 2[\Omega_Z(z)\varpi^2 \!+\! O(\varpi^4)]dt d\varphi
 \!+\![\varpi^2 \!+\! O(\varpi^4)]d\varphi^2
	\nonumber\\
 &&
 + d\varpi^2 \!+\! [1 \!+\! O(\varpi^2)]dz^2.
	\nonumber\\
 \label{eq:251}
 \end{eqnarray}
On the axis the ZAMO {\it is} locally determined,
since $\bl$ is uniquely determined
by the requirements of the previous paragraph
if one has access to the axis, unlike the case
away from the axis, where one does not have local
access to the axis.
In particular,
 \begin{equation}
 \Omega_Z(z)
 \equiv \lim_{\varpi\rightarrow 0}{-{B(\varpi,z)
 \over C(\varpi,z)}}
 = -{1\over 2}k_{\alpha;\beta}l^{\alpha;\beta},
 \label{eq:252}
 \end{equation}
where the last expression is to be evaluated on
the axis, at $\varpi=0$.
However, note that although the ZAMO is locally
defined at the axis, the angular velocity $\Omega_Z$
ascribed to it is not, since the latter depends on
the definition of $\bk$, which is not locally
determined but can be redefined by multiplying it
by a constant and by adding any constant multiple
of $\bl$.

	Similarly,
 \begin{equation}
 A_0(z) \equiv A(\varpi=0,z),
 \label{eq:253}
 \end{equation}
and
 \begin{equation}
 A_2(z) \equiv \lim_{\varpi\rightarrow 0}
	{\nabla A(\varpi,z)\!\cdot\!\nabla C(\varpi,z)
	\over 4C(\varpi,z)}
 = {1\over 8}(2\nabla^2 A - 2A''_0(z) - A'_0(z)^2/A_0(z)),
 \label{eq:254}
 \end{equation}
where in the last expression
the (four-dimensional) Laplacian of $A$
is to be evaluated on the axis,
and I am here using the notation
that a prime denotes a partial derivative
with respect to $z$.
Here I shall also use the convention that when
I give the functional dependence of any quantity
as $(z)$, I mean that the quantity is to be evaluated
at $\varpi=0$.

	If $\bez = \bd z$ is the unit 1-form in the
$z$-direction on the axis,
then one can readily calculate
that the `nonrotating'
$(\Omega = \Omega_{NR}\equiv 0)$
congruence with four-velocity $\bu = A^{-1/2}\bk$
has, on the axis, an acceleration 1-form
 \begin{equation}
 \ba(z) \equiv a(z)\bez = {A'_0(z)\over 2 A_0(z)}\bez
 \label{eq:255}
 \end{equation}
and a normalized rotation 1-form
 \begin{equation}
 \bo_{NR}(z) \equiv \omega_{NR}\bez
 = -A^{-1/2}_0(z)\Omega_Z\bez.
 \label{eq:256}
 \end{equation}
This is negatively rotating, since it is the ZAMO
(and not, for example, the SCAM)
which has the angular velocity $\Omega_Z$
of the Killing vector field $\bK = \bk + \Omega_Z\bl$,
which, when $\Omega_Z$ is held constant,
has zero rotation or vorticity precisely on the axis,
$\bo_Z(z)=0$.
(This is rather opposite to the case at an orbital,
where it is the SCAM rather than the ZAMO
which has the angular velocity of the
Killing vector field that has zero rotation
or vorticity there.
Thus a SCRAM, which has $\Omega$ chosen
at each location to minimize $\omega^2$ there,
should interpolate between the SCAM at any orbitals
and the ZAMO on an axis of symmetry.)

	One should also note that precisely on the axis
where $\bl$ vanishes, $\bu = A^{-1/2}_0(z)\bk$
for any $\Omega$, so the acceleration $\ba(z)$ there
is independent of $\Omega$.
Thus, strictly speaking, the definition of a SCALE
does not work precisely on the axis,
but it does if one moves infinitesimally off,
and then one can define the SCALEs on the axis
to have the limit of $\Omega(\varpi)$
as $\varpi\rightarrow 0$.

	Now when one evaluates the quantity on the
left hand side of Eq. (\ref{eq:30a}) for a SCALE
near the axis of symmetry, one finds that it goes as
$\varpi^2$ plus higher-order terms in $\varpi$,
as the acceleration goes as $\ba(z)$ plus
a correction term that to lowest order
is proportional to $\varpi^2$ with
a coefficient that depends on $\Omega$.
If one divides the left hand side
of Eq. (\ref{eq:30a}) by $\varpi^2$
and takes the limit of $\varpi$ going to zero,
one gets on the axis not a quartic but a cubic
for the SCAM $\Omega_0$ and the other two
SCALES $\Omega_{\pm}$:
 \begin{equation}
 (\Omega - \Omega_Z)^3
 - (A_2 + \Omega^2_Z
 - {A'^2\over 4A})(\Omega - \Omega_Z)
 + {1\over 4}A'\Omega'_Z = 0.
 \label{eq:257}
 \end{equation}
Here all the quantities are to be evaluated on the axis
$\varpi=0$ and so are functions purely of $z$,
and for compactness I have used $A$ and its
$z$-derivative $A'$ instead of $A_0$ and $A'_0$,
since they are the same quantities on the axis.

	There are three distinct real solutions
of this equation, giving one local maximum
(the angular velocity $\Omega_0$ of the SCAM)
and two local minima (the angular velocities
$\Omega_{\pm}$ of the remaining two SCALEs)
of the magnitude of the acceleration,
if the discriminant of the cubic,
which is proportional to
 \begin{equation}
 27A'^2\Omega'^2_Z
 - 16(A_2 + \Omega^2_Z - {A'^2\over 4A})^3,
 \label{eq:258}
 \end{equation}
is negative.
If, on the other hand, this quantity is positive,
there is only one real solution, corresponding
to the minimum value of the acceleration,
and thus the SCAM does not exist there.

	One notes that the coefficient of the
quadratic term in $\Omega - \Omega_Z$ is zero.
Thus when all three solutions are real
and hence denote the angular velocities
of the three SCALEs,
 \begin{equation}
 \Omega_0 +\Omega_- +\Omega_+ = 3\Omega_Z,
 \label{eq:259}
 \end{equation}
so the ZAMO angular velocity is the average
of that of the three SCALES on the axis.

	Instead of writing the cubic Eq. (\ref{eq:257})
for the SCALEs in terms of their angular velocities
(which depend on the choice of $\bk$),
it may be more illuminating to write it in terms
of the orthonormal $z$-component $\omega$,
on the axis, of the normalized rotation 1-form $\bo$
of the rigidly rotating congruence
with the corresponding angular velocity $\Omega$
at the location in question on the axis,
 \begin{equation}
 \bo(z) \equiv \omega\bez
 = A^{-1/2}_0(z)(\Omega - \Omega_Z)\bez.
 \label{eq:260}
 \end{equation}

	Then a bit of algebra, including the use of the
standard formula \cite{Wald}
 \begin{equation}
 \xi_{\alpha;\beta\gamma}
 = R_{\alpha\beta\gamma\delta}\xi^{\delta}
 \label{eq:261}
 \end{equation}
for any Killing vector field $\bxi$ to eliminate its
covariant derivatives of order higher than one,
gives the following simple
cubic for the orthonormal rotation component
of the limit of a SCALE on the axis:
 \begin{equation}
 \omega^3 - p \omega - q = 0,
 \label{eq:262}
 \end{equation}
where
 \begin{equation}
 p = R_{\hat{t}\varpi\hat{t}\varpi} - a^2
 = {1\over 2}(R_{\hat{t}\hat{t}} - 3a^2 - a')
 \label{eq:263}
 \end{equation}
and
 \begin{equation}
 q = {1\over 2}aR_{\hat{\varphi}\varpi z \hat{t}}
 = {1\over 2}a(\omega'_R + a\omega_R).
 \label{eq:264}
 \end{equation}

	Here all of the quantities
(such as the orthonormal Riemann and Ricci
curvature components, and
the orthonormal $z$-component $a$
of the acceleration---its only nonzero component)
are to evaluated in the limit of going onto the axis,
the prime denotes a derivative
with respect to proper length $z$ along the axis, and
 \begin{equation}
 \omega_R = A^{-1/2}(\Omega_R - \Omega_Z)
 \label{eq:265}
 \end{equation}
is the $z$-dependent rotation of any
rigidly rotating ($\Omega_R = {\rm const.}$)
congruence, e.g., the nonrotating congruence
with $\Omega = 0$, though to define this particular
congruence requires a specific choice
of $\bk$ that is not
required in Eq. (\ref{eq:265}).
One can easily see that $(\omega'_R + a\omega_R)$
is invariant under changing from one allowed
rigidly rotating congruence to another by changing
the constant $\Omega_R$ in Eq. (\ref{eq:265}),
so the solutions of the cubic Eq. (\ref{eq:262})
do not depend on this choice.
In other words, the coefficients $p$ and $q$
are both invariant under the allowed
transformations (\ref{eq:30c}) and (\ref{eq:30d}),
which on the axis are restricted to have $\gamma = 0$
and $\delta =1$ so that only $\bk$, but not $\bl$,
may be changed.

	The explicit
solutions of the cubic Eq. (\ref{eq:262})
may be written as
 \begin{equation}
 \omega_0 = -2\sqrt{{p\over 3}}
 \sin{[{1\over 3}\sin^{-1}{(q\sqrt{{27\over 4p^3}})}]}
 \label{eq:266}
 \end{equation}
for the normalized rotation rate of the SCAM, and
 \begin{equation}
 \omega_{\pm} = 2\sqrt{{p\over 3}}
	\sin{[\pm{2\pi\over 3}
	 - {1\over 3}\sin^{-1}{(q\sqrt{{27\over 4p^3}})}]}.
 \label{eq:267}
 \end{equation}
for the rotations of the other two SCALEs on the axis.

	After solving the cubic Eq. (\ref{eq:262})
for the invariant normalized rotation rates $\omega$,
one can use Eq. (\ref{eq:260}) to solve for
the angular velocities $\Omega_0$ and $\Omega_{\pm}$,
if one wishes, for a particular
choice of $\bk$ and hence of $A$ and of $\Omega_Z$.
However, these angular velocities $\Omega$
are not invariant under the allowed
transformations (\ref{eq:30c}) and (\ref{eq:30d})
that change $\bk$, whereas the normalized
local rotation rates $\omega$
(of congruences with constant angular velocities
$\Omega$ that match those of the SCALEs
at the chosen location on or infinitesimally near the axis)
are invariant under these transformations
of the Killing vector field $\bk$.

	To first order in $q(27/4p^3)^{1/2}$
when $q^2 \ll p^3$,
the explicit solutions (\ref{eq:266}) and (\ref{eq:267})
reduce to the approximations
 \begin{equation}
 \omega_0 \approx -{q\over p}
 = {aR_{\hat{\varphi}\varpi z \hat{t}} \over
	R_{\hat{t}\varpi\hat{t}\varpi} - a^2},
 \label{eq:268}
 \end{equation}
 \begin{equation}
 \omega_{\pm} \approx \pm\sqrt{p} + {q\over 2p}.
 \label{eq:269}
 \end{equation}
For example, in the far-field limit outside
an isolated stationary source of mass $M$
and intrinsic angular momentum $J$
at rest at the origin,
with $z \gg M + \sqrt{J}$
being the positive proper distance
along the axis from the source,
in the direction of the angular momentum vector,
and with the stress-energy tensor
(and hence $R_{\hat{t}\hat{t}}$,
assuming Einstein's equations)
being negligible there,
one has on the axis $A = -g_{tt} \approx 1 - 2M/z$,
$a = A'/2A \approx M/z^2$,
$a'\equiv da/dz \approx -2M/z^3 \ll -a^2$,
$R_{\hat{t}\varpi\hat{t}\varpi}
 = {1\over 2}(R_{\hat{t}\hat{t}} - a^2 - a') \approx M/z^3$,
so $p \approx M/z^3$,
and then choosing $\Omega_R = 0$ gives
$\omega_R = -A^{-1/2}\Omega_Z \approx - \Omega_Z
\approx -2J/z^3$,
$\omega'_R \equiv d\omega_R/dz \approx 6J/z^4
\gg |a\omega_R|$, so $R_{\hat{\varphi}\varpi z \hat{t}}
 = \omega'_R + a\omega_R \approx 6J/z^4$,
and $q \approx 3MJ/z^6$.
Then Eqs. (\ref{eq:268}) and
(\ref{eq:269}) give
 \begin{equation}
 \omega_0 \approx - 3J/z^3
 \label{eq:270}
 \end{equation}
and
 \begin{equation}
 \omega_{\pm} \approx
 \pm\sqrt{M/z^3} + (3/2)J/z^3
 \approx \pm\sqrt{R_{\hat{t}\varpi\hat{t}\varpi}}.
 \label{eq:271}
 \end{equation}
Inserting Eq. (\ref{eq:270}) into Eq. (\ref{eq:260})
then gives $\Omega_0 \approx -J/z^3$,
which agrees with the general far-field
Eq. (\ref{eq:182}) with $r=z$.

	One can iterate the approximations of
Eqs. (\ref{eq:268}) and (\ref{eq:269}) by
rewriting the cubic Eq. (\ref{eq:262}) as
 \begin{equation}
 \omega_0 = {-q + \omega_0^3 \over p}
 \label{eq:272}
 \end{equation}
and as
 \begin{equation}
 \omega_{\pm} = \pm\sqrt{p + {q\over\omega_{\pm}}}.
 \label{eq:273}
 \end{equation}
Then one starts with setting the terms involving
$\omega_0$ or $\omega_{\pm}$ equal to zero
on the right hand sides, evaluates the right hand
sides to get the first approximations for the left
hand sides, enters these approximations back into the
right hand sides to get the second approximations,
and iterates to get the desired accuracy.
This procedure is often a better way to proceed
when $q^2 \ll p^3$
(particularly when one has functional expressions
for $p$ and $q$ in terms of some coordinate
like $z$ along the axis) than to use the explicit
solutions (\ref{eq:266}) and (\ref{eq:267})
of the cubic Eq. (\ref{eq:262}).

	The discriminant of the cubic Eq. (\ref{eq:262}),
which is proportional to
 \begin{equation}
 108q^2 - 16p^3 = 27a^2(R_{\hat{\varphi}\varpi z \hat{t}})^2
 - 16(R_{\hat{t}\varpi\hat{t}\varpi} - a^2)^3,
 \label{eq:274}
 \end{equation}
is negative when the magnitude
of the argument of the inverse sine
in the exact solutions (\ref{eq:266}) and (\ref{eq:267}),
$(27q^2/4p^3)^{1/2}$, is less than unity, leading to
three real solutions for $\omega$.
When one gets sufficiently deep into a strong
rotating gravitational field that $27q^2$ exceeds $4p^3$,
the roots $\omega_0$ and $\omega_-$ merge
at $-\sqrt{p/3}$ and then go off into the complex plane,
leaving no real $\omega_0$ for a SCAM but only the single
remaining real root for the SCALE
that has a global minimum of the acceleration,
with normalized rotation
 \begin{equation}
 \omega_+
 = ({q\over 2})^{1/3}[(1+\sqrt{1-{4p^3\over 27q^2}})^{1/3}
	+ (1-\sqrt{1-{4p^3\over 27q^2}})^{1/3}].
 \label{eq:275}
 \end{equation}

\section{The SCAM and other SCALEs \newline
in the Kerr-Newman metric}

	Now let us evaluate the properties
discussed above of the SCAM
(Stationary Congruence Accelerating Maximally)
and other SCALEs (Stationary Congruences
Accelerating Locally Extremely)
in the Kerr-Newman metric
 \begin{equation}
 ds^2 = -{\Delta\over \rho^2}[dt-a(1-c^2)d\varphi]^2
		+{1-c^2\over \rho^2}[(r^2+a^2)d\varphi - adt]^2
		+{\rho^2\over \Delta}dr^2
		+ {\rho^2\over 1-c^2}dc^2,
 \label{eq:301}
 \end{equation}
where, to keep everything algebraic,
I have used $c\equiv\cos{\theta}$ instead of $\theta$
in what are otherwise Boyer-Lindquist coordinates.
Here, as usual
(see, e.g., \cite{MTW,Carter72,Wald,Chandra,NF})
 \begin{equation}
 a\equiv J/M
 \label{eq:302}
 \end{equation}
is such a standard Kerr parameter that I shall continue
to use it in expressions directly involving the metric,
even though elsewhere I use it for the acceleration,
 \begin{equation}
 \Delta \equiv r^2-2Mr+a^2+Q^2,
 \label{eq:303}
 \end{equation}
which is $D/(1-c^2)$ in terms of
$D = g_{01}^2-g_{00}g_{11}$
defined by (\ref{eq:14b}) and (\ref{eq:27b}) above,
and
 \begin{equation}
 \rho^2 \equiv r^2 + a^2 c^2.
 \label{eq:304}
 \end{equation}

	The Kerr-Newman metric thus has
 \begin{equation}
 A \equiv -g_{tt} = 1 - {2Mr-Q^2\over \rho^2} = 1 - 2U,
 \label{eq:305}
 \end{equation}
 \begin{equation}
 B \equiv -g_{t\varphi}
 = {a(2Mr-Q^2)(1-c^2)\over \rho^2} = 2aU(1-c^2),
 \label{eq:306}
 \end{equation}
 \begin{eqnarray}
 C &\equiv& -g_{\varphi\varphi}
 = - (r^2+a^2)(1-c^2) - {a^2(2Mr-Q^2)(1-c^2)^2\over \rho^2}
	\nonumber\\
 &=& - (r^2+a^2)(1-c^2) - 2U a^2 (1-c^2)^2,
 \label{eq:307}
 \end{eqnarray}
where
 \begin{equation}
 U \equiv {2Mr-Q^2\over 2\rho^2}
 \equiv {Mr-Q^2/2\over r^2 + a^2 c^2}
 \label{eq:308}
 \end{equation}
is a particular generalization of the Newtonian potential
(with the sign reversed to make it positive
for $r>Q^2/(2M)$).
In fact, another way to write the Kerr-Newman metric above is
 \begin{eqnarray}
 ds^2 &=& -dt^2 +(r^2+a^2)(1-c^2)d\varphi^2
		+ {r^2+a^2c^2\over r^2+a^2}dr^2
		+ {r^2+a^2c^2\over 1-c^2}dc^2
		\nonumber \\
	&+& 2U\{[dt-a(1-c^2)d\varphi ]^2
	+{(r^2+a^2c^2)^2 dr^2\over (r^2+a^2)(r^2+a^2-2Mr+Q^2)}\},
 \label{eq:309}
 \end{eqnarray}
where the first line is simply flat spacetime
in spheroidal coordinates \cite{Chandra}
 \begin{equation}
 r \equiv \sqrt{{1\over 2}[x^2+y^2+z^2-a^2
	+\sqrt{(x^2+y^2+z^2-a^2)^2+4a^2z^2}]},
 \label{eq:310}
 \end{equation}
 \begin{equation}
 c \equiv \cos{\theta} \equiv z/r.
 \label{eq:311}
 \end{equation}

	One can in principle obtain
the angular velocity $\Omega$ of the SCAM
in the Kerr-Newman metric
as a function of the coordinates $r$ and $c$
by evaluating Eq. (\ref{eq:30a}) and setting it equal to zero,
which gives a quartic equation for $\Omega$.
However, when this equation is rationalized and
expanded out in powers of $a$, $M$, $Q$, $r$, $c$,
and $\Omega$, one gets literally hundreds of terms.
Thus it seems likely that
the explicit solution would take more
space to print than the entire rest of this paper,
so I have not bothered to do that.

	In this way a SCAM, though it is simple to specify
implicitly, is not nearly so simple to specify explicitly
(e.g., by $\Omega(r,c)$) as a Zero Angular Momentum
Observer (ZAMO \cite{Bardeen,MTW}) or even as an
Extremely Accelerated Observer (EAO \cite{Sem93,Sem94}),
which is a Stationary Observer whose angular
velocity extremizes the cylindrical radial component
of the acceleration, and which can be given by
a seven-line explicit expression for the Kerr metric
($Q=0$).

	However, one can specify explicitly the SCAM
in the three limiting cases described above for
a general stationary axisymmetric metric.

\subsection{In the slow-rotation limit}

	First, consider the case in which the Kerr parameter
$a$ is much smaller than $M$.
This is the slow-rotation limit, and evaluating
Eq. (\ref{eq:181}) to first order in $a$
(but including the lowest-order term in $1/r$ which is cubic
in the Kerr parameter $a\equiv J/M$, the first term that
depends on the angular variable $c\equiv \cos{\theta}$)
gives the angular velocity of the SCAM as
 \begin{eqnarray}
 \Omega_0
 &=& {-a(Mr^2-Q^2r+3a^2c^2)\over
	r^3(r^2-3Mr+2Q^2)}
	+ O\Bigl({a^3 M^2\over r^6},{a^3 Q^2\over r^6}\Bigr)
	\nonumber\\
 &=& -{aM\over r^3}\Bigl[1+{3M^2-Q^2\over Mr}
	+{9M^2-5Q^2+3a^2c^2\over r^2}
	+O\Bigl({a^2 M\over r^3},
	{a^2 Q^2\over Mr^3}\Bigr)\Bigr].
 \label{eq:312}
 \end{eqnarray}
This requires that
 \begin{equation}
 r > {1\over 2}(3M+\sqrt{9M^2-8Q^2})
 \label{eq:313}
 \end{equation}
in order that the denominator of the first term
on the right hand side
of Eq. (\ref{eq:312}) not have changed
sign as one comes in from infinity,
but the denominator may be arbitrarily small,
so long as $a$ is sufficiently small
(e.g., not only small compared with $M$,
but also much smaller than the square root
of the denominator divided by $Mr^2$).

	This has the consequence that,
in the slow-rotation limit,
the SCAM for Kerr-Newman is defined for values
of the radial variable $r$ obeying the inequality
(\ref{eq:313}), and its angular velocity is given
by Eq. (\ref{eq:312}) to good accuracy
so long as the magnitude of this expression
for $\Omega_0$ is much smaller than $1/a$.

	By comparison, in the Kerr-Newman metric
the ZAMO has angular velocity given explicitly by
 \begin{equation}
 \Omega_Z = {a(2Mr-Q^2) \over
 (r^2 + a^2)^2 - a^2 \Delta (1-c^2)}
 = {a(2Mr-Q^2) \over r^4} + O(a^3).
 \label{eq:314}
 \end{equation}

	Obviously, in the far-field limit,
$r^2 \gg M^2 + Q^2$, Eqs. (\ref{eq:312}) and (\ref{eq:314})
agree with Eqs. (\ref{eq:182}) and (\ref{eq:183}).
However, outside the far-field limit but still within
the slow-motion limit, one does not have the simple
relation $\Omega_Z = -2\Omega_0$ that one has
in the far-field limit.

\subsection{At the orbitals (on the equatorial plane)}

	Second, consider the case of the orbitals
(locations of pairs of stationary geodesic observers,
corotating and counterrotating
timelike circular Keplerian orbits).
These all occur on the equatorial plane ($c=0$)
of the Kerr-Newman metric.
There the angular velocities of the speed of light,
given in general by Eq. (\ref{eq:30b}), are
 \begin{equation}
 \Omega_{\pm 1}
 = {\pm r^2\sqrt{r^2-2Mr+a^2+Q^2}+a(2Mr-Q^2)
	 \over r^2(r^2+a^2) + a^2(2Mr-Q^2)},
 \label{eq:316a}
 \end{equation}
and the angular velocities of the SCALEs that
are the Keplerian orbiting stationary geodesics,
given in general by Eq. (\ref{eq:204}), are
 \begin{equation}
 \Omega_{\pm K} = {\pm\sqrt{Mr-Q^2}
	\over r^2\pm a\sqrt{Mr-Q^2}}.
 \label{eq:316b}
 \end{equation}

	On the equatorial plane the other SCALE, the SCAM
which gives a local maximum of the acceleration,
reduces to what Semer\'{a}k \cite{Sem93,Sem94} calls an
Extremely Accelerated Observer (EAO),
with Eq. (\ref{eq:206}) giving its angular velocity as
 \begin{eqnarray}
 \Omega_0 \!\!\!\!&=&\!\!\!\!
-{r^2(r^2\!-\!3Mr\!+\!2Q^2)\!-\!2a^2(Mr\!-Q^2)
	\!-\!r^2\sqrt{(r^2\!-\!3Mr\!+\!2Q^2)^2\!-\!4a^2(Mr\!-\!Q^2)}
	\over 2a[r^2(3Mr-2Q^2)+a^2(Mr-Q^2)]}
	\nonumber \\
 &=&\!\!\!\!{-2a(Mr-Q^2)\over
	r^2(r^2\!-\!3Mr\!+\!2Q^2)\!-\!2a^2(Mr\!-\!Q^2)
	\!+\!r^2\sqrt{(r^2\!-\!3Mr\!+\!2Q^2)^2\!-\!4a^2(Mr\!-\!Q^2)}}
	\nonumber \\
	&=&\!\!\!\!{-a(Mr-Q^2)\over r^2(r^2-3Mr+Q^2)}
	[1+{a^2(Mr-Q^2)(2r^2-3Mr+2Q^2)
	\over r^2(r^2-3Mr+2Q^2)^2}
		+O(a^4)],
	\nonumber \\
 &=&\!\!\!\!-{a\over r^3}[M\!+\!{3M^2\!-\!Q^2\over r}
	\!+\!{9M^3\!-\!5MQ^2\over r^2}
	\!+\!{27M^4\!-\!21MQ^2\!+\!2Q^4\!+\!2a^2M^2\over r^3}
	\!+\!O({1\over r^4})],
	\nonumber\\
	&&
 \label{eq:316}
 \end{eqnarray}
a straightforward extension to $Q\neq 0$ of the result
in the Kerr metric \cite{Sem93,Sem94,deF95}.

	The velocities of the circular Keplerian orbits in the SCAM
frame (both of equal magnitudes but of opposite signs,
as Semer\'{a}k found was the case in the Kerr equatorial plane
\cite{Sem94} and later found in general \cite{Sem96})
have magnitude given by Eq. (\ref{eq:212})
specialized to the Kerr-Newman metric:
 \begin{eqnarray}
 v_K &=& {r(r-M) - \sqrt{(r^2-3Mr+Q^2)^2-4a^2(Mr-Q^2)}
	\over 2\sqrt{(Mr-Q^2)(r^2-2Mr+a^2+Q^2)} }
	\nonumber \\
	&=&{ 2\sqrt{(Mr-Q^2)(r^2-2Mr+a^2+Q^2)}\over
	r(r-M) + \sqrt{(r^2-3Mr+Q^2)^2-4a^2(Mr-Q^2)}}
	\nonumber \\
	&=&\sqrt{{Mr-Q^2\over r^2-2Mr+Q^2}}
	[1+{a^2 r(r-M)\over 2(r^2-2Mr+Q^2)(r^2-3Mr+2Q^2)^2}
		+O(a^4)].
	\nonumber \\
	&&
 \label{eq:317}
 \end{eqnarray}
Similarly, Eq. (\ref{eq:219})
specialized to the Kerr-Newman metric
gives the magnitude of the acceleration of
the equatorial SCAM as
 \begin{eqnarray}
 a_0 &=&{\sqrt{Mr-Q^2}\over r^2}v_K
	\nonumber \\
	&=& {r(r-M) - \sqrt{(r^2-3Mr+Q^2)^2-4a^2(Mr-Q^2)}
	\over 2r^2\sqrt{r^2-2Mr+a^2+Q^2} }
	\nonumber \\
	&=&{ 2(Mr-Q^2)\sqrt{r^2-2Mr+a^2+Q^2}\over
	r^2[r(r-M) + \sqrt{(r^2-3Mr+Q^2)^2-4a^2(Mr-Q^2)}]}
	\nonumber \\
	&=&{Mr-Q^2\over r^2\sqrt{r^2-2Mr+Q^2}}
	[1+{a^2 r(r-M)\over 2(r^2-2Mr+Q^2)(r^2-3Mr+2Q^2)^2}
		+O(a^4)]
	\nonumber \\
	&&
 \label{eq:318}
 \end{eqnarray}
Then one can use Eq. (\ref{eq:220}) to get the acceleration
of a stationary observer at any speed $v$ relative to that
of the SCAM on the equatorial plane.

	Many of these formulas for the properties of the SCAM
on the equatorial plane of the Kerr-Newman metric are simpler
in the case of extreme Kerr-Newman, $Q^2 = M^2 - a^2$,
so that $\Delta = (r-M)^2$ and the event horizon is at $r=M$.
Then, taking $a$ to be positive,
one can calculate that the SCAM exists for values
of $r$ down to $2M+2a$, where $\sigma$
defined by Eq. (\ref{eq:213}) passes through zero
and goes negative, making $\Omega_0$ and various
other quantities complex for smaller $r$.  Hence I shall
first give the value at general $r$ when $Q^2=M^2-a^2$, and then,
after the first arrow in each equation, the limiting value
for each quantity when one sets $r=2M+2a$.
Finally, after a second arrow, I shall give the
limiting value in extreme Kerr $a=M$ (so $Q=0$)
at $r=4M$, the radial inner boundary on the equatorial plane
of the region where the SCAM exists as a real congruence:
 \begin{equation}
 \Omega_{+1} = {r^2(r\!-\!M)\!+\!a(2Mr\!-\!M^2\!+\!a^2)
	 \over r^2(r^2\!+\!a^2) \!+\! a^2(2Mr\!-\!M^2\!+\!a^2)}
 \rightarrow {M+3a\over 4M^2\!+\!9aM\!+\!7a^2}
 \rightarrow {1\over 5M},
 \label{eq:326a}
 \end{equation}
 \begin{equation}
 \Omega_{-1} = {-r^2(r\!-\!M)\!+\!a(2Mr\!-\!M^2\!+\!a^2)
	 \over r^2(r^2\!+\!a^2) \!+\! a^2(2Mr\!-\!M^2\!+\!a^2)}
 \rightarrow {-1\over 4M+3a}
 \rightarrow -{1\over 7M},
 \label{eq:326b}\\
 \end{equation}
 \begin{equation}
 \Omega_{+K} = {\sqrt{Mr-M^2+a^2}
	\over r^2 + a\sqrt{Mr-M^2+a^2}}
 \rightarrow {1\over 4M+5a}
 \rightarrow {1\over 9M},
 \label{eq:326c}
 \end{equation}
 \begin{equation}
 \Omega_{-K} = {-\sqrt{Mr-M^2+a^2}
	\over r^2 - a\sqrt{M^2+a^2}}
 \rightarrow {-1\over 4M+3a}
 \rightarrow -{1\over 7M},
 \label{eq:326d}
 \end{equation}
 \begin{eqnarray}
 \Omega_0 \!\!\!\!&=&\!\!\!\!{-2a(Mr-M^2+a^2)\over
	r^2(r\!-\!M)(r\!-\!2M)\!-\!2a^2(r^2\!+\!Mr\!-\!M^2\!+\!a^2)\!
	+\!r^2(r\!-\!M)\sqrt{(r\!-\!2M)^2\!-\!4a^2)}}
	\nonumber \\
 &\rightarrow& {-1\over 4M+3a}
 \rightarrow -{1\over 7M},
 \label{eq:326e}
 \end{eqnarray}
 \begin{equation}
 \Omega_Z \!=\! {a(2Mr-M^2+a^2) \over
 (r^2\!-\!ar\!+\!aM\!+\!a^2)(r^2\!+\!ar\!-\!aM\!+\!a^2)}
 \!\rightarrow\! {a(3M+a)\over (4M\!+\!3a)(4M^2\!+\!9aM\!+\!7a^2)}
 \!\rightarrow\! {1\over 35M},
 \label{eq:326f}
 \end{equation}
 \begin{equation}
 v_K \!=\! {r \!-\! \sqrt{(r\!-\!2M\!-\!2a)(r\!-\!2M\!+\!2a)}
	\over 2\sqrt{Mr-M^2+a^2} }
	\!\!=\!\!{ 2\sqrt{Mr-M^2+a^2}\over
	r \!+\! \sqrt{(r\!-\!2M\!-\!2a)(r\!-\!2M\!+\!2a)}}
 \rightarrow 1 \rightarrow 1,
 \label{eq:327}
 \end{equation}
 \begin{equation}
 a_0 ={r-\sqrt{(r-2M-2a)(r-2M+2a)} \over 2r^2}
 \rightarrow {1\over 4(M+a)}
 \rightarrow {1\over 8M},
 \label{eq:328}
 \end{equation}

	One can thus see that at the inner boundary of the SCAM,
at $r=2(M+a)$ on the equatorial plane of extreme Kerr-Newman
($Q^2=M^2-a^2$), both the counterrotating
circular Keplerian orbit and the SCAM have angular velocities,
$\Omega_{-K}$ and $\Omega_0$ respectively,
that approach the angular velocity $\Omega_{-1}$
of the counterrotating speed of light.
Since at this inner boundary $v_K = 1$,
Eq. (\ref{eq:220}) says that the acceleration there
is independent of the (signed) speed $v$ of a stationary
observer relative to the SCAM if $|v| < 1$.
However, one must take care,
since this formula is then degenerate at $|v|=1$,
and since Eq. (\ref{eq:211}) with
$\Omega_1 = \Omega_0 = \Omega_{-1}$
gives $v(\Omega_2,\Omega_1)=1$
for any angular velocity $\Omega_2 > \Omega_1$.
For $\Omega > \Omega_0 = \Omega_{-1}$,
one should instead return to Eq. (\ref{eq:29}),
which gives at $r=2(M+a)$ in extreme Kerr-Newman
the acceleration
 \begin{equation}
 \ba ={(M+2a)[1-(4M+5a)\Omega] \over
	4(M+a)[(M+3a)-(4M^2+9aM+7a^2)\Omega]}\ber
 \rightarrow {3\over 32M}
	\biggl({1-9M\Omega\over 1-5M\Omega}\biggr)\ber,
 \label{eq:332}
 \end{equation}
with the expression after the arrow being
that at $r=4M$ when $a=M$ (and hence $Q=0$).
Note that Eq. (\ref{eq:332}) is one where the $\ba$
on the left hand side is the acceleration, whereas
all the $a$'s on the right hand side denote the
Kerr parameter $a=J/M$, as I have warned.

\subsection{On the axis of rotation}

	The third limiting case where one can give the SCAM
explicitly for the Kerr-Newman metric without solving
a very messy quartic equation is on one of the axes
of symmetry, say, for concreteness, the one at $\theta=0$
($c\equiv \cos{\theta}=1$).
There the rotation $\omega_0$ of the SCAM is given by
the solution (\ref{eq:266}) of the cubic Eq. (\ref{eq:262}),
$\omega^3-p\omega-q=0$,
where the coefficients $p$ and $q$,
given by Eqs. (\ref{eq:263}) and (\ref{eq:264})
in the general case, take on, in the Kerr-Newman metric,
the values
 \begin{equation}
 p = {Mr^3-Q^2r^2-3a^2Mr+a^2Q^2 \over (r^2+a^2)^3}
 - {(Mr^2-Q^2r-a^2M)^2  \over (r^2+a^2)^3(r^2-2Mr+a^2+Q^2)}
 \label{eq:333}
 \end{equation}
and
 \begin{equation}
 q = {a(Mr^2-Q^2r-a^2M)(3Mr^2-2Q^2r-a^2M) \over
	(r^2+a^2)^{9/2}(r^2-2Mr+a^2+Q^2)^{1/2}}.
 \label{eq:334}
 \end{equation}

	Once one has the normalized rotation $\omega_0$,
which is independent of the choice of the Killing vector
field $\bk$ (though near the axis $\bl$ is uniquely
determined, up to sign, by its property of having
a magnitude which goes as the proper distance
from the axis), one can calculate, for $\bk = \partial/\partial t$,
the angular velocity $\Omega_0$ of the SCAM
arbitrarily near the axis by solving Eq. (\ref{eq:260}).
On the axis of the Kerr-Newman metric, this gives
 \begin{eqnarray}
 \Omega_0 \!\!&=&\!\! \Omega_Z + \sqrt{A}\omega_0
	\nonumber\\
 &=&\!\! {q(2Mr-Q^2)\over r^2(r^2+a^2)}
	-2\sqrt{{r^2-2Mr+a^2+Q^2\over r^2+a^2}}
	\sqrt{{p\over 3}}
	\sin{[{1\over 3}\sin^{-1}{(q\sqrt{{27\over 4p^3}})}]}.
 \label{eq:335}
 \end{eqnarray}

	In the case of the extreme Kerr metric, $a=M$
and $Q=0$, if we let $x\equiv r/M$ and
$y\equiv (r^2+M^2)^{3/2}\omega/M^2$,
then the cubic Eq. (\ref{eq:262}) becomes
 \begin{equation}
 y^3 - (x^3-x^2-5x-1)y - (3x^3+3x^2-x-1) = 0.
 \label{eq:336}
 \end{equation}
The discriminant of this cubic is proportional to
$31+114x+115x^2+56x^3+12x^4-4x^5$,
which is negative for
 \begin{equation}
 x\equiv {r\over M}
	{\ \lower-1.2pt\vbox{\hbox{\rlap{$>$}
	\lower5pt\vbox{\hbox{$\sim$}}}}\ }
 6.15862999016071260705,
 \label{eq:337}
 \end{equation}
the region where the cubic has three real roots
(one for the SCAM and one for each of the
other two SCALEs that are local minima
of the acceleration).
Therefore, near the axis of an extreme Kerr
black hole, the SCAM exists only outside
$r\approx 6.15862999M$,
whereas on the equatorial plane it exists
outside $r = 4M$.

	Although it is rather messy to do for
general $a$ and $Q$, for extreme Kerr
one can also readily compare expansions
of $\Omega_0$ in inverse powers of $r$
in the equatorial plane and on the axis.
In the equatorial plane for $a=M$ and $Q=0$
one gets
 \begin{equation}
 \Omega_0 = -{M^2\over r^3}[1+{3M\over r}
	+{9M^2\over r^2}+{29M^3\over r^3}+O({M^4\over r^4})],
 \label{eq:338}
 \end{equation}
whereas on the axis one gets
 \begin{equation}
 \Omega_0 = -{M^2\over r^3}[1+{3M\over r}
 +{12M^2\over r^2}+{53M^3\over r^3}+O({M^4\over r^4})].
 \label{eq:339}
 \end{equation}
The difference in the two expressions, starting
at the second-order correction, persists even when one
changes the radial variable from $r$ to
$\rho\equiv\sqrt{r^2+a^2\cos^2{\theta}}$
(which is the same as $r$ on the axis),
since then the series on the equatorial plane becomes
 \begin{equation}
 \Omega_0 = -{M^2\over \rho^3}[1+{3M\over \rho}
	+{21M^2\over 2\rho^2}+{35M^3\over \rho^3}
	+O({M^4\over \rho^4})],
 \label{eq:340}
 \end{equation}
whereas on the axis it has the same form as
Eq. (\ref{eq:339}) but with $r$ replaced by $\rho$.
The first three terms of Eqs. (\ref{eq:338}) and (\ref{eq:339})
can be readily be seen to agree with Eq. (\ref{eq:312})
when $a=M$, $Q=0$, and either $c=0$ ($\theta=\pi/2$)
or $c=1$ ($\theta=0$).

\section{Stationary Worldlines Accelerating \newline
Radially Maximally (SWARM)}

	Because the explicit expression for the angular velocity
$\Omega_0$ of the SCAM (a root of a messy quartic equation)
appears to be generally rather intractable except in the special
cases discussed above (slow rotation, at an orbital, and on an axis),
it might be useful to define other congruences that
have simpler explicit expressions and which have some of
the properties of the SCAM.  For example, one might define
Stationary Worldlines Accelerating Radially Maximally (SWARM)
as those that have $\Omega$ chosen to maximize the radial
component of the acceleration.

	Then the question arises as to how to define the
radial component (or direction, since the radial component
is needed only up to a positive constant of proportionality
at each location in order to be able to define its maximum as
a function of the angular velocity).
Semer\'{a}k \cite{Sem93,Sem94}
defined a maximum of the cylindrical radial component
of the acceleration in the Kerr metric
as an Extremely Accelerated Observer (EAO),
but I prefer to define something different here
and leave that name for what he defined there.
(And to avoid confusion I propose that the EAO
retain its original definition in \cite{Sem94},
rather than being redefined to be the SCAM that I invented,
despite Semer\'{a}k's proposal to do that in \cite{Sem96}.)

	Alternatively, if one defined
the radial direction to be that of
the acceleration of the SCAM, then of course
the SWARM would simply be the SCAM,
but then finding the radial direction would involve
solving a quartic equation, and there would be
no advantage to defining a SWARM.

	Therefore, I propose that a SWARM be defined
so that in the Kerr-Newman metric the radial direction
is that of $\nabla r$, the gradient of the Boyer-Lindquist
radial coordinate $r$.  This direction has several remarkable
properties in Kerr-Newman, connected with the existence
of Carter's `fourth constant of motion' \cite{Carter68,MTW}.

	For example, there is the fact that for a timelike geodesic,
out of the set of four parameters that govern the orbit
in the (2,3) or $(r,c\equiv\cos{\theta})$ coordinates
(e.g., the value of $r$ at $c=0$, the value of $dr/dc$ there,
the value of the conserved energy per rest mass
$-u_0\equiv -u_t$,
and the value of the conserved
angular momentum per rest mass $u_1\equiv u_{\varphi}$),
a two-parameter subset leads to orbits with constant $r$
(e.g., $r$ and $u_1$, setting $dr/dc=0$ at $c=0$ and choosing
$-u_0$ as a function of $u_1$ so that $d^2r/dc^2=0$ there).
For a generic stationary axisymmetric metric,
one could define a radial coordinate so that
a one-parameter set of geodesics have constant $r$
(e.g., $r$ along some one-dimensional line in the (2,3)
plane analogous to the $c=0$ line, by defining
$r$ so that it is constant along the orbit that one gets
with a specific choice of $-u_0$ and $u_1$),
but if one tried varying a second parameter, e.g. $u_1$,
then there would be no choice of $-u_0$
for different values of $u_1$ that would give
other orbits along the same constant $r$ line.

	In the Kerr-Newman metric, the geodesic orbits with
constant $r$ typically oscillate in $c$ (unless they
are circular Keplerian orbits in the equatorial plane,
which stay at fixed $c=0$ as well as constant $r$ and hence
are stationary geodesics), going between a positive maximum
where $dc/d\tau=0$ as well as $dr/d\tau=0$
(so that the velocity is momentarily zero in the (2,3)
plane at that turning point, though of course
the velocity is not zero in the (0,1) plane),
and a negative minimum of equal magnitude
(because of the symmetry with
respect to reflections about the equatorial plane)
where also the velocity in the (2,3) plane is momentarily
zero, $dc/d\tau=0$ as well as $dr/d\tau=0$.

	However, in a generic stationary axisymmetric metric,
an orbit that starts with zero velocity
somewhere in the (2,3) plane will not generically
have zero velocity elsewhere on its orbit in that plane.
Nevertheless, requiring zero velocity in the (2,3) plane
at one point there puts only one restriction on the
conserved quantities $-u_0$ and $u_1$, so by choosing
$u_1$, say, appropriately (which then leaves $-u_0$
determined as a function of $u_1$ so that the velocity
in the (2,3) plane is zero there), one has the right number
of free parameters to be able to get zero velocity in the (2,3)
plane
at some other location.  If this requirement of a second
turning point of zero velocity in the (2,3) plane
uniquely fixes $u_1$ (and hence also $-u_0$) at the original
turning point, then the direction in which the orbit starts out
(determined by $d^2x^a/d\tau^2$ at the turning point
where $dx^a/d\tau=0$)
could be defined as the angular direction,
the direction of constant radius $r$
(after defining the radius $r$ suitably).
Then the orthogonal direction
could be defined as the radial direction.

	This is admittedly a nonlocal definition of the radial
direction, and it might not always give a unique answer,
but it is a procedure that would give the $\nabla r$ direction
in the Kerr-Newman metric and presumably would give
a unique direction for small perturbations of that metric.
So let me define the radial direction by this procedure
when it works, and then define a prime as
denoting a partial derivative in that direction.

	Then Eq. (\ref{eq:206}) gives the angular velocity
of what is now the SWARM,
which maximizes the radial component
(rather than the entire magnitude) of the acceleration.
Eq. (\ref{eq:204}) gives the angular velocities, not of
circular Keplerian orbits, but of stationary worldlines
with zero radial component of the acceleration.
These will also be the angular velocities
at that location of the nonstationary geodesic
orbits that have a turning point in the (2,3) plane
at that location and also have $d^2r/d\tau^2=0$ there.
By the same argument as before, these have equal
but opposite velocities in the frame of the SWARM,
of magnitude given by Eq. (\ref{eq:212}).
Furthermore, Eq. (\ref{eq:219}) gives, not the total
acceleration, but the radial component of the acceleration
of the SWARM at that location, and Eq. (\ref{eq:220})
gives the part of the acceleration in the radial direction
for a stationary worldline at speed $v$ in the SWARM frame.

	In the Kerr-Newman metric (\ref{eq:301}),
one can write out all these expressions
explicitly as functions of $r$ and $c$,
but they don't easily fit into single lines, so I won't bother
doing that straightforward calculation here.
One can see that in the Kerr-Newman metric,
which has a two-parameter family of timelike
geodesics with constant $r$ as mentioned above,
the SWARM four-velocity is the normalized
average of the four-velocities of the two constant-$r$
geodesics that have their turning points
in the (2,3) plane at that location
(i.e, have their maximum of $|c|$ there).
In other words, if one takes two equal-mass particles
moving along constant-$r$ geodesics
that at a certain location in the (2,3) plane
have their four-velocity entirely in the (0,1) plane,
with one moving forward in $\varphi$ and the
other moving backward, then if one makes a totally
inelastic collision between these particles,
immediately after the collision the resulting
particle will have the four-velocity of
the SWARM at that location.

	One can also say that on the equatorial plane
in the Kerr-Newman metric,
as at any orbital in a generic stationary axisymmetric metric,
the total acceleration is in the radial direction,
so the SWARM
has the same four-velocity as the SCAM there.  
The four-velocity (and acceleration) also agrees on an axis,
trivially, since there the four-velocity is
independent of the angular velocity,
but the angular velocity of the SWARM is not the same
as that of the SCAM near the axis of the Kerr-Newman metric.
In the limit of going onto the axis itself, Eq. (\ref{eq:335})
gives the angular velocity of the SCAM,
and the angular velocity of the SWARM there is
 \begin{eqnarray}
 \tilde{\Omega}_0 &=& {-a(Mr^2-Q^2r-a^2M)\over
	(r^2+a^2)(r^3-3Mr+2Q^2r+a^2r+a^2M)}
	\nonumber\\
 &=& -{aM\over r^3}\Bigl[1+{3M^2-Q^2\over Mr}
	+{9M^2-5Q^2-2a^2\over r^2}
	+O\Bigl({a^2 M\over r^3},{a^2 Q^2\over Mr^3}\Bigr)\Bigr].
\label{eq:341}
 \end{eqnarray}

	By comparing with the series expansion of Eq. (\ref{eq:312})
at the axis ($c^2=1$), one sees that, to lowest order
in $1/r$, the angular velocity of the SWARM on the axis
is less negative than that of the SCAM by $5a^3M/r^5$,
so if $a/M$ is small, the SWARM is a very good
approximation for the SCAM on the axis, and,
presumably, at all other angles or values of $c$.
This agreement is a consequence of the fact,
that for small $a/M$ at least, the direction
of the acceleration of the SCAM is nearly radial
in the Kerr-Newman metric,
so maximizing the radial component of the acceleration
for the SWARM (Stationary Congruence Accelerating
Radially Maximally) gives very nearly the same
angular velocity as finding the local maximum
of the magnitude of the total acceleration
for the SCAM (Stationary Congruence Accelerating
Maximally).

\section{Application to Maximally Rotating Stars}

	One application of the fact that the SCAM is
usually counterrotating relative to a rotating source
(e.g., a black hole or star) is an explanation of the fact
that corotating Keplerian orbits
in an equatorial plane of the source may have periods
that are longer than one would get from a na\"{\i}ve
application of Kepler's third law.
In certain cases Kepler's third law gives a better
approximation to the orbital frequency $\Omega-\Omega_0$
relative to the SCAM,
rather than the orbital frequency $\Omega$
relative to a nonrotating observer with $\Omega_{NR}=0$.
Since $\Omega_0$ is typically negative
(when the coordinate system is chosen so that the source
is rotating positively),
the orbital frequency relative to a nonrotating observer
will be slower than that relative to the SCAM,
so its period will be greater.

	Physically,
this effect may be explained by the same mechanism
used above to explain the counterrotation of the SCAM:
A corotating orbiting observer will see the part of the source
nearest her as partially moving with her and hence
as having a lower energy density in her frame
than the part of the source farthest away from her,
which is moving in the opposite direction.
Hence she will see the energy distribution shifted slightly
away from her, where it will have a weaker net gravitational
attraction on her than a source
of the same energy density distribution
(as seen in a nonrotating frame) that is not rotating.
Therefore, she will orbit more slowly by this relativistic effect
that can be ascribed to the angular momentum of the source.

	Realistic situations are further complicated by the fact
that the source (e.g., a star) will not generally have the same
energy density distribution in a nonrotating frame
when it is rotating as when it is nonrotating.
One effect is that if a source is spun up,
it will gain rotational energy and hence total mass-energy.
However, this effect can easily be compensated for
by removing from the source an amount of energy
equal to that given it in spinning it up,
say by removing an appropriate number of baryons.
In any case, if one is using Kepler's third law,
the mass in that formula should be the total mass-energy
in the source, so a change in the total mass is already
taken into account.

	However, another effect that is not taken into account
by Kepler's third law is that the shape of the source
generally changes when it is spun up.
For example, if the source is a self-gravitating fluid,
such as a star,
the centrifugal forces of the rotation will generally
cause the star to become oblate.
Then as the energy becomes more concentrated
upon the equatorial plane, where it is on average nearer
to the observer orbiting in that plane,
it will exert a greater gravitational attraction upon the observer,
causing her to orbit faster than she would have
in the absence of the oblateness.
Equivalently, the greater gravitational attraction
at a fixed radius in the equatorial plane
is caused by the quadrupole moment of the source.

	This oblateness or quadrupole effect
on the orbital frequency
is opposite in sign to the relativistic effect
of the source angular momentum discussed above.
At low source angular velocities, the relativistic effect is
linear in the source angular velocity, whereas the oblateness
effect is quadratic.  However, the quadrupole effect
persists even in the Newtonian limit, so it can be larger
than the relativistic angular momentum effect.

	As an example in which the corotating orbital period
in the equatorial plane
can be calculated exactly and compared with Kepler's third
law, consider the Kerr metric with zero charge. 
Then from Eq. (\ref{eq:316b}),
and temporally restoring Newton's constant $G$
and the speed of light $c$ that have been set equal to unity,
one can readily get the period as
 \begin{equation}
 P = {2\pi \over \Omega_{+K}} = 2\pi\sqrt{r^3\over GM}\:
 + {2\pi J\over Mc^2}. 
 \label{eq:401}
 \end{equation}
The second term on the right hand side can be identified with
the linear increase in the period with the angular momentum $J$,
and the square of the speed of light in the denominator
shows that it is a relativistic effect.
(It is interesting that Newton's constant does not appear in
this term.  If one takes the quantum-mechanical phase
of a system with energy $E=Mc^2$ and angular momentum
$J$ to be $e^{\,(-iEt + iJ\varphi)/\hbar}$, then the time period
for this phase to rotate around once is precisely this second term. 
However, Newton's constant does appear
in the ratio of the second term to the first term,
so the increase in the period with the angular momentum really
does involve both special relativity and gravity and is thus
a general relativistic effect rather than purely
a special relativistic effect.)

	It is tempting to identify the first of the two terms
on the right hand side of Eq. (\ref{eq:401}) as being precisely
Kepler's third law for a circular orbit of radius $r$,
but there is the question of whether $r$
is the most natural measure of the the radius.
In the nonrotating limit (Schwarzschild, $J = Ma = 0$),
$r$ is the circumferential radius, $1/2\pi$ times
the proper circumference of a closed circle
in the equatorial plane with fixed $r$ and $t$,
which is a fairly simple geometric definition
of a radius that makes Kepler's third law exact
for circular orbits in the Schwarzschild metric.
If in this section one takes $R$
to be the circumferential radius
in the equatorial plane around
a source of total mass $M$, then one can define
 \begin{equation}
 \Gamma \equiv {R^3\Omega_{+K}^2\over GM}
 \label{eq:402}
 \end{equation}
as a measure of how closely Kepler's third law holds,
which would state that $\Gamma=1$.

	Although $\Gamma=1$ for the Schwarzschild metric,
in the equatorial plane of the Kerr metric with $J\neq 0$
the circumferential radius is
 \begin{equation}
 R = \sqrt{-C}
 = \sqrt{r^2 + {J^2\over M^2c^2} + {2GJ^2\over c^4 Mr}},
 \label{eq:403}
 \end{equation}
so
 \begin{eqnarray}
 \Gamma &
 =& \left( 1+{J\over c^2}\sqrt{G\over Mr^3}\ \right)^{-2}
	\left( 1+{J^2\over M^2 c^2 r^2}
	 + {2GJ^2\over Mc^2 r^3} \right)^{3/2}
	\nonumber\\
	&=& 1 - {2J\over c^2}\sqrt{G\over Mr^3}\:
	 + {3J^2\over 2M^2 c^2 r^2} + O(r^{-3}).
 \label{eq:404}
 \end{eqnarray}
Alternatively, one can solve Eq. (\ref{eq:403})
for $r$ as a function of $R$
and insert this into Eq. (\ref{eq:401})
to get the orbital period
(as seen from radial infinity) as
 \begin{equation}
 P = {2\pi \over \Omega_{+K}}
 = 2\pi\sqrt{R^3\over GM} + {2\pi J\over Mc^2}
 - {3\pi J^2 \over 2M^2 c^2 \sqrt{GMR}} + O(R^{-3/2}). 
 \label{eq:405}
 \end{equation}
The third term on the right hand side
represents the effect of the quadrupole moment
of the Kerr metric.

	Since in the Kerr metric a stationary observer
must have $r > GM/c^2$,
and since the dimensionless Kerr rotation parameter,
 \begin{equation}
 \alpha \equiv {a \over M} \equiv {cJ \over GM^2},
 \label{eq:406}
 \end{equation}
is less than or equal to unity,
the negative of the ratio of the second term
to the third term in the last expression for $\Gamma$
in Eq. (\ref{eq:404}),
or in the last expression for $P$ in Eq. (\ref{eq:405}), is
 \begin{equation}
 \sqrt{16GM^3 r \over 9J^2}
 = {4\over 3\alpha}\sqrt{c^2 r\over GM} \: > 1.
 \label{eq:407}
 \end{equation}
Therefore, at least at large $r$
where one can drop the $O(r^{-3})$
term in Eq. (\ref{eq:404})
or the $O(R^{-3/2})$ term in Eq. (\ref{eq:405}),
the second term of either of these equations,
which represents
the relativistic period-increasing effect
that is linear in the angular momentum $J$,
dominates over the third term,
which represents the effect of the period-decreasing
quadrupole moment that is
quadratic in the angular momentum.

	However,
if the dimensionless Kerr rotation parameter
$\alpha$ is larger than about $0.952518$,
then there are stable circular corotating orbits in Kerr
(which exist for \cite{Chandra}
 \begin{equation}
 r^2 - 6Mr + 8a\sqrt{Mr} - 3a^2 \geq 1,
 \label{eq:408}
 \end{equation}
temporarily reverting to units in which $G = c = 1$)
for which the effect of the quadrupole moment dominates
so that $\Gamma > 1$.
For example, for the extreme Kerr metric
($\alpha = 1$ or $a = M$),
$\Gamma > 1$ for $r$ less than about $2.01186 \, GM/c^2$,
and at the smallest innermost stable circular corotating orbit
at $r = GM/c^2$, one has $R = 2GM/c^2$,
$\Omega_{+K} = c^3/2GM$, and hence $\Gamma = 2$.

	If we turn to models of maximally rotating stars
(stars with the equatorial surface rotating at
the Keplerian velocity and hence just marginally bound),
typically the quadrupole moments are larger
than they are for the Kerr metric
with the same stellar mass and angular momentum,
because stars are not so gravitationally concentrated.
Thus one gets a larger radius at which the relativistic
period-increasing effect (linear in the angular momentum)
balances the effect
of the period-decreasing quadrupole moment.
For nonrelativistic maximally rotating stars
this radius tends to be outside the surface of the star,
so that one usually gets $\Gamma > 1$
from Eq. (\ref{eq:402})
when $R$ is set to be the circumferential equatorial radius
of the star in its rotating frame, and $\Omega_{+K}$ is
both the angular velocity of the
corotating circular equatorial orbit
at the surface of the star,
and the angular velocity of the star itself.

	However, for certain relativistic maximally rotating stars,
the relativistic period-increasing effect can exceed
the period-decreasing effect of the quadrupole moment,
making $\Gamma < 1$.
For example, in the numerical models of rapidly rotating
polytropes in general relativity by
Cook, Shapiro, and Teukolsky \cite {CST},
the maximum uniform rotation models of the
``supramassive'' sequence given in their Table 2
have $\Gamma < 1$ for values
of the polytropic index $n \leq 1.5$, namely
$\Gamma = 0.958$ for $n=0.5$,
$\Gamma = 0.988$ for $n=1.0$, and
$\Gamma = 0.999$ for $n=1.5$.
Higher values of the polytropic index, $n \geq 2$,
generally seem to give $\Gamma > 1$, namely
$\Gamma = 1.005$ for $n=2.0$ and
$\Gamma = 1.009$ for $n=2.5$.
The data from \cite{CST} for $n=2.9$
na\"{\i}vely gives $\Gamma = 0.999 < 1$,
but since the data are given only to three places,
and since this large value of $n$ gives a highly
nonrelativistic model ($2GM/Rc^2 = 0.0109$,
as opposed to the highly relativistic value
$2GM/Rc^2 = 0.578$ for $n=0.5$, for example),
I would be sceptical that actually $\Gamma < 1$
for this large value of $n$.

	Similarly, one might doubt that
$\Gamma < 1$ for $n=1.5$,
where the three-place data give a result
also just slightly below unity,
but presumably there is a value
of the polytropic index $n$ fairly near 1.5
such that
$\Gamma \equiv R^3\Omega_{+K}^2 / (GM) < 1$
for maximally uniformly rotating
supramassive polytropic models
with smaller $n$
but such that $\Gamma > 1$
for similar models with larger $n$.
The small values of $n$ give relativistic stellar models
for which the counterrotation of the SCAM is sufficient
to make it so that the corotating Keplerian orbital velocity
(which matches the stellar rotation rate at the equatorial
surface in these maximally rotating models)
relative to a nonrotating observer at infinity
is slowed down by this relativistic effect
more than it is sped up by the quadrupole moment.

\section{Application to the gravitational field
\newline
of the Sun and Solar System}

Another application
of the stationary congruences defined above
and of the related deviations from Kepler's third law
is to the gravitational field of the Sun.
Since the Sun is nearly spherical and is not very relativistic,
the metric for its external gravitational field may be given,
to an accuracy of about one part in $10^{15}$,
and using units in which $G=c=1$,
as \cite{HarTho}
 \begin{eqnarray}
 ds^2 = &-& [1-{2M_{\odot}\over r}
	+{2Q_{\odot}\over r^3}P_2(\cos{\theta})]dt^2
		+[1-{2M_{\odot}\over r}
	+{2Q_{\odot}\over r^3}P_2(\cos{\theta})]^{-1}dr^2
		\nonumber\\
	&+& [1-{2Q_{\odot}\over r^3}P_2(\cos{\theta})]r^2
		[d\theta^2+\sin^2{\theta}(d\varphi
		 - {2J_{\odot}\over r^3}dt)^2]. \label{eq:501}
 \end{eqnarray}
Here $P_2(\cos{\theta}) = (3\cos^2{\theta} - 1)/2$ is the standard
second-order Legendre polynomial,
and $Q_{\odot}$ is the quadrupole
moment of the Sun
(not to be confused with the previous use
of $Q$ to denote the charge
of the Kerr-Newman black hole;
in this section the charge
will always taken to be zero and $Q$
will always denote a quadrupole moment,
with its sign chosen
so that an oblate spheroid,
such as a rotating body, has positive $Q$),
 \begin{equation}
 Q_{\odot} \equiv J_2 M_{\odot} R_{\odot}^2
 = -\int{\rho r^2 P_2(\cos{\theta})
	dr \sin{\theta} d\theta d\varphi}
 = \int{\rho ({1\over 2}x^2 + {1\over 2}y^2 - z^2) dx dy dz}.
 \label{eq:502}
 \end{equation}
Here $R_{\odot}$ is the radius of the Sun,
and $J_2$ is its dimensionless
quadrupole moment parameter.

	In metric length units (meters),
the 1996 {\it Review of Particle Physics} \cite{RPP}
gives the mass and radius of the Sun as
 \begin{equation}
 GM_{\odot}/c^2 = 1\,476.625\,04 \; {\rm m}
 = 0.9137 \times 10^{38} \: \ell_P,
 \label{eq:503}
 \end{equation}
 \begin{equation}
 R_{\odot} = 6.96 \times 10^8 \: {\rm m}
 = 4.31 \times 10^{43} \: \ell_P,
 \label{eq:504}
 \end{equation}
where $\ell_P \equiv \sqrt{\hbar G/c^3}
 = (1.616\,05 \pm 0.000\,10)\times 10^{-35}
 \: {\rm m}$ is the Planck length,
using data from the same source \cite{RPP}.

	A recent helioseismic determination of the solar
gravitational quadrupole moment \cite{Pij},
which is consistent with a less-precise
direct measurement of the solar oblateness \cite{PSD},
gives the dimensionless solar quadrupole parameter as
 \begin{equation}
 J_2 = (2.18 \pm 0.06) \times 10^{-7}.
 \label{eq:505}
 \end{equation}
Therefore, the solar quadrupole moment is
 \begin{equation}
 Q_{\odot} = J_2 M_{\odot}R_{\odot}^2
 = (2.10 \pm 0.06) \times 10^{41} \: {\rm kg \; m}^2,
 \label{eq:506a}
 \end{equation}
or in length units (cubic meters) it is
 \begin{eqnarray}
 GQ_{\odot}/c^2 &=& GJ_2 M_{\odot}R_{\odot}^2/c^2
 = (1.60 \pm 0.04) \times 10^{14} \: {\rm m}^3
 = 160\,000 \pm 4\,000 \: {\rm km}^3
 \nonumber\\
 &=& (53.8 \pm 0.5 \: {\rm km})^3
 = (3.69 \pm 0.11) \times 10^{117} \: \ell_P^3.
 \label{eq:506}
 \end{eqnarray}
From this one can define
an effective quadrupole radius
of the Sun as
 \begin{eqnarray}
 r_{Q_{\odot}} &=& \sqrt{2Q_{\odot}/M_{\odot}}
 = 460 \pm 6 \; {\rm km}
 \nonumber\\
 &=& (3.07 \pm 0.04) \times 10^{-6} \: {\rm AU}
 = (2.84 \pm 0.04)\times 10^{40} \: \ell_P,
 \label{eq:507}
 \end{eqnarray}
the radius of a solar-mass ring
of the same quadrupole moment as the Sun.
Here $1 \: {\rm AU} = 149\,597\,870\,660 \pm 20 \; {\rm m}
= 0.9257 \times 10^{46} \: \ell_P$
is the astronomical unit \cite{RPP}.

	The same helioseismic measurements \cite{Pij}
also give the angular momentum of the Sun as
 \begin{equation}
 J_{\odot} = (1.900 \pm 0.015) \times 10^{41}
	 \: {\rm kg \: m}^2 \: {\rm s}^{-1}
 = (1.801 \pm 0.014) \times 10^{75} \: \hbar,
 \label{eq:508}
 \end{equation}
which when converted to length and area units gives
 \begin{eqnarray}
 GJ_{\odot}/c^3 &=& (4.705 \pm 0.037) \times 10^5 \: {\rm m}^2
 = (686 \pm 3 \: {\rm m})^2
 \nonumber\\
 &=& 47.05 \pm 0.37 \; {\rm hectares}
 = 116 \pm 1 \; {\rm acres}.
 \label{eq:509}
 \end{eqnarray}
From this and the solar mass one can readily calculate that
the Kerr parameter $a$ of Eq. (\ref{eq:302})
and the dimensionless Kerr rotation parameter $\alpha$
of Eq. (\ref{eq:406}) have the values
 \begin{equation}
 a_{\odot} \equiv {J_{\odot}\over M_{\odot}c}
 = 318.65 \pm 2.52 \; {\rm m}
 = (1.972 \pm 0.016)\times 10^{37} \: \ell_P,
 \label{eq:510}
 \end{equation}
 \begin{equation}
 \alpha_{\odot} \equiv {a_{\odot}c^2 \over GM_{\odot}}
 \equiv {cJ_{\odot} \over GM_{\odot}^2} = 0.2158 \pm 0.0017.
 \label{eq:511}
 \end{equation}
The fact that $\alpha_{\odot} < 1$ means that if the Sun
were able to undergo gravitational collapse to become
a black hole (which it is not, since it is too light,
except for some {\it extremely} tiny tunneling probability
or possibly some artificial compressional procedure),
it would not need to lose any angular momentum to do so.

	By comparing the metric (\ref{eq:501})
with the uncharged Kerr metric (\ref{eq:301}),
one may deduce \cite{HarTho}
that the quadrupole moment of the Kerr metric
is $Q = J^2/Mc^2 = Ma^2$, so $Mc^2 Q/J^2 = 1$.
(Remember that in this Section,
$Q$ denotes a quadrupole moment
and not the charge,
which we are here setting to zero.)
For the Sun we find that
 \begin{equation}
 {Q_{\odot}\over J_{\odot}^2 / M_{\odot} c^2}
 = (1.04 \pm 0.05)\times 10^6,
 \label{eq:511b}
 \end{equation}
so the quadrupole moment of the Sun
is about a million times
larger than that of a Kerr metric
with the same mass and angular momentum.
This is reasonable,
since the dimensionless quadrupole moment $J_2$
is two-thirds of the solar oblateness \cite{Pij},
which one expects to be of the order of
the square of the angular velocity of the Sun
divided by the Kepler orbital velocity
at the surface of the Sun, which is
$\Omega_{\odot}^2/\Omega_{+K}^2
\approx \Omega_{\odot}^2 R_{\odot}^3/GM_{\odot}$.
Therefore, one expects (to order of magnitude)
$Q_{\odot} = J_2 M_{\odot} R_{\odot}^2
\sim R_{\odot}^5 \Omega_{\odot}^2/G$.
Then since
$J_{\odot} \sim M_{\odot} R^2 \Omega_{\odot}$,
one gets
$M_{\odot} c^2 Q_{\odot}/J_{\odot}^2
\sim R_{\odot} c^2 /GM_{\odot}
= 0.471 \times 10^6$,
within about a factor of two of the correct answer.

	From the angular momentum $J_{\odot}$ of the Sun,
one can calculate that
the Stationary Congruence Accelerating Maximally
(SCAM) in the equatorial plane rotates around the Sun with
an angular velocity
(as seen by a nonrotating observer at infinity)
 \begin{eqnarray}
 \Omega_0 &\approx&  -{GJ_{\odot}\over c^2 r^3}
 = - (4.213 \pm 0.033) \times 10^{-20}
	\: \left({{\rm AU}\over r} \right)^3
	 \: {\rm s}^{-1}
 \nonumber\\
 &=& - (4.184 \pm 0.033) \times 10^{-13}
	 \: \left({{\rm R_{\odot}}\over r} \right)^3
	 \: {\rm s}^{-1}
 \nonumber\\
 &=& - {2\pi \over 476\,000 \pm 4\,000	\: {\rm yr}}
	\left({R_{\odot}\over r} \right)^3,
 \label{eq:511c}
 \end{eqnarray}
and a linear velocity
 \begin{eqnarray}
 v_0 &\approx&  -{GJ_{\odot}\over c^2 r^2}
 = (6.303 \pm 0.050) \times 10^{-9}\:
	 \left({{\rm AU}\over r} \right)^2
	\: {\rm m \: s}^{-1}
 \nonumber\\
 &=& - (2.932 \pm 0.023) \times 10^{-4}
	 \:  \left({{\rm R_{\odot}}\over r} \right)^2
	 \: {\rm m \: s}^{-1}
 \nonumber\\
 &=& - (9.253 \pm 0.073)
	 \: \left({{\rm R_{\odot}}\over r} \right)^2
	 \: {\rm km/yr}.
 \label{eq:511d}
 \end{eqnarray}
This rotation rate is extremely slow,
though not quite glacially slow,
but in the $4.6 \pm 0.1$ billion year
age of the Solar System,
the member of the SCAM at the surface
of the Sun would have made
almost ten thousand backward revolutions
relative to the distant stars,
assuming that the far-field Eq. (\ref{eq:182})
applies all the way down to the surface of the Sun.

	Since the SCAM maximizes the magnitude
of the acceleration for all stationary observers
at a given location, it has a larger acceleration
than that of a nonrotating observer.
For the weak field of the Sun,
the difference is very tiny:
 \begin{eqnarray}
 \Delta a &\equiv& a_0 - a_{NR} \approx {v_0^2 \over r}
	\approx  -{G^2 J_{\odot}^2 \over c^4 r^5}
 \nonumber\\
 &=& (2.656 \pm 0.042) \times 10^{-28}\:
	 \left({{\rm AU}\over r} \right)^5
	\: {\rm m \: s}^{-2}
 \nonumber\\
 &=& (4.775 \pm 0.075) \times 10^{-80}\:
	 \left({{\rm AU}\over r} \right)^5 \: c^2 \ell_P^{-2}.
 \label{eq:511e}
 \end{eqnarray}
Over the lifetime of the Solar System,
this tiny acceleration at $r = 1 \: {\rm AU}$
would, starting from rest, give a spatial motion
${1\over 2}\Delta a \, t^2$ of about 2800 km.

	If, despite its rotation, the Sun were perfectly spherical
and hence had no quadrupole moment,
then Kepler's third law would apply to high accuracy
to the orbital angular velocity relative to the SCAM.
Since the SCAM is counterrotating, the corotating
Keplerian orbit has a lower angular frequency,
and hence a longer period, relative to a static observer
than relative to the SCAM.  However, for the realistic Sun,
the quadrupole moment from the oblateness, also caused
by the rotation, increases the equatorial gravitational
attraction, and hence also the orbital angular frequency there.
This effect thus decreases the orbital period as measured
by a nonrotating observer at infinity.
These two changes in the period may be calculated as follows:

	In the metric Eq. (\ref{eq:501}) for the gravitational field
of the Sun, in the equatorial plane
($\theta = \pi/2$ or $P_2(\cos{\theta}) =  - 1/2$)
the circumferential radius is
 \begin{equation}
 R \equiv \sqrt{-C} \equiv \sqrt{g_{\varphi\varphi}}
	= r\sqrt{1+{Q_{\odot}\over r^3}}\: 
	\approx r + {Q_{\odot}\over 2r^2}.
 \label{eq:512}
 \end{equation}
The corotating Keplerian orbital velocity is then
 \begin{equation}
 \Omega_{+K} = \sqrt{M_{\odot}\over R^3}
	\left(1 - {J_{\odot}\over \sqrt{M_{\odot}R^3}}
	+ {3Q_{\odot}\over 4M_{\odot}R^2}
	 + O(R^{-3}) \right) ,
 \label{eq:513}
 \end{equation}
and so the orbital period
(as seen by a nonrotating observer at infinity) is
 \begin{eqnarray}
 P \equiv {2\pi \over \Omega_{+K}}
 &=& P_K + \Delta P_{J_{\odot}}
 + \Delta P_{J_2}  + O(R^{-3/2})
 \nonumber\\
 &=& 2\pi\sqrt{R^3\over GM_{\odot}}
 + {2\pi J_{\odot}\over M_{\odot}c^2}
 - {3\pi Q_{\odot} \over 2\sqrt{GM_{\odot}^3 R}}
 + O(R^{-3/2}), 
 \label{eq:514}
 \end{eqnarray}
where Newton's gravitational constant $G$
and the speed of light $c$
have been restored in Eq. (\ref{eq:514}).

	Here
 \begin{equation}
 P_K  = 2\pi\sqrt{R^3\over GM_{\odot}}
 = 31\,558\,196.0 \: \left({R \over {\rm AU}} \right)^{3/2} \: {\rm s}
 \label{eq:515}
 \end{equation}
is Kepler's third law for the period of a test body
in the field of the Sun
(ignoring the gravitational effects of the planets for now),
 \begin{eqnarray}
 \Delta P_{J_{\odot}}
 &=& {2\pi J_{\odot}\over M_{\odot}c^2}
 = 2\pi a_{\odot}/c = (2\,002 \pm 16 \; {\rm m})/c
 \nonumber\\
 &=& (6.678 \pm 0.053)\times 10^{-6} \: {\rm s}
 = 6.678 \pm 0.053 \: \mu{\rm s}
 \label{eq:516}
 \end{eqnarray}
is the increase in the period due to the linear effect
of the Sun's angular momentum, and
 \begin{equation}
 \Delta P_{J_2}
 = - {3\pi Q_{\odot} \over 2\sqrt{GM_{\odot}^3 R}}
 = - (1.117 \pm 0.032)\times 10^{-4}
	\: \left({{\rm AU} \over R} \right)^{1/2} \: {\rm s} 
 \label{eq:517}
 \end{equation}
is the decrease in the period due to the quadrupole moment
of the Sun.

	One can see that these corrections to Kepler's third law
in the gravitational field of the Sun are very small
and currently unmeasurable,
but it is amusing to calculate them
as an academic exercise.
It is also amusing to note that
the effect of the quadrupole moment
(which is essentially quadratic in the angular momentum)
dominates over that linear in the angular momentum
for radii $r < r_{K_{\odot}}$, where
 \begin{equation}
 r_{K_{\odot}}
 = {9c^4 Q_{\odot}^2 \over 16 G J_{\odot}^2 M_{\odot}}
	= 280 \pm 20 \: {\rm AU}
 \label{eq:518}
 \end{equation}
is the orbital radius at which Kepler's third law would be exact,
at about seven times the orbital radius (semimajor axis) of Pluto.

	One can use the order-of-magnitude estimates above
for the angular momentum
$J_{\odot} \sim M_{\odot} R_{\odot}^2 \Omega_{\odot}$
and the quadrupole moment
$Q_{\odot} \sim R_{\odot}^5 \Omega_{\odot}^2/G$
to estimate
 \begin{equation}
 r_{K_{\odot}}
 \sim \left( v_r \over c \right)^2
 \left( c \over v_e \right)^6 R_{\odot}, 
 \label{eq:519}
 \end{equation}
where $v_r = R_{\odot} \Omega_{\odot}$
is the linear rotation velocity
of the equatorial surface of the Sun
and $v_e = \sqrt{2GM_{\odot}/R_{\odot}}
 = 2.060 \times 10^{-3}\: c$
is the escape velocity from the surface of the Sun.
The Sun is not rotating rigidly,
so its angular velocity $\Omega_{\odot}$
is not constant, but one can take,
as a sort of averaged value
for $\Omega_{\odot}$,
twice $T_{\odot}$,
the total kinetic energy in rotation of the Sun,
divided by the Sun's angular momentum $J_{\odot}$.
Since the total kinetic energy in rotation is \cite{Pij}
 \begin{equation}
 T_{\odot}
 = (2.534 \pm 0.072) \times 10^{35}
 \: {\rm kg \: m}^2 \: {\rm s}^{-2}, 
 \label{eq:520}
 \end{equation}
one gets an effective averaged
angular velocity of the Sun as
 \begin{equation}
 \Omega_{\odot} = {2T_{\odot} \over J_{\odot}}
 = (2.67 \pm 0.10) \times 10^{-6} \: {\rm s}^{-1}
 = {2\pi \over 27.3 \pm 1.0 \: {\rm days}} \; . 
 \label{eq:521}
 \end{equation}
(Here I have simply linearly added
the relative errors given \cite{Pij}
for $J_{\odot}$, about 0.0079, and for $T_{\odot}$, about 0.0284,
to get a conservative relative error estimate of 0.0363
for $\Omega_{\odot}$.)
Multiplying $\Omega_{\odot}$
by the radius $R_{\odot} = 6.96 \times 10^8 \: {\rm m}$
of the Sun gives
$v_r = 1860 \pm 70 \: {\rm m/s}
 = (6.19 \pm 0.22) \times 10^{-6} \: c$.
Inserting this linear surface velocity $v_r$
and the escape velocity $v_e$
above into Eq. (\ref{eq:519})
gives the order-of-magnitude estimate
of Eq. (\ref{eq:519})
$r_{K_{\odot}}$ as roughly $500\,000 \: R_{\odot}$,
which is approximately $3.5 \times 10^{14} \: {\rm m}$
or 2300 AU.
This is about a factor of eight larger
than what Eq. (\ref{eq:518}) gives,
which is not too surprising, because of the neglect of all
numerical factors and details of the structure of the Sun
in Eq. (\ref{eq:519}),
and because of the high powers of the
velocities that enter into that estimate.

	For a self-gravitating rotating fluid object (e.g., a star)
which has its linear rotational velocity $v_r/c$
at its surface (in units of the speed of light)
greater than roughly the cube of the escape velocity $v_e/c$
from its surface (again in units of the speed of light),
as indeed is the case for the Sun,
we can see from the estimate of Eq. (\ref{eq:519})
that the radius $r_K$, where Kepler's third law
is exact for corotating circular orbits in the equatorial plane,
is greater than the radius $R$ of the object.
If the object is maximally rotating,
so that its rotational velocity $v_r$ is comparable
to its escape velocity $v_e$,
then $r_K$ will be outside the object
(leading to $\Gamma \equiv R^3\Omega_{+K}^2 / (GM) > 1$
and hence a shorter orbital period at the surface
than what Kepler's third law would give),
unless possibly the rotational and escape velocities
are close to the velocities of light
(i.e., unless possibly the object is highly relativistic,
as we found was necessary for the polytropic models
with $\Gamma < 1$ in the previous Section).

	In particular, one might ask what different
effective angular velocity the Sun
would need in order that its new $r_K$
would then coincide with the solar radius $R_{\odot}$.
To calculate this, instead of using Eq.  (\ref{eq:519}),
use the precise formula (\ref{eq:518})
(precise to the extent that it gives, as it does for the Sun,
an $r_{K_{\odot}}$ that is in the far-field region
where the field is both weak and is entirely dominated
by the monopole and quadrupole contributions).
Suppose that the effective moment of inertia
of the Sun about its axis,
 \begin{eqnarray}
 I_{\odot} &=& {J_{\odot}^2 \over 2T_{\odot}}
 = (7.12 \pm 0.31) \times 10^{48}  \: {\rm kg \: m}^2
 = (c^2/G) (5.29 \pm 0.23) \times 10^{21} \: {\rm m}^3
 \nonumber\\
 &=& (c^2/G) (1.25 \pm 0.06) \times 10^{126} \: \ell_P^3
 \approx 3.4 \times 10^7 \: Q_{\odot}, 
 \label{eq:521b}
 \end{eqnarray}
would stay constant as its angular velocity were changed.
More precisely, assume that the angular momentum of the Sun
would be linearly proportional to its effective angular velocity,
and that the quadrupole moment of the Sun would be
proportional to the square of this angular velocity,
with the proportionality constants staying fixed.
Then in order to get $r_K = R_{\odot}$, one would
need to change the Sun's angular velocity
from $\Omega_{\odot}$ to
 \begin{equation}
 \Omega
 = \left({R_{\odot}\over
	 r_{K_{\odot}}} \right)^{1/2}\Omega_{\odot}
 = {8 T_{\odot} \sqrt{GM_{\odot}R_{\odot}}
	 \over 3 c^2 Q_{\odot}}
 = (1.088 \pm 0.061) \times 10^{-8}\: {\rm s}^{-1}
 = {2\pi \over 18.3 \pm 1.0 \: {\rm yr}}. 
 \label{eq:521c}
 \end{equation}
In other words, if the Sun were rotating with a period of greater
than about 18.3 years, then the linear term in the angular velocity
(the relativistic effect linear in the angular momentum $J$)
would dominate over the quadratic term in the angular velocity
(the Newtonian quadrupole effect) at all radii outside the Sun,
and so the corotating circular orbital period would be slightly
increased everywhere outside such a slowly rotating Sun.

	Of course, the numerical result of Eq. (\ref{eq:518})
for the actual radius $r_{K_{\odot}}$,
where Kepler's third law would
be exact for a corotating test body in a circular equatorial orbit
in the gravitational field of the Sun as it is actually rotating,
is entirely hypothetical, since the planets
would exert perturbations on the orbital period
far larger than those of $\Delta P_J$ and $\Delta P_{J_2}$
of Eq. (\ref{eq:514}).
For orbits at $r_{K_{\odot}} \sim 280$ AU or greater,
one might suppose that a reasonable estimate
for some sort of averaged corotating period
in the equatorial plane would be to use Eq. (\ref{eq:514})
but with the solar mass $M_{\odot}$,
angular momentum $J_{\odot}$,
and quadrupole moment $Q_{\odot}$
replaced by the analogous quantities
$M_{SS}$, $J_{SS}$, and $Q_{SS}$ for the entire Solar System.

	Combining the data in \cite{Allen} with that in \cite{RPP}
and in \cite{Pij} gives directly
 \begin{equation}
 GM_{SS}/c^2 = 1.001\,346 \: GM_{\odot}/c^2
 = 1\,478.612 \; {\rm m} = 0.9138 \times 10^{38} \: \ell_P
 \label{eq:522}
 \end{equation}
and
 \begin{equation}
 J_{SS} = 3.148 \times 10^{43} \: {\rm kg \: m}^2 \: {\rm s}^{-1}
 = 2.985 \times 10^{77} \: \hbar = (165.7 \pm 1.3) \: J_{\odot},
 \label{eq:523}
 \end{equation}
which when converted to length and area units gives
 \begin{eqnarray}
 GJ_{SS}/c^3 &=& 7.796 \times 10^7 \: {\rm m}^2
 = (8\,829 \: {\rm m})^2
 = 30.10 \; {\rm mi}^2
 \nonumber\\
 &=& 7\,800 \; {\rm hectares}
 = 19\,260 \; {\rm acres}.
 \label{eq:524}
 \end{eqnarray}

	From these data one can readily calculate that
the Kerr rotational length parameter $a$,
and the corresponding
dimensionless rotation parameter $\alpha$,
take on the values for the entire Solar System of
 \begin{equation}
 a_{SS} \equiv {J_{SS}\over M_{SS}c}
 = 52.72 \: {\rm km} = 3.263 \times 10^{39} \: \ell_P
 = (165.5 \pm 1.3) \: a_{\odot},
 \label{eq:524b}
 \end{equation}
 \begin{equation}
 \alpha_{SS} \equiv {a_{SS}c^2 \over GM_{SS}}
 \equiv {cJ_{SS} \over GM_{SS}^2} = 35.66
 = (165.3 \pm 1.3) \: \alpha_{\odot}.
 \label{eq:524c}
 \end{equation}
The fact that $\alpha_{SS} > 1$,
unlike the case for the Sun,
means that the Solar System
would have to give up angular momentum
(in fact, give up more than 97\% of its angular momentum)
before it could possibly become a black hole.

	By adding up the time-averaged quadrupole moments of
each planet and the Sun,
around their common center of mass, from the data
on the planetary masses (including their moons)
and the semimajor axes and eccentricities of their orbits
in \cite{Allen},
I obtained a quadrupole moment for the Solar System of
 \begin{equation}
 Q_{SS} = 2.576 \times 10^{51} \: {\rm kg \: m}^2
 = 1.23 \times 10^{10} \: Q_{\odot},
 \label{eq:525}
 \end{equation}
of which about 40\% came from Neptune,
23\% came from Saturn,
22\% came from Jupiter,
14\% came from Uranus,
0.67\% came from Pluto,
0.0026\% came from Earth,
0.0011\% came from Venus,
0.00066\% came from Mars,
and 0.000023\% came from Mercury.
There is a positive error in my estimate
from neglecting the fact that the orbits
are not all in the same plane,
and a negative error from neglecting
the quadrupole moment contributions
of the asteroids and comets,
but I have not attempted to estimate these errors.
Almost certainly not all of the four digits given above
are correct, but I have given them just to show
the answer I got for the planets and Sun if their orbits
were coplanar.

	Converting the quadrupole moment of the Solar System
to length units gives
 \begin{equation}
 GQ_{SS}/c^2 = 1.91 \times 10^{24} \: {\rm m}^3
 = (124\,000 \: {\rm km})^3 = 4.53 \times 10^{128} \: \ell_P^3,
 \label{eq:526}
 \end{equation}
The effective quadrupole radius
of the Solar System is then
 \begin{equation}
 r_{Q_{SS}} = \sqrt{2Q_{SS}/M_{SS}}
 = 111\,000 \: r_{Q_{\odot}}
 = 5.09 \times 10^{10} \: {\rm m}
 = 0.340 \: {\rm AU}
 = 3.15 \times 10^{45} \: \ell_P,
 \label{eq:527}
 \end{equation}
meaning that one would get the same quadrupole moment
if one placed all the mass
(of which 99.8656\% comes from the Sun)
into a ring at radius $r_{Q_{SS}} = 0.34$ AU,
about 88\% of the semimajor axis
of the orbit of Mercury \cite{Allen}.
Incidentally, for such a ring to give the angular momentum
of the Solar System, it would have to rotate around with
a period of
 \begin{equation}
 P_{SS} = {4\pi Q_{SS} \over J_{SS}} = 1.03 \times 10^9 \: {\rm s}
 = 32.6 \: {\rm yr}.
 \label{eq:528}
 \end{equation}
The corresponding period
for the 460 km solar-mass ring
that gives the quadrupole moment
of the Sun in Eq. (\ref{eq:507})
is about 14 seconds
to give the solar angular momentum
given in Eq. (\ref{eq:508}).

	As we did for the Sun in Eqs. (\ref{eq:511b}),
we can calculate that for the Solar System
 \begin{equation}
 {Q_{SS}\over J_{SS}^2 / M_{SS} c^2}
 = 4.65 \times 10^{11},
 \label{eq:528b}
 \end{equation}
so the quadrupole moment of the Solar System
is about 465 billion times
larger than that of a Kerr metric
with the same mass and angular momentum.

	Now if we insert these data for the Solar System
in place of the corresponding data for the Sun alone
in Eqs. (\ref{eq:514}) - (\ref{eq:517}),
we get some sort of averaged deviations
from Kepler's third law for very distant orbits
around the Solar System as follows:
 \begin{eqnarray}
 P \equiv {2\pi \over \Omega_{+K}}
 &=& P_{K_{SS}} + \Delta P_{J_{SS}}
 + \Delta P_{Q_{SS}} + O(R^{-3/2})
 \nonumber\\
 &=& 2\pi\sqrt{R^3\over GM_{SS}} \:
 + {2\pi J_{SS}\over M_{SS}c^2}
 - {3\pi Q_{SS} \over 2\sqrt{GM_{SS}^3 R}} + O(R^{-3/2}), 
 \label{eq:529}
 \end{eqnarray}
 \begin{equation}
 P_{K_{SS}} = 2\pi\sqrt{R^3\over GM_{SS}}
 = 31\,536\,986 \: \left({R \over {\rm AU}} \right)^{3/2}
 \: {\rm s},
 \label{eq:530}
 \end{equation}
 \begin{eqnarray}
 \Delta P_{J_{SS}}
 &=& {2\pi J_{SS}\over M_{SS}c^2}
 = 2\pi a_{SS}/c = (331 \: {\rm km})/c
 \nonumber\\
 &=& 1.105 \times 10^{-3} \: {\rm s} = 1.105 \: {\rm ms}
 \label{eq:531}
 \end{eqnarray}
due to the linear effect
of the angular momentum of the Solar System, and
 \begin{equation}
 \Delta P_{Q_{SS}}
 = - {3\pi Q_{SS} \over 2\sqrt{GM_{SS}^3 R}}
 = - 1.367 \times 10^6 \: \left({{\rm AU} \over R} \right)^{1/2}
 \: {\rm s}
 = - 15.82 \: \left({{\rm AU} \over R} \right)^{1/2} \: {\rm days}
 \label{eq:532}
 \end{equation}
due to the quadrupole moment of the Solar System,
e.g., about -2.5 days for the orbit of Pluto
at 39.481\,686\,77 AU \cite{Pluto},
assuming that the effect of the planets is merely
to provide mass, angular momentum,
and quadrupole moment for the Solar System.
This is actually not a very good approximation
for Pluto, since the formula above gives
an orbital period of about 90\,550 days for Pluto,
whereas the sidereal period is actually
90\,465 days \cite{Allen, Pluto}, about 85 days shorter.

	One can then calculate
that the effect of the quadrupole moment
dominates over that linear in the angular momentum for
the Solar System for
 \begin{equation}
 r < r_{K_{SS}}
 = {9c^4 Q_{SS}^2 \over 16 G J_{SS}^2 M_{SS}}
 = 2.29 \times 10^{29} \: {\rm m}
 = 1.53 \times 10^{18} \: {\rm AU}
 = 7.42 \times 10^6 \: {\rm Mpc}. 
 \label{eq:533}
 \end{equation}
Since $r_{K_{\odot}}$ defined by Eq. (\ref{eq:518})
for the Sun alone is larger than the Solar System,
and since the radius $r_{K_{SS}}$
at which Kepler's third law would be exact
for a test body orbiting the entire Solar System
in otherwise flat spacetime
is far larger than the presently observable universe,
we can conclude that for realistic orbits
(i.e., at orbital distances less than that to the next nearest star),
the quadrupole moment of the Sun or of the planets always
dominates over the linear effect of the angular momentum
in changing the period of circular orbits from
the value given by Kepler's third law,
so that the period is always smaller than
that given by Kepler's third law.
(This is under the approximation that the effect
of the planets is merely to give a quadrupole moment to the
Solar System and ignores more complicated effects when
the orbital radii, and hence periods,
of the planets are non-negligible
fractions of the radius and period
of the orbit of the test body.)

	However, we saw in the previous Section
that for certain relativistic polytropic star models
with polytropic index not too large
(not too soft an equation of state),
it is possible to have the linear effect
of the angular momentum dominate
over that of the quadrupole moment,
even at the surface of the star,
so that the corotating Keplerian orbits
at the equatorial surface
of the star can have a longer period
(as seen by a nonrotating observer at infinity)
than what Kepler's third law would give.
As discussed above, this effect
is related to the fact that the angular momentum
of a source generally causes the
Stationary Congruence Accelerating Maximally (SCAM)
to be counterrotating,
so that for a given positive
orbital angular velocity relative
to the SCAM, the angular velocity relative to infinity
is less, giving rise to a longer period.

\section*{Acknowledgments}

	I am grateful for discussions, mostly by e-mail, with
Marek Abramowicz, Brandon Carter, Gregory Cook,
Fernando de Felice, Valeri Frolov, Eric Gourgoulhon,
Pawel Haensel, Werner Israel, John Leibacher,
Frank Pijpers, Fred Rasio, Old\v{r}ich Semer\'{a}k,
and Saul Teukolsky.
Semer\'{a}k corrected a missing 2 in my Eq. (\ref{eq:316})
and sent me a closely related manuscript \cite{Sem97},
enabling me to add nine references and correct two others.
Pijpers kindly sent me his manuscript \cite{Pij}
with the latest solar data.
Trying to summarize my results to my wife Cathy
led me to the Delphic statements of the MAIN theorem.
This work was supported in part by the
Natural Sciences and Engineering Research Council
of Canada.


\end{document}